\documentclass[traditabstract]{aa} 
\usepackage{graphicx,amsmath,mathtools}
\usepackage{txfonts}
\usepackage{natbib}
\usepackage{multibib}
\usepackage{siunitx}
\usepackage{float}

\usepackage{wrapfig}
\usepackage{lscape}
\usepackage{rotating}
\usepackage{epstopdf}
\usepackage{longtable}

\usepackage{footnote}
\makesavenoteenv{tabular}
\makesavenoteenv{table}

\usepackage{tablefootnote}

\usepackage{threeparttable}
\usepackage{multirow}

\bibpunct{(}{)}{;}{a}{}{,} 

\def\gtrsim{\mathrel{\hbox{\rlap{\hbox{\lower4pt\hbox{$\sim$}}}\hbox{$>$}}}}
\def\lesssim{\mathrel{\hbox{\rlap{\hbox{\lower4pt\hbox{$\sim$}}}\hbox{$<$}}}}
\newcommand{\hi}{H{\sc i} }

\newcommand{\hidef}{H{\sc i}-deficient}
\newcommand{\hii}{H{\sc ii} }

\newcommand{\kms}{km s$^{-1}$}

\newcommand{\tmsun}{\text{M}_{\odot}}

\newcommand{\tlsun}{\text{L}_{\odot}}

\newcommand{\arsec}{$^{\prime\prime}\!$}
\newcommand{\micron}{$\mu$m}

\newcommand{\myprime}{$^{\prime}$}

\newcommand{\D}{{$\mathcal D$}}
\newcommand{\hers}{{\it Herschel }}
\newcommand{\iras}{{\it IRAS}}

\newcommand{\mdust}{$M_{d}$ }
\newcommand{\mstar}{$M_{\star}$ }
\newcommand{\rr}{\raggedright}

\defcitealias{2013MNRAS.428.1880A}{A13}
\defcitealias{2005ApJ...631..231P}{PT05}
\defcitealias{2013MNRAS.433.2946W}{W13}

\begin{document}

\title{The {\em Herschel} Virgo Cluster Survey. XVIII. Star-forming dwarf galaxies in a cluster environment}

\subtitle{}

\author{M. Grossi\inst{1,2,3}
\and L. K. Hunt\inst{4}
\and S. C. Madden\inst{5}
\and T. M. Hughes\inst{6}
\and R. Auld\inst{7}
\and M. Baes\inst{6}
\and G. J. Bendo\inst{8}
\and S. Bianchi\inst{4}
\and L. Bizzocchi\inst{1,9}
\and M. Boquien\inst{10}
\and A. Boselli\inst{11}
\and M. Clemens\inst{12}
\and E. Corbelli\inst{4}
\and L. Cortese\inst{13}
\and J. Davies\inst{7}
\and I. De Looze\inst{6}
\and S. di Serego Alighieri\inst{4}
\and J. Fritz\inst{6}
\and C. Pappalardo\inst{1,2,3}
\and D. Pierini\inst{14}
\and A. R\'emy-Ruyer\inst{5}
\and M. W. L. Smith\inst{7}
\and J. Verstappen\inst{6}
\and S. Viaene\inst{6}
\and C. Vlahakis\inst{15}
}
\institute{
Centro de Astronomia e Astrof\'isica da Universidade de Lisboa, OAL, Tapada da Ajuda, PT1349-018, Lisbon, Portugal\\
\email{grossi@oal.ul.pt}
\and
Instituto de Astrof\'isica e Ci\^{e}ncias do Espa\c{c}o, Universidade de Lisboa, OAL, Tapada da
Ajuda, PT1349-018 Lisbon, Portugal
\and
Departamento de F\'isica, Faculdade de Ci\^{e}ncias, Universidade de Lisboa, Edif\'icio C8, Campo Grande, PT1749-016 Lisbon, Portugal
\and
INAF-Osservatorio Astrofisico di Arcetri, Largo Enrico Fermi 5, 50125 Firenze, Italy
\and
Laboratoire AIM, CEA/DSM - CNRS - Université Paris Diderot, Irfu/Service d'Astrophysique, CEA Saclay, 91191 Gif-sur-Yvette, France
\and
Sterrenkundig Observatorium, Universiteit Gent, Krijgslaan 281, B-9000 Gent, Belgium
\and
School of Physics and Astronomy, Cardiff University, Queens Buildings, The Parade, Cardiff CF24 3AA, UK
\and
UK ALMA Regional Centre Node, Jodrell Bank \\
Centre for Astrophysics, School of Physics and Astronomy, University of
Manchester, Oxford Road, Manchester M13 9PL, UK
\and
Center for Astrochemical Studies, Max-Planck-Institut f{\"u}r extraterrestrische Physik (MPE),
Giessenbachstra{\ss}e, 85748 Garching, Germany
\and
Institute of Astronomy, University of Cambridge, Madingley Road, Cambridge CB3 0HA, UK
\and
Laboratoire d'Astrophysique de Marseille, UMR 6110 CNRS, 38 rue F. Joliot-Curie, F-13388 Marseille, France
\and
INAF-Osservatorio Astronomico di Padova, Vicolo dell'Osservatorio 5, I-35122 Padova, Italy
\and
Centre for Astrophysics and Supercomputing, Swinburne University of Technology, Mail H30 - PO Box 218, Hawthorn, VIC 3122, Australia
\and
Max-Planck-Institut f{\"u}r extraterrestrische Physik (MPE), Postfach 1312,
Giessenbachstra{\ss}e, 85748 Garching, Germany
\and
Joint ALMA Observatory / European Southern Observatory, Alonso de Cordova 3107, Vitacura, Santiago, Chile
}

   \date{}

\abstract{
To assess the effects of
the cluster environment on the different components of the interstellar
medium,
we analyse the far-infrared (FIR) and submillimetre (submm) properties of a sample of star-forming dwarf galaxies
detected by the \hers Virgo Cluster Survey (HeViCS).
We determine dust masses and dust temperatures by fitting a modified
black body function to the spectral energy distributions (SEDs).
Stellar and gas masses, star formation rates (SFRs), and metallicities are obtained
from the analysis of a set of ancillary data.
Dust is detected in 49 out of a total 140 optically identified dwarfs covered by the
HeViCS field; considering only dwarfs brighter than $m_B = 18$ mag,
this gives a detection rate of 43\%.
After evaluating different emissivity indices, we find
that the FIR-submm SEDs are
best-fit by $\beta\,=\,1.5$, with a median dust temperature 
$T_d$\,=\,22.4\,K.  Assuming $\beta\,=\,1.5$,
67\% of the 23 galaxies detected in all five \hers\ bands show
emission at 500 $\mu$m in excess of the modified black-body model.
The fraction of galaxies with a submillimetre excess decreases for lower values
of $\beta$, while a similarly high fraction (54\%) is found if a $\beta$-free SED modelling is applied. The excess is
inversely correlated with SFR and stellar masses.
To study the variations in the global properties of our sample that come from
environmental effects, we compare the Virgo dwarfs to other \hers surveys,
such as the Key Insights into Nearby Galaxies: Far-Infrared Survey
with \hers\ (KINGFISH), the Dwarf Galaxy Survey (DGS), 
and the HeViCS bright galaxy catalogue (BGC).
We explore the relations between stellar mass and \hi fraction, specific star formation rate, dust fraction, gas-to-dust ratio over a wide range of stellar masses (from $10^{7}$ to $10^{11}$ M$_{\odot}$) for both dwarfs and spirals.
Highly \hidef\ Virgo dwarf galaxies are mostly characterised by quenched star formation activity and lower dust fractions
giving hints for dust stripping in cluster dwarfs. However,
to explain the large dust-to-gas mass ratios observed in these systems, we find
that the fraction of dust removed has to be less than that of the \hi component.
The cluster environment seems to mostly affect the gas component and star formation
activity of the dwarfs.
Since the Virgo star-forming dwarfs are likely to be
crossing the cluster for the first
time, a longer timescale might be necessary to strip the more centrally
concentrated dust distribution.}

\titlerunning{The {\em Herschel} Virgo Cluster Survey XVIII}
\authorrunning{M. Grossi et al.}

\maketitle

\section{Introduction}

Dust, gas, and star formation activity are tightly linked in galaxies,
implying that detailed investigation of
these components and of their mutual relation is fundamental for our understanding of galaxy evolution.
It is known that one of the main roles of dust in the star formation cycle of galaxies is the formation
of molecular hydrogen
\citep{1963ApJ...138..393G,1971ApJ...163..155H}.
As galaxies form stars, their interstellar medium (ISM) becomes
enriched in dust, and galaxies with a higher star formation
rate are found to host a more massive dust component \citep{2010MNRAS.403.1894D}.
Dust is observed to be well mixed with gas
\citep{1978ApJ...224..132B,1996A&A...312..256B},
and dust formation models show that the dust-to-gas ratios, \D, 
should be tied to the oxygen abundance of a galaxy
\citep{1998ApJ...501..643D,2013ApJ...777....5S}.

It is still not clear, however, how dust properties and their link with gas and star formation activity vary
when we consider galaxies in a dense cluster, where external perturbations can affect the ISM content and
star formation activity.
Indeed the evolution of galaxies in clusters is driven by interactions between their ISM and the surrounding environment:
ram pressure stripping \citep{1972ApJ...176....1G,2000Sci...288.1617Q,2007ApJ...671.1434T}, harassment
\citep{1996Natur.379..613M,1998ApJ...495..139M}, tidal interactions
\citep{2004MNRAS.349..357B}, and strangulation \citep{1980ApJ...237..692L,2008ApJ...672L.103K} are among the processes
that can be responsible for removing the ISM and quenching star formation.
Studies of nearby rich clusters have shown that ram pressure stripping can be the dominant transformation process
of star-forming galaxies into quiescent systems
\citep{2005AJ....130...65C,2006PASP..118..517B,2013A&A...553A..89G}.

It is well established  that late-type galaxies in dense environments tend to have less \hi than their field
counterparts and that there is an anticorrelation between the \hi deficiency and the distance to the
cluster centre \citep{1983ApJ...267...35G,1984AJ.....89..758H,2009AJ....138.1741C}.
On the other hand, it is
debated whether this is not also true for the molecular gas component that is usually
more centrally concentrated \citep{2009ApJ...697.1811F,2012A&A...545A..75P,2014A&A...564A..67B} and for
the dust that is supposed to be more closely linked to
the molecular than to the atomic gas phase.
Before the launch of the \hers Space Observatory \citep{2010A&A...518L...1P},
the influence of the environment on the removal of dust in H{\sc i}-deficient spirals
has been addressed in studies using observations with both the Infrared Astronomical Satellite \citep[{\em IRAS},][]{1989MNRAS.239..347D} and the Infrared Space Observatory
 \citep[{\em ISO},][]{2006PASP..118..517B}.
However, the small number of studied objects and the lack of an unperturbed reference sample prevented drawing
conclusions on dust stripping in high-density environments.
Only recent observations with \hers were able to show that dust can
be stripped from Virgo cluster galaxies \citep{2010A&A...518L..49C,2010A&A...518L..45G}, providing conclusive evidence that it is significantly reduced in the discs of very \hidef\ cluster spirals \citep{2012A&A...540A..52C,2012A&A...542A..32C}.

The Virgo cluster, at a distance of approximately 17 Mpc \citep{1999MNRAS.304..595G,2007ApJ...655..144M} and comprising $\sim$ 1300 confirmed members \citep{1985AJ.....90.1681B}, is indeed the nearest example of a high-density environment.
It contains about two hundred star-forming dwarf (SFD) galaxies  --
i.e. classified as Sm, Im, and blue compact dwarfs (BCDs) according to the Virgo Cluster Catalogue \citep{1985AJ.....90.1681B} and
GOLDMine \citep{2003A&A...400..451G,2014arXiv1401.8123G}.
Because of their lower gravitational potentials and less dense ambient ISM \citep{2008ApJ...686..948B}, dwarfs
are more sensitive to their surroundings than more massive galaxies, which makes them excellent targets
for investigating the environmental effects on a weakly bound ISM \citep{2008ApJ...674..742B}.

Through the \hers Virgo Cluster Survey
\citep[HeViCS;][]{2010A&A...518L..48D,2012MNRAS.419.3505D},
a \hers Open Time Key Project
that covers $\sim$ 80 square degrees of the Virgo cluster from
100 $\mu$m to 500 $\mu$m, we present an analysis of the far-infrared (FIR) and submillimetre
 (submm) observations of a sample of SFDs in this cluster. We discuss their FIR properties, the relation between dust and other global galaxy parameters (i.e. stellar mass, star formation rate, and gas content), and analyse the effects of the environment on the dust component.

Previous Virgo surveys with {\em IRAS} \citep{1984ApJ...278L...1N} and {\em ISO} \citep{1996A&A...315L..27K} also targeted the SFD population. About one third of the cluster BCDs were detected at 60 and 100 $\mu$m; their dust content, compared to their stellar and gas masses, is only a factor 2 to 3 smaller than normal spiral galaxies. The warm \iras\ colours also suggested that
the FIR luminosity was dominated by the emission from star-forming regions \citep{1989ApJ...339..812H}.
\citet{2002ApJ...567..221P} and \citet{2002ApJS..139...37T} analysed a small sample of late-type Virgo galaxies
including irregulars and BCDs with ISOPHOT, finding very cold dust temperatures (a median value 
of 15.9 K), and extended dust distributions similar to the size of the \hi discs.
However,
given the small number of objects investigated, the lack of coverage beyond 200 $\mu$m where cold dust emission is predominant, and the large beam size of the ISOPHOT instrument at 170 $\mu$m (FWHM  $\sim$ 1\arcmin),
further investigations over a larger sample and a
wider spectral coverage is required to better assess the dust content of Virgo SFDs.

The paper is organised as follows.
In Section 
\ref{sec:herschel_obs}
we briefly describe the HeViCS survey observations and data reduction,
and in Section \ref{sec:sample_and_phot_main} the sample selection and the photometry. In Section \ref{sec:comparison_sample} we present the samples that
will be used as a comparison throughout the paper: 1)
the Key Insights into Nearby Galaxies: Far-Infrared Survey with \hers
\citep[KINGFISH,][]{2011PASP..123.1347K,2012ApJ...745...95D}; 2) the Dwarf Galaxy Survey
\citep[DGS;][]{2013PASP..125..600M,2013A&A...557A..95R}, both targeting systems in lower density environments; 3) the brightest galaxies in the HeViCS survey \citep{2012MNRAS.419.3505D}.
We list the ancillary data available in the literature for all these surveys in Section \ref{sec:ancillary}.
In Section \ref{sec:mbb}
we analyse the FIR-submm SEDs of the detected Virgo SFDs, and infer dust temperatures and masses, using different values for the emissivity index $\beta$. The properties of FIR-detected and FIR-undetected Virgo SFDs are compared in Section \ref{sec:det_vs_nondet}.
The presence of a submm excess emission at 500 $\mu$m is discussed in Section \ref{sec:excess}.
The global properties of the ISM and dust-scaling relations are investigated in Section \ref{sec:probe},
comparing Virgo SFDs to the other \hers surveys.
Finally, in Section \ref{sec:conclusions} we summarise our conclusions.

\section{{\em Herschel} observations}
\label{sec:herschel_obs}

The HeViCS survey consists of four fields with a size of $\sim$ 4$^{\circ} \times 4^{\circ}$ each, covering the main structures of the cluster:
the M87 and M49 subgroups, the W, W$^{\prime}$, and M clouds
\citep{1987AJ.....94..251B,2007ApJ...655..144M}. 
\hers Photodetecting Array Camera and Spectrometer \citep[PACS;][]{2010A&A...518L...2P} and
Spectral and Photometric Imaging Receiver \citep[SPIRE;][]{2010A&A...518L...3G} observations of Virgo were taken between December 2009 and
June 2011. A more detailed description  of the observing strategy and data reduction process is given in the HeViCS overview
and catalogue papers \citep[][hereafter A13]{2012MNRAS.419.3505D,2013MNRAS.428.1880A}, and a brief summary of the main steps followed are given below.

\hers observations were carried out using the SPIRE/PACS parallel scan-map mode with a fast scan speed
of 60\arsec/sec over two orthogonal crossed-linked scan directions.
A total of 8 scans was then obtained for each field, with overlapping regions
between the four tiles being covered by 16 scans.

Regarding the PACS data release, we used a more recent version compared to that described in
\citetalias{2013MNRAS.428.1880A}. Data at 100 and 160~$\mu$m were reduced within the {\em Herschel} Interactive Processing Environment \citep[version 11.0;][]{2010ASPC..434..139O}, and maps were created with the {\em Scanamorphos} task \citep[version 23;][]{2013PASP..125.1126R} with a pixel size of 2$^{\prime\prime}$ and 3$^{\prime\prime}$, respectively. 
The angular resolution for PACS in fast scan parallel mode is 9\farcs4 and 13\farcs4,
at 100, and 160 $\mu$m, respectively.
Maps attain noise levels of 1.9 and 1.2 mJy pixel$^{-1}$ which decrease to 1.3 and 0.8 mJy pixel$^{-1}$ in the regions covered by 16 scans.
A calibration uncertainty of 5\% is assumed for both 100 and 160 $\mu$m channels \citep{2013ExA...tmp...38B}.

SPIRE data reduction was carried out up to Level 1 adapting the standard pipeline ({\em POF5 pipeline.py}, dated 8 Jun
2010) provided by the SPIRE Instrument Control Service \citep{2010A&A...518L...3G,2010SPIE.7731E.101D},
while temperature drift correction and residual baseline subtraction were performed using the BriGAdE method \citep{SmithPhd12}.
Final maps were created with the naive mapper provided by the standard pipeline (\texttt{naiveScanmapper} task in HIPE v9.0.0), with pixel sizes of  6\arsec,
8\arsec, and 12\arsec$\,$ at 250, 350, and 500 $\mu$m, respectively.
The global noise level in the SPIRE images is 4.9, 4.9, and 5.7 mJy beam$^{-1}$
\citepalias[at 250, 350, 500 $\mu$m;][]{2013MNRAS.428.1880A}.
The calibration uncertainty for SPIRE flux densities is 6\% for each band\footnote{http://herschel.esac.esa.int/twiki/bin/view/Public/SpireCalibrationWeb}, and
the beam size full width at half maximum (FWHM) in the three channels is 17\farcs6, 23\farcs9,  and
35\farcs2.

Given the new analysis of the SPIRE beam profile
we adopt the revised beam areas of
465.4, 822.6, 1768.7 square arcseconds (SPIRE Handbook, version 2.5)\footnote{herschel.esac.esa.int/Docs/SPIRE/spire\_handbook.pdf}
to derive flux densities at SPIRE wavelengths.
We applied the updated $K_{PtoE}$ conversion factors to optimise the data for extended source photometry, i.e. 91.289, 51.799, 24.039  MJy sr$^{-1}$ (Jy beam$^{-1}$)$^{-1}$, as indicated in the SPIRE handbook,
and the latest calibration correction factors (1.0253$\pm$0.0012, 1.0250$\pm$0.0045, and 1.0125$\pm$0.006 at 250, 350, and 500 $\mu$m, respectively.)

\begin{figure}[h!]
   \centering
 \includegraphics[bb=0 -40 535 270,width=8cm]{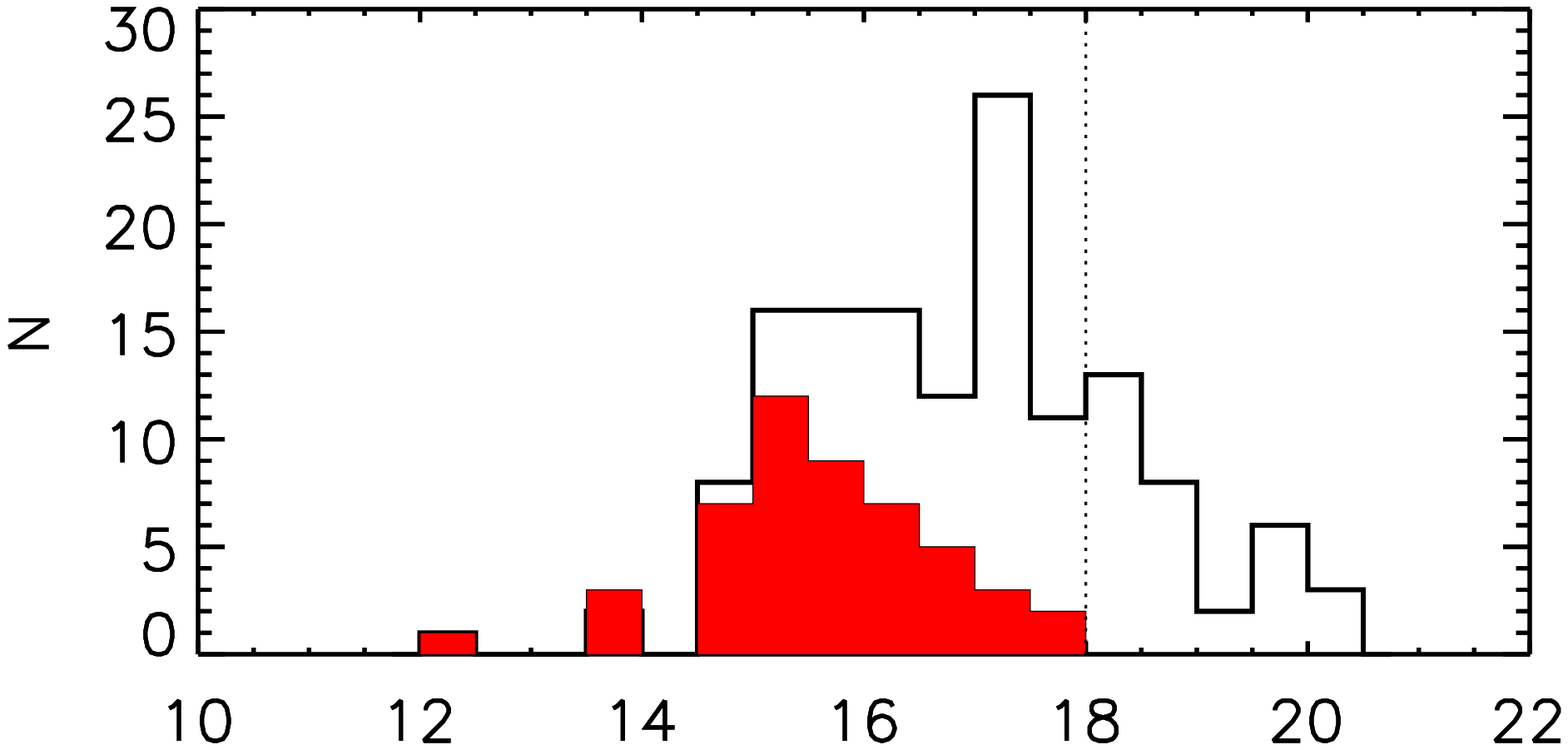}
 \includegraphics[bb=70 5 400 555,width=8cm]{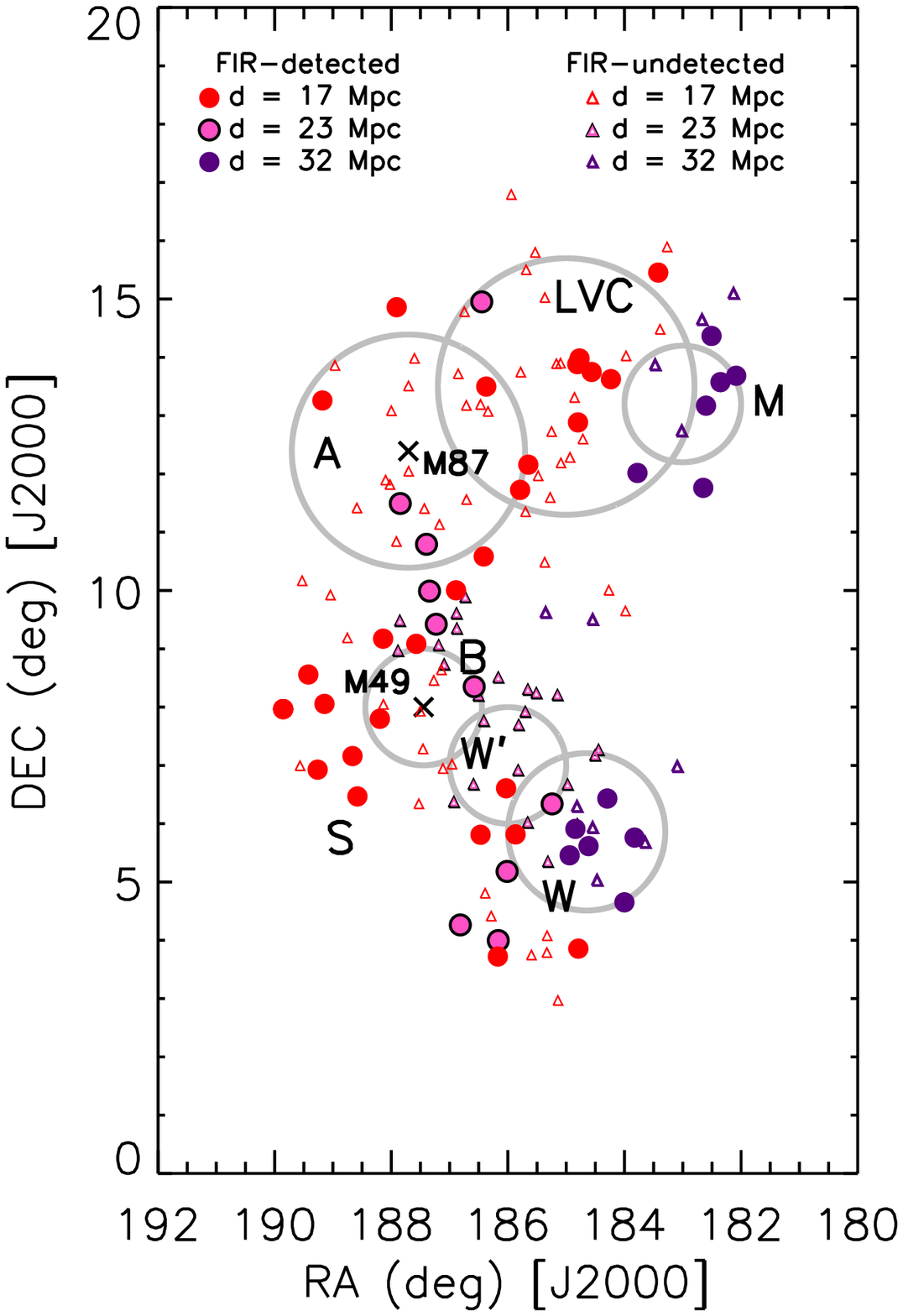}
 \caption{{\em Upper panel:} Distribution of apparent $B$ magnitudes of the Virgo Sm, Im, and BCD galaxies in the four HeViCS fields. The red filled histogram shows the galaxies detected by {\em Herschel}.
 The dotted line corresponds to the completeness limit of the VCC catalogue. {\em Lower panel:} Spatial distribution of the Sm, Im, and BCD galaxies in the four HeViCS fields. Grey circles
 show the main substructures within the cluster. Filled red,  pink, and purple dots indicate FIR detections in at least two \hers bands at distances of 17, 23, and 32 Mpc, respectively, that will be analysed in this work. Triangles with the same colour codes
 correspond to FIR non-detections at the three distance ranges.}
   \label{fig1}%
    \end{figure}

\section{Virgo star-forming dwarfs: sample selection and photometry}
\label{sec:sample_and_phot_main}

\subsection{Sample selection}
\label{sec:sample}

\begin{figure*}
   \centering
 \includegraphics[bb=60 40 530 560,width=13cm]{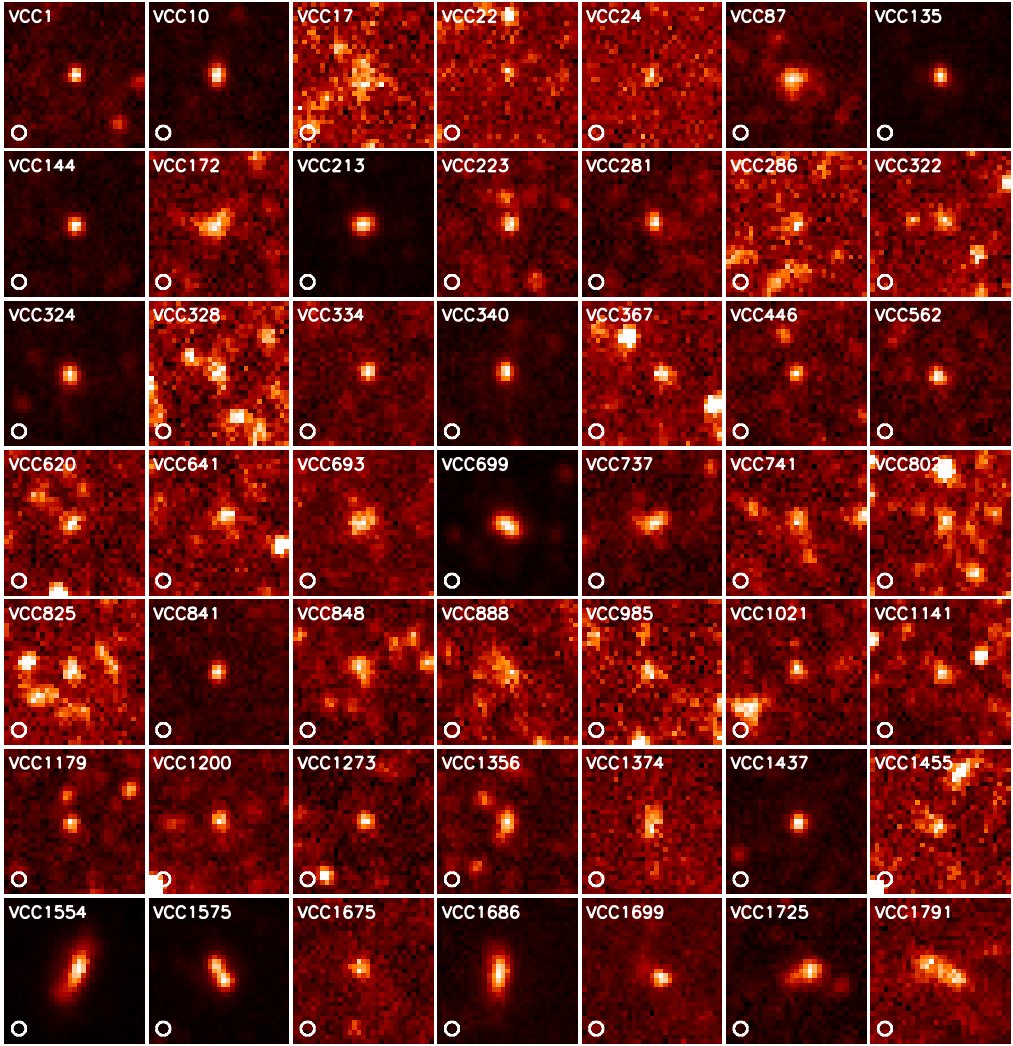}
 \caption{250 \micron\ image stamps of the sample of star-forming dwarf galaxies detected by HeViCS. The field size is 180$^{\prime\prime}$. The SPIRE beam size at 250 $\mu$m is shown at the lower-left corner of each image stamp.}
   \label{250_stamps}%
    \end{figure*}

The HeViCS fields contain 140 galaxies classified in the Virgo Cluster Catalogue \citep[VCC;][]{1985AJ.....90.1681B} and in the GOLDMine database \citep{2003A&A...400..451G,2014arXiv1401.8123G}
as Sm, Im, BCD/dIrr\footnote{This corresponds to GOLDMine morphological types from 11 to 17.} with radial velocity $V < 3000$~km s$^{-1}$.
The galaxies span a varied range of $B$ magnitudes and radial velocities.
$B$ magnitudes of the selected objects are between 12 and 21~mag ({upper panel} of Fig. \ref{fig1}),
the radial velocity distribution of the galaxies extends from -200~\kms\ to 2700~\kms. Virgo SFDs are spread along the different substructures within the cluster: a)
the main body of the cluster  centred on the cD galaxy M87 (cluster A, $V \sim$ 1100~\kms); b) the smaller subcluster  centred on the elliptical
galaxy M49 roughly at the same distance as M87 (cluster B, $V \sim$ 1000~\kms); c) the so-called {\em low-velocity cloud} (LVC), a subgroup of galaxies at $V \lesssim 0$~\kms\ superposed to the M87 region which is thought
to be infalling towards the cluster core from behind \citep{1989ApJ...339..812H}; d) the Virgo Southern extension (S), a filamentary structure that extends to the south of the cluster; e) the W and M clouds, to the southwest and to the northwest of
the cluster core respectively,  at roughly twice the distance of M87 \citep[$V \sim$ 2200~\kms; ][]{1984ApJ...282...19F,1987AJ.....94..251B}; the W\myprime\ cloud, a substructure which connects
the W cloud to the M49 subgroup.
Following GOLDMine we assume three main values for the distances to the objects of the sample:
17 Mpc,  whether they belong to the M87 and M49 subclusters, the LVC, and Virgo Southern extension; 23 Mpc for the W\myprime\ cloud and the substructure rich in late-type galaxies between cluster A and B; 32 Mpc for galaxies in the M and W clouds.
We note that distance assignment to individual objects of the Virgo cluster can be highly uncertain, and according to \citet{2005A&A...429..439G}
errors on distances to Virgo members can be as high as 30\%.

The distribution of SFD galaxies within the cluster and the substructures at larger distances is shown in Fig. \ref{fig1}.
Galaxies in each subgroup are probably at different stages of interaction with the surrounding environment,
and it is likely that a fraction of the SFDs at $d \sim$ 17 Mpc
are entering the cluster for the first time \citep{1993A&AS...98..275B,2002A&A...396..449G,2003AJ....126.2774H}.

It is important to note that the W\myprime, W, and M structures are outside the virial radius of the cluster and represent an intermediate density environment between the cluster and the field.
Significant \hi deficiencies were
identified in galaxies even at large distances from the Virgo core, well
beyond the extension of the hot X-ray intracluster medium \, mainly in correspondence with the W\myprime\ and W clouds \citep{2002AJ....124.2440S}.
However, in a recent analysis of the \hi content of Virgo late-type galaxies, \citet{2013A&A...553A..89G} reported that
the W and M cloud population do not appear to have a large atomic hydrogen deficit.

\subsection{{\em Herschel} detections: SPIRE/PACS photometry}
\label{sec:photometry}

Within the initial sample of 140 SFDs, 57 have a FIR-submm detection in the HeViCS catalogue \citepalias{2013MNRAS.428.1880A}, with at least one detection in one \hers band with $S/N > 3$.
Because we used an updated release of the PACS maps compared to that in
\citetalias{2013MNRAS.428.1880A},
we remeasured the photometry at PACS wavelengths. We also recalculated the photometry at 250, 350, and 500 $\mu$m in order to have an homogeneous set of measurements obtained with the same method, despite having used the same data release as \citetalias{2013MNRAS.428.1880A}.

Flux densities of extended sources were measured through elliptical apertures defined on the basis of the isophotal semi-major ($a_{25}$) and semi-minor axis ($b_{25}$) at the 25th $B$-magnitude arcsec$^{-2}$, which were taken from the GOLDMine database. Apertures were chosen to be $\sim$1.5 times the optical radii. For a few galaxies (VCC1, VCC24, VCC322, VCC1021, VCC1179, VCC1200, VCC1273), smaller apertures were adopted ($\sim$1.0 times the optical radii).
For the most compact dwarfs, i.e. with $a_{25}$ smaller or comparable to the \hers\ resolution at 500 $\mu$m, we used circular apertures with 30\arsec\ radii (VCC22, VCC223, VC281, VCC334, VCC367, VCC1141, VCC1437).
These same apertures
were applied to derive flux densities at all wavelengths. However, to measure PACS 100 $\mu$m photometry we tended to
use smaller apertures (by a factor $\sim$0.65) because of the reduced extent of the dust emission at this wavelength compared to the stellar disc (see also Table \ref{tab:fir_fluxes}). This choice allowed us to prevent an artificial increase of the error associated with our measurement.
The background was measured following the approach of \citetalias{2013MNRAS.428.1880A},
i.e. the estimate was achieved with a 2-D polynomial fit over an area of 180\arcsec\ around the aperture defined to extract the galaxy emission, after having masked the galaxy. Following \citetalias{2013MNRAS.428.1880A}, a
fifth order polynomial was used to determine the background in SPIRE images, while a second order polynomial was sufficient for PACS data. To reduce the contribution of possible contaminating sources
a 95\% flux clip was applied before estimating the background.
As a comparison we also estimated the background in fixed annuli with a 60\arcsec\ width, and found that on average we obtained a better curve of growth convergence with the 2-D polynomial fit. The difference in the final flux densities between the two methods is less than 10-15\%, which is close to the relative error at all wavelengths.

Uncertainties were calculated following \citet{2012A&A...543A.161C}, adding in quadrature
the instrumental error, $\sigma_{inst}$, the sky background error, $\sigma_{sky}$, the confusion noise due to the presence of faint background sources, $\sigma_{conf}$ \citep[calculated only for SPIRE images;][]{2010A&A...518L...5N}, and the error on the calibration, $\sigma_{cal}$, assumed to be 5\% and  6\% for PACS and SPIRE channels, respectively (see 
Sect. ~\ref{sec:herschel_obs}):

\begin{equation}\label{eq:toterr}
\sigma_{tot} = \sqrt{ \sigma_{inst}^2 + \sigma_{sky}^2 + \sigma_{conf}^2 + \sigma_{cal}^2}
\end{equation}

The instrumental error, $\sigma_{inst}$, depends on the number of scans crossing a pixel, and it
was obtained by summing in quadrature the values on the error map provided by the pipeline within the chosen aperture.
The sky background error, $\sigma_{sky}$, results from the combination of the uncorrelated
uncertainty on the mean value of the sky ($\sigma_{skypix}$ i.e., the pixel-to-pixel
variation in the region where we derived the sky background), and the correlated noise ($\sigma_{skymean}$)
due to
large scale structures present in the image such as the Galactic cirrus \citep{2012A&A...543A.161C,2013PASP..125.1126R,2014MNRAS.440..942C}.
To estimate $\sigma_{skymean}$ we defined 24 apertures around each galaxy with the same number of pixels $N$  used to measure the galaxy flux density, and we calculated the standard deviation of the mean value of the sky. The sky background uncertainty was then given by

\begin{equation}
\sigma_{sky} = \sqrt{N \sigma_{skypix}^2 + N^2 \sigma_{skymean}^2}
\end{equation}

The confusion noise $\sigma_{conf}$ was determined using Eq. 3 of \citet{2012A&A...543A.161C} and the estimates given by \citet{2010A&A...518L...5N}.

\begin{figure}[h]
 \includegraphics[bb=-20 0 370 480,width=7.5cm]{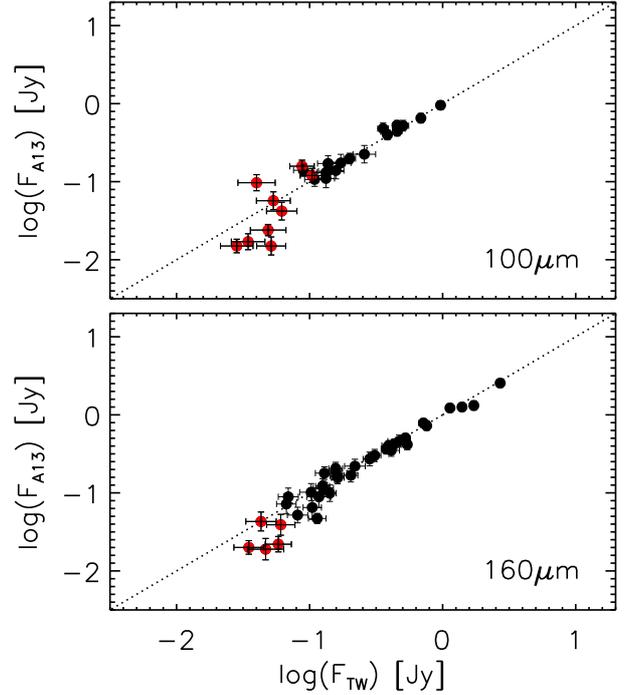}
   \caption{Comparison between PACS 100 and 160 $\mu$m flux densities derived in this work ($F_{TW}$) and in A13 ($F_{A13}$). The one-to-one relation is given by the dotted line. At $S/N > 5$ there is good agreement between our values and A13 at both wavelengths (black dots). Red dots show lower $S/N$ detections ($3 < S/N < 5$).
   }
   \label{fig:compare_PACS}%
    \end{figure}

For some of the dwarfs the extent of the emission at 500 $\mu$m is comparable to the FWHM of SPIRE and appear as marginally resolved. The flux density of point-like sources can be extracted directly from the timeline data using a PSF fitting method \citep{2013MNRAS.433.3062B}. This method provides a more reliable estimate than the aperture photometry technique of unresolved sources \citep{2013ExA...tmp...37P}, especially in the case of faint detections ($\sim$ 20 -- 30 mJy). To check whether some of the dwarfs of our sample could be treated as point-like sources we cross-correlated our list of detected galaxies with the HeViCS point-source catalogue \citep{Pappalardo:submitted}, finding 21 matches.
For these objects flux densities were estimated with a timeline-based point source fitter that fits a Gaussian function to the timeline data. More detail about the catalogue and the source extraction technique can be found in \citep{Pappalardo:submitted}.
Errors on the flux densities of point sources were determined directly from the timeline fitting technique.

We decided to include in our final sample only galaxies with at least a detection in two bands with $S/N > 3$,
with a total of
{\em 49} objects satisfying this criterion. Compared to the A13 catalogue we do not take into account the following galaxies: VCC309, VCC331, VCC410, VCC793, VCC890, VCC1654, VCC1750, because they have a detection in only one band. 
We also rejected VCC83 and VCC512 because of possible contamination from background galaxies which may affect the correct
assessment of the FIR-submm flux densities. Finally we added to the list of detections VCC367 which appears to be missing from the A13 catalogue.
{\em Herschel}/SPIRE cut-out images of the final sample at 250~$\mu$m are shown in Fig. \ref{250_stamps}.

Comparison with PACS photometry derived in A13 (Fig. \ref{fig:compare_PACS}) shows a good agreement between our and previous measurements at least for sources with $S/N > 5$ (black dots). At lower $S/N$, and especially at 100 $\mu$m, there is a larger discrepancy.
This could be due to both the better performances of {\em Scanamorphos} compared to HIPE at preserving low level flux densities, and to our choice of using apertures smaller than 1.4 times the optical extent of the galaxy to reduce the contribution of the background to the measured flux densities of low $S/N$ 100 $\mu$m detections.

\subsection{Stacking of non-detections}
\label{sec:stacking}

\begin{figure}[h!]
   \centering
 \includegraphics[bb= 50 25 570 570, width=6cm]{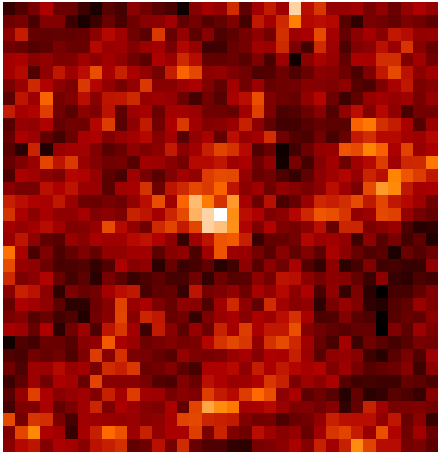}
   \caption{Mean stacked image at 250 $\mu$m of 64 dwarf galaxies with $m_B < 18$ undetected by HeViCS. The image has a rms of 0.85 mJy/beam. The 3.5$\sigma$ detection at the centre has
   a flux density of $\sim$ 4 mJy.}
   \label{fig:stacking}%
    \end{figure}

The mean FIR emission of the undetected galaxy population can be explored to deeper levels by stacking 250 \micron\
images of the dwarfs at their optical positions.
Among the FIR non-detections we selected galaxies with $m_B < 18$ mag, according to the completeness limit of the VCC catalogue. We excluded VCC169 and VCC217 because they were too close to the edges of the HeViCS map, and four objects showing nearby background sources which could affect the result of the stacking process (VCC83, VCC168, VCC468, and VCC512). The final list of undetected galaxies to stack includes 64 dwarfs.
For each sub-image with a size of 50$\times$50 pixels we computed the root mean square (rms) with iterative sigma clipping, and masked all sources above 5$\sigma$ in the region outside a circular aperture of 5 pixel radius (30\arcsec) around the position of the galaxy. Then we derived the mean of each pixel weighted by the square of the inverse of the background rms of the corresponding sub-image.
The rms of the stacked image, shown in Fig. \ref{fig:stacking} is 0.85 mJy/beam, about 8 times lower than the mean rms of the 64 input sub-images (6.7 mJy/beam).
This offers a
significant improvement over the original data set, giving evidence for a 3.5$\sigma$ detection with a flux density of 4.5 mJy within a circular aperture of 4 pixel radius.
For comparison we repeated the same procedure median combining the images without masking the brighter sources scattered around the sub-images, and obtained a slightly higher rms (1.0 mJy/beam) with a final $S/N$ ratio of 3.1.

We estimate the average dust mass of undetected galaxies in Sect. ~\ref{sec:dustmass} and we discuss their properties in Sect. ~\ref{sec:det_vs_nondet}.

\section{Selection of comparison samples}
\label{sec:comparison_sample}

To assess the effects of the cluster environment on the dust content of the dwarf galaxies in Virgo we use, as a comparison sample, dwarfs extracted from other {\em Herschel} surveys targeting lower density environments.

The Dwarf Galaxy Survey \citep[DGS,][]{2013PASP..125..600M} is a
photometric and spectroscopic survey of 50 dwarf galaxies, which aims at studying the gas and dust properties in low-metallicity systems. Among these galaxies we selected a subset of objects which have been detected by {\em Herschel} in at least three bands (100, 160, and 250 $\mu$m), so that we can determine dust temperatures and masses in the same way as we have done for the Virgo galaxies (see Sect. ~\ref{sec:dustmass}). Haro11 was excluded from the final list because its properties are remarkably different from our sample of Virgo dwarfs, being a merger with a SFR of tens of solar masses per year. Therefore the final subset of selected DGS galaxies includes 27 objects. {\em Herschel} photometry
for this sample was taken from \citet{2013A&A...557A..95R}. To take into account the updated SPIRE calibration we multiplied their flux densities for the correction factors given in Sect.  \ref{sec:herschel_obs}.

KINGFISH is an imaging and spectroscopic survey of 61 nearby (d $<$ 30 Mpc) galaxies, chosen to cover a wide range of morphological types and ISM properties \citep{2011PASP..123.1347K}.
Among the 61 KINGFISH objects, there are 12 Irregular/Magellanic-type (Im/Sm) galaxies, and 39 spirals ranging from early to late types, that we will use throughout the rest of this work. We used flux densities given by \citet{2012ApJ...745...95D}, corrected for the revised SPIRE beam areas and calibration, and we applied the updated $K_{PtoE}$ conversion factors as we did for the HeViCS data (Sect. ~\ref{sec:herschel_obs}).

Finally, to compare the properties of low-mass systems to the more massive galaxies within Virgo we include to our list of comparison samples 68 spiral galaxies (from Sa to Sd) from the HeViCS Bright Galaxy Catalogue \citep[BGC,][]{2012MNRAS.419.3505D}. FIR-submm photometry was taken from \citetalias{2013MNRAS.428.1880A} and corrected for the updated SPIRE beam sizes and calibration (see Sect.  \ref{sec:herschel_obs}).

\section{Ancillary data and analysis}
\label{sec:ancillary}

We have assembled several sets of additional data in order to
derive other properties of the Virgo SFDs and the comparison samples.
These include stellar masses, atomic gas masses, star formation rates, and gas metallicities
which will be incorporated in the subsequent analysis together with
dust masses to better assess the effect of environment.

\begin{table}
\caption{Coefficients used to derive stellar masses in Equation \ref{eq:Wen13}.}
\label{Tab:W1_coeff}
\begin{tabular}{p{3cm}cc}
\hline \hline
Sample & $a$ & $b$ \\
\hline \hline
\rr Virgo SFDs, DGS, KINGFISH dwarfs & \multirow{2}{*}{0.779$\pm$ 0.002} & \multirow{2}{*}{1.019 $\pm$ 0.001} \\ \cline{1-1} \hline 
\rr HeViCS BGC, KINGFISH spirals & \multirow{2}{*}{0.679$\pm$ 0.002} & \multirow{2}{*}{1.033 $\pm$ 0.001} \\ \cline{1-1}
\hline \hline
\end{tabular}
\end{table}

\subsection{Stellar masses}
\label{sec:ancillary_stellmass}

\subsubsection{Virgo SFDs}
\label{subsec:ancillary_stellmass_Virgo_SFD}

Stellar masses were calculated following the approach of
\citet[][hereafter W13]{2013MNRAS.433.2946W}
which is based on 3.4~\micron\ photometry from
the Wide-field Infrared Survey Explorer ({\em WISE})  all-sky catalogue \citep{2010AJ....140.1868W}
\footnote{The \citet{2013MNRAS.433.2946W} relation between stellar mass and 3.4 $\mu$m luminosity was derived by matching the {\em WISE} All-Sky Release Catalogue (http://wise2.ipac.caltech.edu/docs/release/allsky/) and the MPA-JHU Sloan Digital Sky Survey catalogue (http://www.mpa-garching.mpg.de/SDSS/DR7/), where the stellar
masses were calculated by fitting the $u,g,r,i,z$ photometry with a large number of model SEDs constructed from the \citet{2003MNRAS.344.1000B} population synthesis code which assumes a \citet{2003PASP..115..763C} initial mass function (IMF).}.
{\em WISE} has mapped the full sky in four bands centred at
3.4, 4.6, 12, and 22~$\mu$m (W1, W2, W3, W4),
achieving 5$\sigma$ point-source sensitivities
of 0.08, 0.11, 1, and 6 mJy, respectively.

We performed aperture photometry on Band 1 {\em WISE} Atlas Images with SEXTRACTOR using the prescription given by the {\em WISE} team\footnote{
http://wise2.ipac.caltech.edu/staff/fmasci/SEx\_WPhot.html}, applied aperture and colour corrections as indicated in the {\em WISE} Explanatory Supplement\footnote{http://wise2.ipac.caltech.edu/docs/release/allsky/expsup/}.
Because of the potential importance of nebular continuum and line emission in
the near-infrared wave bands
\citep[e.g.,][]{2009AJ....138..130S}
we calculated and subtracted the expected nebular contribution in the {\em WISE} band 1
according to \citet{2012MNRAS.427..906H} to obtain a star-only flux. Nonetheless, because of the relatively low star-formation rates (SFRs) for the HeViCS dwarfs (see Sect. ~\ref{sec:ancillary_sfr}), the nebular contribution to W1 for these galaxies is low, $\sim$1\% on average.
Stellar masses were estimated from the relation for star forming galaxies provided in
\citetalias{2013MNRAS.433.2946W},
\begin{equation}\label{eq:Wen13}
 \log ( M_{\star}/ \tmsun )  = a + b \, \log [ \nu L_{\nu} (3.4 \mu\text{m})/ \tlsun ]
\end{equation}
\noindent
where the $a$ and $b$ coefficients are given in Table \ref{Tab:W1_coeff}.

The errors include the uncertainties in the photometric errors and
in the coefficients of the \citet{2013MNRAS.433.2946W} relation.
Because of the large uncertainties in the distance to the
Virgo galaxies, they are not included in the error calculation of stellar masses and of other parameters derived in this section.

We found that the approach of \citetalias{2013MNRAS.433.2946W} gives
stellar masses to within 10-20\% of those derived with the method of
\citet{2006ApJ...647..970L} which relies on a variable mass-to-light ratio. In Fig. \ref{app_fig:compare_mstar},
we show that our stellar masses are also in good agreement with those provided by GOLDMine, which are
derived from the $i$ magnitude and
($g - i$)$_0$ colour, and calibrated on the MPA-JHU sample \citep{2013A&A...553A..89G}, similarly to that done in \citetalias{2013MNRAS.433.2946W}.
The residual distribution between the two estimates is displayed in the bottom panel (blue histogram), with the result of the gaussian fitting which
peaks at 0.05 dex and it has a dispersion of 0.08 dex.
Virgo dwarf stellar masses are listed in Table \ref{Tab:VirgoSFDs}.

\subsubsection{Comparison samples}
\label{sec:comp_sample_stellmass}

To avoid systematics due to the choice of different stellar mass estimates we derived
$M_{\star}$ for the comparison samples with the same method adopted for the Virgo SFDs.
We chose not
to derive the stellar masses with methods using optical photometry such as $i$-band luminosity and the $(g - i)$
colour-dependent stellar mass-to-light ratio relation \citep{2009MNRAS.400.1181Z,2013A&A...553A..89G},
because most of the DGS galaxies do not have optical photometry measurements in the literature, and only 24 KINGFISH galaxies are in the area covered by the SDSS. Therefore this would have inevitably created a systematic offset between the stellar masses of the DGS/KINGFISH and those of the other samples.

Regarding the DGS and HeViCS BGC galaxies we measured {\em WISE} W1 photometry from the {\em WISE} Atlas Images as explained in the previous section, we subtracted the expected nebular contribution to the 3.4 $\mu$m emission, and then applied Eq. \ref{eq:Wen13} to derive $M_{\star}$.

The KINGFISH galaxies have IRAC 3.6 $\mu$m flux measurements in the literature. In this case we  derived a conversion factor between IRAC 3.6 $\mu$m and $WISE$ W1 flux densities and then we calculated stellar masses with Eq. \ref{eq:Wen13}. To derive the conversion factor we used the
atlas of 129 spectral energy distributions for nearby galaxies \citep{2014ApJS..212...18B}, which includes measurements from both {\em Spitzer} and {\em WISE}. The atlas contain 23 spirals and 1 Sm galaxy from the KINGFISH sample; for these objects we found that the mean ratio between the two bands is $F_{3.4}/F_{3.6}$ = 1.020$\pm$0.035. We applied this conversion factor to the IRAC fluxes, subtracted the expected nebular contribution, and estimated stellar masses with Eq. \ref{eq:Wen13}.
Comparison with stellar mass estimates obtained with different methods for these three samples is discussed in Appendix~\ref{app_sec:compare_mstar}. Stellar masses of the DGS, KINGFISH, and HeViCS BGC galaxies are listed in Tables~\ref{tab:DGS}, \ref{tab:King_dwarfs}, \ref{tab:King_spirals}, and \ref{tab:BGC}.

Although it is often assumed that the 3.4/3.6 $\mu$m band is dominated by
starlight we cannot rule out that a source of possible contamination to this emission could be provided by polycyclic aromatic hydrocarbons (PAH) and hot dust \citep{2010ApJ...725.1971M,2014ApJ...788..144M}.
The issue of this possible contamination is not discussed or taken into account in \citet{2013MNRAS.433.2946W}.
Analysis in a small sample of disc galaxies in the Spitzer Survey of Stellar Structure in Galaxies show that hot dust and PAH can contribute between 5\% and 13\% of the total integrated light at 3.6 $\mu$m \citep{2014ApJ...788..144M}.
In a sample of local dwarf galaxies, comparison with stellar population synthesis models shows that starlight alone can account, within the uncertainties, for the 3.6 $\mu$m emission \citep{2009AJ....138..130S}.
Comparison to \citet{2013A&A...553A..89G} stellar mass estimates for Virgo galaxies (see also Appendix~\ref{app_sec:compare_mstar}) suggests that the possible contamination of hot dust will not significantly influence the results discussed in the rest of this work.

\subsection{\hi masses}
\label{sec:ancillary_himass}

\subsubsection{Virgo SFDs}

The atomic hydrogen (H{\sc i}) content of Virgo dwarf galaxies was derived from the Arecibo Legacy Fast ALFA (ALFALFA) blind \hi survey \citep{2005AJ....130.2598G}.
The latest catalogue release, the $\alpha$.40 catalogue \citep{2011AJ....142..170H}, covers the cluster at declinations
$4^{\circ} < \delta < 16^{\circ}$, almost the whole extent of the HeViCS fields. With a mean rms of  2 mJy/beam, the survey detection limit for a dwarf galaxy with $S/N$ = 6.5 and a typical \hi line width of 40 km s$^{-1}$ at a distance of 17 Mpc, is $M_{H\mathsc{i}} \approx 10^{7.5}$ M$_{\odot}$.
For those galaxies not included in the ALFALFA catalogue, \hi mass measurements were obtained
from the literature: VCC1 \citep{2005A&A...429..439G};  VCC286, VCC741 \citep{1987ApJS...63..247H};
VCC135 \citep{2005ApJS..160..149S}.
Only five galaxies have not been detected at 21-cm (see Table \ref{Tab:VirgoSFDs}).

Following \citet{1984AJ.....89..758H} and \citet{2013A&A...553A..90G}, we estimated the \hi deficiency parameter defined as the logarithmic difference between the \hi mass of a reference sample of isolated galaxies for a given morphological type and the observed
\hi mass:  $Def_{HI}$ = log $M_{HI}^{ref}$ -
log $M_{HI}^{obs}$. The reference \hi mass is derived as
log $M_{HI}^{ref}$ = $C_1$ + $C_2$ log $d$, where $d$ is the galaxy linear diameter in kpc
at the 25$^{\text{th}}$ mag arcsec$^{-2}$ $B$-band isophote, and the $C_1$ (7.51) and $C_2$ (0.68) coefficients have been rederived by \citet{2013A&A...553A..90G}
for all late-type galaxies (independently of the Hubble type) using a sample of isolated objects from the ALFALFA survey. A threshold of $Def_{HI}$ = 0.5 is adopted to distinguish H{\sc i}-deficient from H{\sc i}-normal systems,
corresponding to galaxies with at least 70\% less atomic hydrogen than expected for isolated objects
of the same optical size and morphology. Galaxies with $Def_{HI} > 0.9$ are considered highly
H{\sc i}-deficient \citep{2013A&A...553A..90G}. \hi masses and H{\sc i}-deficiency of the Virgo dwarfs are given in Table \ref{Tab:VirgoSFDs}.

 \begin{figure}
   \centering
\includegraphics[bb=70 10 550 550,width=8cm]{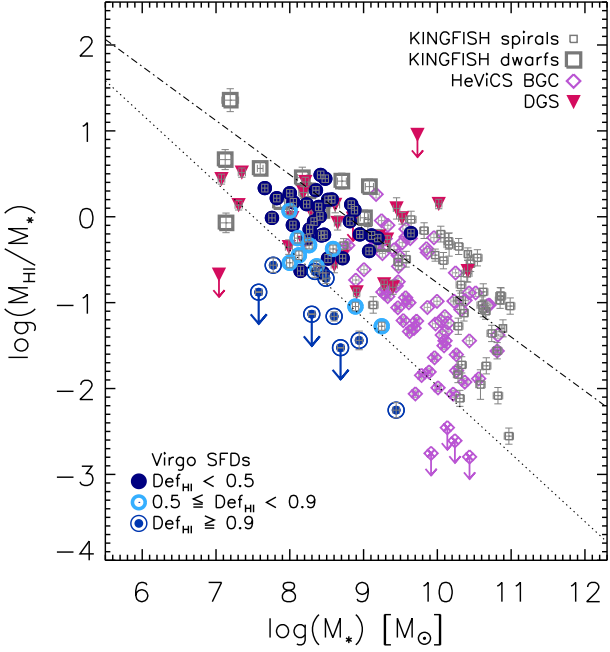} 
  \caption{\hi gas fraction (M$_{\text{HI}}/M_{\star}$) as a function of stellar mass. Blue symbols correspond to the Virgo SFDs, with the different shapes indicating the atomic
  hydrogen content of the galaxies as given by the H{\sc i} deficiency parameter: H{\sc i}-normal ({filled dots}), H{\sc i}-deficient ({rings}), highly H{\sc i}-deficient ({ringed dots}). Red-purple {triangles} represent the DGS sample, grey {squares} show the spiral and dwarf galaxies
  of the KINGFISH sample, and purple {diamonds} correspond to the HeViCS BGC. \hidef\ HeViCS BGC galaxies ($Def_{HI} \geq 0.5$) are indicated by a diamond with a cross. Gas-scaling relations from \citet{2013A&A...553A..90G} are overlaid for normal (dash-dotted line), and highly deficient (dotted line) galaxies.
  }
   \label{fig:hi_mstar}%
    \end{figure}

\subsubsection{Comparison samples}

\hi masses for the DGS galaxies were obtained from
\citet{2014A&A...563A..31R}.
Only four galaxies do not have a 21-cm detection (see Table \ref{tab:DGS}).
Sixteen out of 27 galaxies have a CO detection in the literature, and H$_2$ masses have been calculated by \citet{2014A&A...563A..31R} using the Galactic CO-to-H$_2$ conversion factor, $X_{CO}^{MW} = 2.1 \times 10^{20}$
cm$^{-2}$/ K km s$^{-1}$ \citep{2011ApJ...726...81A}
and a metallicity dependent $X_{CO}$ scaling with $(O/H)^{-2}$ \citep{2012AJ....143..138S}.

Atomic hydrogen masses for the KINGFISH galaxies were also taken from \citet{2014A&A...563A..31R} where they
combined literature measurements from \citet{2007ApJ...663..866D} and \citet{2011A&A...532A..56G}.
CO observations are available in the literature for 33 out of 51 galaxies and they have been assembled by \citet{2014A&A...563A..31R}. H$_2$ masses were derived using two $X_{CO}$ factors similarly to the DGS sample. KINGFISH gas masses are displayed in Table \ref{tab:King_dwarfs} and \ref{tab:King_spirals}.

H{\sc i} masses for the HeViCS BGC sample were obtained from the $\alpha$.40 catalogue and the
GOLDMine database. Only four galaxies have not been detected at 21-cm: VCC341, VCC362, VCC1190, VCC1552
(see Table \ref{tab:BGC}). For a subset of HeViCS BGC galaxies, H$_2$ masses are available from the Herschel Reference Survey  \citep[HRS; ][]{2014A&A...564A..65B}, and are also listed in Table \ref{tab:BGC}, calculated for both a Galactic CO-to-H$_2$ conversion factor and a $H$-band luminosity dependent conversion factor $\log \frac{X_{CO}}{\text{cm}^{-2} \,(\text{K km s}^{-1})^{-1}}$ = -0.38  $\log\, \frac{L_H}{\text{L}_{\odot}} + 24.23$ \citep{2002A&A...384...33B}.

Figure \ref{fig:hi_mstar} displays the \hi fraction
$f_{{HI}} = M_{{HI}}/M_{\star}$ against the stellar mass for Virgo galaxies and the comparison samples.
The \hi content of the Virgo dwarfs, as given
by the H{\sc i} deficiency parameter, is highlighted by the different shapes of the circles and shades of blue: galaxies with $Def_{HI} < 0.5$ have a normal \hi content
({filled dots}), galaxies with $0.5 \leq Def_{HI} < 0.9$ are {H\sc i}-deficient ({rings}),
and those with $Def_{HI} > 0.9$ ({ringed dots}) are extremely poor in atomic hydrogen.
The \hi content of DGS, KINGFISH spirals (from Sa to Sd types), KINGFISH dwarfs (objects later than Sd), and the HeViCS BGC is also shown.
Gas-scaling relations of the H$\alpha$3 sample from \citet{2013A&A...553A..90G}
are overlaid for comparison for two classes of H{\sc i}-deficiency: normal ({\em dash-dotted} line),
and highly deficient ({dotted} line).

The \hi fraction decreases by approximately
4 orders of magnitude with stellar mass,
from $\log(M_{\star}$/M$_{\odot}$) $\sim$ 7  to 11.
As expected, more massive galaxies are characterised by lower gas fractions,
while low-mass galaxies retain larger quantities of \hi compared to their stellar masses
\citep{2011MNRAS.415.1797C,2012ApJ...756..113H,2013A&A...553A..89G}.

Most of the Virgo dwarf galaxies with a normal atomic hydrogen content ($Def_{HI} < 0.5$) show similar gas fractions to the
KINGFISH and DGS dwarfs with comparable stellar masses. Among the H{\sc i}-normal Virgo SFDs, about a third fall in the region
of higher \hi\-deficiency defined by the gas scaling relations of \citet{2013A&A...553A..90G}, and they do show
gas fractions similar to dwarfs with $0.5 \leq Def_{HI} < 0.9$. It is possible the $Def_{HI}$ is not well assessed for this subset.
Approximately 20\% of Virgo SFDs show a large gas deficit relative to other dwarfs, as Fig. \ref{fig:hi_mstar} illustrates, giving a clear signature of the interaction occurring between these systems and the surrounding environment.
The figure also shows the well-known decrease in the \hi fraction of Virgo late-type spiral galaxies compared to galaxies with similar stellar mass and morphological type but evolving in less dense environments such as KINGFISH objects \citep{2011MNRAS.415.1797C}.

\subsection{Star formation rates}
\label{sec:ancillary_sfr}

\subsubsection{Virgo SFDs}

We estimated the global star-formation rate starting from
H$\alpha$ photometry which was obtained from the GOLDMine data base.
H$\alpha$ fluxes were corrected for
Galactic extinction with the \citet{2011ApJ...737..103S} extinction curve ($R_V = 3.1$) using $A(H\alpha) = 0.81 A_V$.  Correction for [NII] deblending was obtained calculating the [NII]$\lambda$6584/H$\alpha$ ratio with line fluxes extracted from the SDSS MPA-JHU DR7 release\footnote{http://www.mpa-garching.mpg.de/SDSS/DR7/raw\_data.html}.
A ratio of [NII]$\lambda$6548/[NII]$\lambda$6584 = 0.34 was assumed to take into account the contribution of both lines to the H$\alpha$ flux \citep{2012A&A...545A..16G}.
When [NII]$\lambda$6584 line flux was not available we derived the
([NII]/H$\alpha$) ratio using the relation calibrated on the
 absolute $i$-band magnitude, ([NII]/H$\alpha$) = -0.0854 $\times$ $M_i$ - 1.326. \citep{2012A&A...545A..16G}.

To account for both unobscured and obscured star formation we followed two procedures.
First, we searched for
mid-IR emission using the {\em WISE} All-Sky Survey at 22~\micron, and found
30 dwarfs with a mid-IR counterpart. For these galaxies we performed aperture photometry on the 22~$\mu$m {\em WISE} Atlas Images with SEXTRACTOR in the same way as described in
Sect. ~\ref{sec:ancillary_stellmass},
applied aperture and colour corrections,
and an additional correction factor of 0.92
as recommended in \citet{2013AJ....145....6J}\footnote{This correction is due to a calibration discrepancy between the {\em WISE} photometric standard "blue"
stars and "red" galaxies (e.g., star-forming
systems) and it must be applied only to W4 flux densities.}.
Then we used the relation of \citet{2014MNRAS.438...97W} to derive the SFR\footnote{The relation of \citet{2014MNRAS.438...97W} is calibrated assuming a \citet{2001MNRAS.322..231K} IMF. Because the SFRs calculated with this IMF yields nearly identical results to those derived with a Chabrier (2003) IMF \citep{2011AJ....142..197C,2012ARA&A..50..531K},
we avoid rescaling Eq. \ref{eq:Wen14SFR} to a Chabrier IMF.}:

\begin{equation}\label{eq:Wen14SFR}
\log(SFR) \, [\text{M}_{\odot} \, \text{yr}^{-1}] = \log[L_{H\alpha} + 0.034 \, \nu L_{\nu}(22\mu\text{m})] - 41.27
\end{equation}
\noindent
where $L_{\text{H}\alpha}$  and $\nu L_{\nu}$(22$\mu$m) are the H$\alpha$ and 22~$\mu$m monochromatic luminosity in erg s$^{-1}$, respectively.

For the remaining galaxies without a {\em WISE} band 4 detection,
we calculated the SFR from the H$\alpha$ fluxes only, using
\citet{1998ARA&A..36..189K} for a Kroupa IMF:

\begin{equation}\label{eq:K98}
 SFR \, [\text{M}_{\odot} \, \text{yr}^{-1}] = 5.37 \times 10^{-42} \; L_{\text{H}\alpha} \, [\text{erg s}^{-1}].
\end{equation}

\noindent
after having corrected the H$\alpha$ fluxes for internal extinction using the Balmer decrement measured from SDSS spectra.
We assumed an intrinsic H$\alpha$/H$\beta$ ratio of 2.86  \citep[case B recombination, $T$ = 10 000 K and $n_e$ = 100 cm$^{-3}$][]{2006agna.book.....O}
and adopted the extinction curve of \citet{2000ApJ...533..682C} to be consistent with \citet{2014MNRAS.438...97W}.

However, at low H$\alpha$ luminosities ($L_{H\alpha} < 2.5 \times 10^{39}$ erg s$^{-1}$) both methods described above may underpredict the total SFR, since H$\alpha$ becomes a less reliable SFR indicator compared to the far ultraviolet (FUV) emission \citep{2009ApJ...706..599L}. This discrepancy could be due to effects such as possible leakage of ionizing photons, departures from Case B recombination, stochasticity in the formation of high-mass stars,
or variation in the IMF resulting in a deficiency of high-mass stars \citep{2009ApJ...706..599L,2011ApJ...741L..26F}.  Twentyfour dwarfs in our sample have H$\alpha$ luminosities below this threshold (of which 9 had a mid-IR counterpart). For these galaxies
we used the empirical re-calibration of Eq. \ref{eq:K98} given by \citet{2009ApJ...706..599L}, based on FUV emission:

\begin{equation} \label{eq:Lee2SFR}
\log(SFR) \, [\text{M}_{\odot} \, \text{yr}^{-1}] = 0.62 \, \log ( 5.37 \times 10^{-42} \, L_{H\alpha} \, [\text{erg s}^{-1}])  - 0.57
\end{equation}

\noindent
where $L_{H\alpha}$ is the non-dust corrected H$\alpha$ luminosity.
Uncertainties in the SFR in this case are taken from the 1$\sigma$ scatter between the FUV and H$\alpha$ SFRs listed in Table 2 of \citet{2009ApJ...706..599L}.

Only two galaxies have neither H$\alpha$ measurements available in the GOLDMine database nor a detection at 22 $\mu$m wavelengths (VCC367 and VCC825). SFRs of the Virgo SFDs are given in Table \ref{Tab:VirgoSFDs}.

\begin{figure}
   \centering
\includegraphics[bb=30 10 630 340,width=8cm]{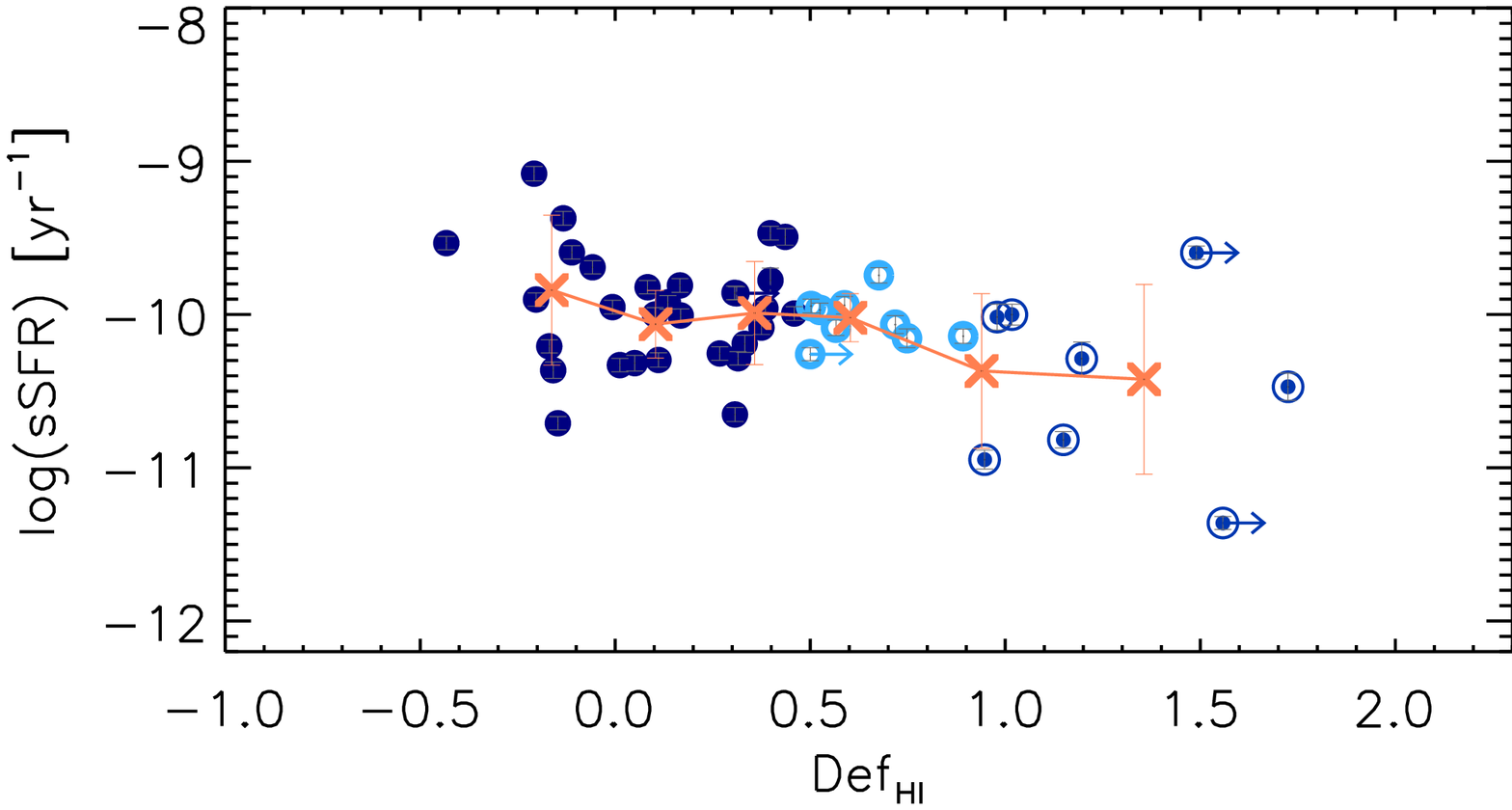}
\includegraphics[bb=70 10 550 550,width=8cm]{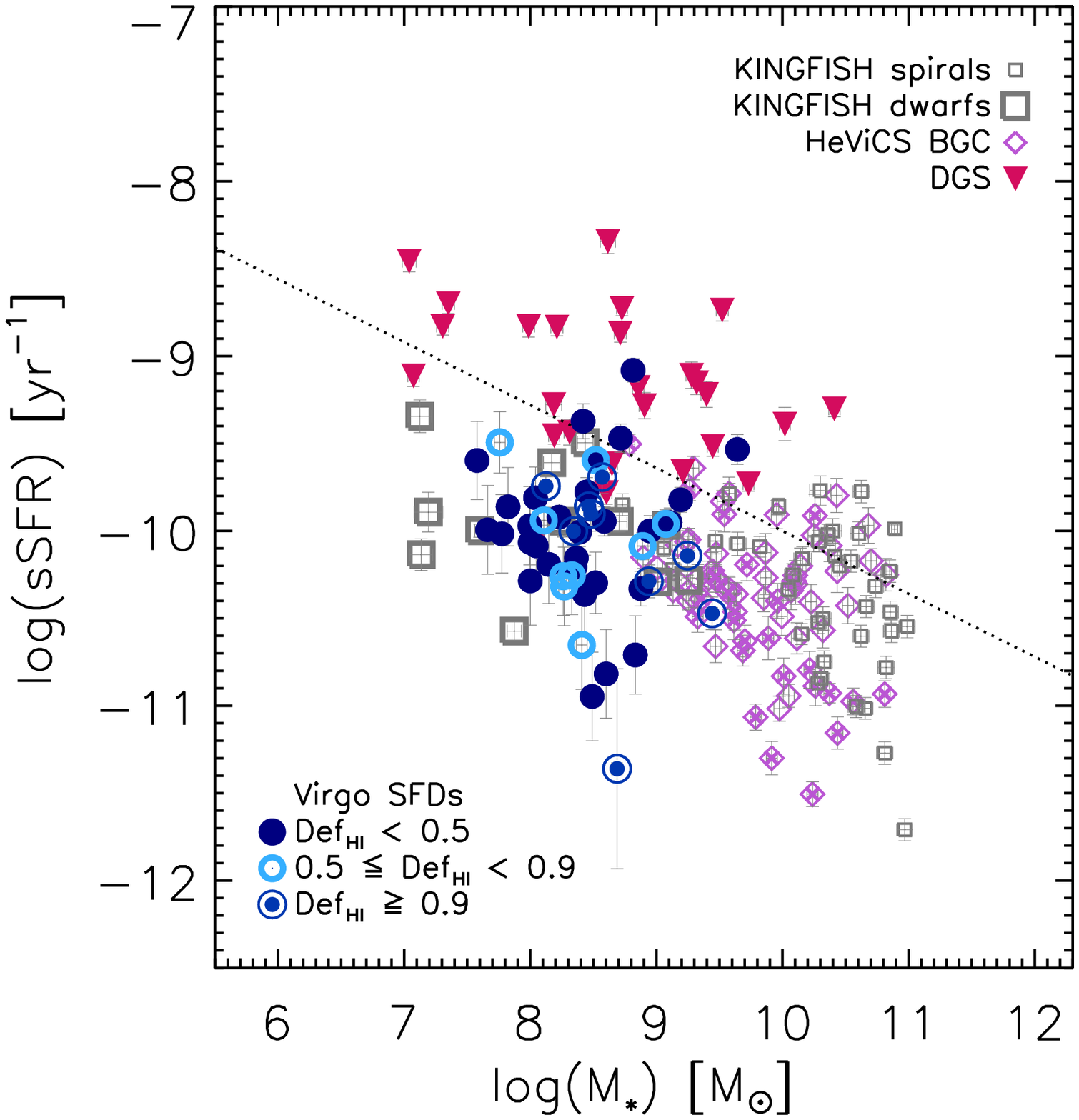}
  \caption{{\em Upper panel:} Specific star formation rate against \hi deficiency for Virgo SFDs. Crosses denote the average value in each bin of $Def_{HI}$. {\em Lower panel:} Specific star formation rates versus stellar masses. Blue symbols correspond to the Virgo SFDs, with the different shapes indicating the atomic
  hydrogen content of the galaxies as given by the H{\sc i} deficiency parameter.
  Symbols of comparison samples are the same used in Fig. \ref{fig:hi_mstar}.
  The dotted line indicates the star formation sequence defined by \citet{2007ApJS..173..315S}.
  }
   \label{ssfr_mstar}%
    \end{figure}

To inspect possible effects of the cluster environment on the dwarf star formation activity,
we plot the sSFR against \hi\-deficiency in the upper panel of Fig. \ref{ssfr_mstar}.
The figure shows that there is an overall decreasing trend of the star formation activity with $Def_{HI}$, confirming that the evolution of these dwarfs in a rich cluster is affecting both their gas content and star formation activity \citep{2002A&A...396..449G}.

\subsubsection{Comparison samples}

KINGFISH SFRs were taken from \citet{2011PASP..123.1347K} and they were derived using the combination of H$\alpha$ and 24 $\mu$m luminosities \citep{2009ApJ...703.1672K,2010ApJ...714.1256C} calibrated for a Kroupa IMF (Tables \ref{tab:King_dwarfs} and \ref{tab:King_spirals}).

Regarding the DGS, we calculated the SFRs in the same way as the KINGFISH sample combining H$\alpha$ measurements  \citep{2003ApJS..147...29G,2006ApJS..164...81M,2006ApJS..164...52S,2008ApJS..178..247K} and 24 $\mu$m flux densities \citep{2012MNRAS.423..197B} from the literature.
H$\alpha$ fluxes were already corrected for foreground galactic extinction and [NII] contamination.
The lack of H$\alpha$ measurements for HS0052+2536 and HS1304+3529 prevented an estimate of the SFR for these two galaxies (see Table \ref{tab:DGS}).

SFRs for the HeViCS BGC galaxies were calculated from Eq. \ref{eq:Wen14SFR} and they are displayed in Table \ref{tab:BGC}. H$\alpha$ fluxes were extracted from GOLDMine, corrected for Galactic extinction and deblending from [NII], using the [NII]$\lambda$6548,$\lambda$6584, and H$\alpha$ equivalent widths given in the database. The 22 $\mu$m photometry was obtained from the {\em WISE} All-Sky Survey in the same way as described for the HeViCS SFDs.

The lower panel of Fig. \ref{ssfr_mstar} illustrates the variation of the specific star formation rate (sSFR) with
stellar mass for the Virgo dwarfs and the comparison samples.
The lower mass galaxies have higher
sSFRs, consistent with the "downsizing" scenario \citep{1996AJ....112..839C} predicting that lower mass
galaxies are more gas-rich and capable to sustain significant star formation activity
at present epoch.
The star formation sequence defined by \citet{2007ApJS..173..315S} clearly separates the different regime of star formation of the DGS galaxies compared to
the majority of Virgo and KINGFISH dwarfs. The scatter between the sSFR of the DGS and of the other samples of dwarfs can reach up to 2 orders of magnitude.

Figure \ref{ssfr_mstar} shows that stellar mass is the main parameter which drives
the scaling relation with star formation activity. The effect of the environment
is then superimposed on this scaling relation and it is evident
in both low- and high-mass Virgo galaxies when compared to systems in lower density environments \citep{2011MNRAS.415.1797C,2012ApJ...756..113H}.

\begin{table*}
\centering
\caption{Stellar masses, \hi masses, dust masses, star formation rates, metallicities, H{\sc i} deficiency, and adopted distances of star-forming dwarf galaxies detected by HeViCS.}
\label{Tab:VirgoSFDs}
\begin{threeparttable}
\begin{tabular}{lcclcccSc}
\hline \hline
    ID   &  log ($M_{\star}$)    & log ($M_{HI})$  &  log ($M_d$)\tnote{$\dagger$}     &  log($SFR$)        &  12 + log(O/H)  &  12 + log(O/H) &    \si{$\: \: Def_{HI}$}     & D           \\
         &  [M$_{\odot}$]  & [M$_{\odot}$]   & $\: \: \:$[M$_{\odot}$]    & [M$_{\odot}$ yr$^{-1}$] &          &   PT05  &        & [Mpc]           \\
 \hline \hline
   VCC1 & 8.94 $\pm$ 0.04 & 7.50 $\pm$ 0.10 & 5.25$_{-0.07}^{+0.07}$           & -1.35 $\pm$ 0.08\tnote{a} & 8.59 $\pm$ 0.10 &  ... &  1.20 & 32.0 \\
  VCC10 & 8.95 $\pm$ 0.04 & 8.74 $\pm$ 0.01 & 6.13$_{-0.06}^{+0.06}$           & -1.05 $\pm$ 0.08\tnote{a} & 8.56 $\pm$ 0.10 &  ... &  0.10 & 32.0 \\
  VCC17 & 8.42 $\pm$ 0.04 & 8.90 $\pm$ 0.01 & 5.86$_{-0.07}^{+0.06}$(${\ast}$) & -0.95 $\pm$ 0.09\tnote{b} & 8.59 $\pm$ 0.10 &  ... & -0.13 & 32.0 \\
  VCC22 & 8.43 $\pm$ 0.04 & 8.21 $\pm$ 0.03 & 5.44$_{-0.07}^{+0.07}$(${\ast}$) & -1.93 $\pm$ 0.22\tnote{c} &       ...       &  ... & -0.16 & 32.0 \\
  VCC24 & 8.83 $\pm$ 0.04 & 8.98 $\pm$ 0.01 & 5.59$_{-0.10}^{+0.10}$           & -1.88 $\pm$ 0.22\tnote{c} & 8.31 $\pm$ 0.10 &  ... & -0.15 & 32.0 \\
  VCC87 & 8.39 $\pm$ 0.04 & 8.51 $\pm$ 0.01 & 5.91$_{-0.06}^{+0.07}$           & -1.62 $\pm$ 0.07\tnote{a} & 8.25 $\pm$ 0.10 &  ... &  0.17 & 17.0 \\
 VCC135 & 9.44 $\pm$ 0.04 & 7.19 $\pm$ 0.08 & 6.19$_{-0.06}^{+0.06}$           & -1.03 $\pm$ 0.08\tnote{a} & 8.65 $\pm$ 0.10 & 8.47 &  1.73 & 32.0 \\
 VCC144 & 8.81 $\pm$ 0.05 & 8.76 $\pm$ 0.01 & 5.70$_{-0.06}^{+0.06}$           & -0.27 $\pm$ 0.05\tnote{a} & 8.21 $\pm$ 0.10 & 8.30 & -0.21 & 32.0 \\
 VCC172 & 8.88 $\pm$ 0.04 & 8.95 $\pm$ 0.01 & 6.04$_{-0.08}^{+0.08}$           & -1.45 $\pm$ 0.09\tnote{b} & 8.58 $\pm$ 0.10 &  ... &  0.01 & 32.0 \\
 VCC213 & 8.89 $\pm$ 0.04 & 7.84 $\pm$ 0.03 & 5.84$_{-0.06}^{+0.06}$           & -1.20 $\pm$ 0.06\tnote{a} & 8.77 $\pm$ 0.12 & 8.27 &  0.57 & 17.0 \\
 VCC223 & 8.45 $\pm$ 0.04 & 7.79 $\pm$ 0.07 & 5.61$_{-0.07}^{+0.08}$           & -1.33 $\pm$ 0.07\tnote{a} & 8.20 $\pm$ 0.10 &  ... &  0.40 & 32.0 \\
 VCC281 & 8.15 $\pm$ 0.04 & 7.51 $\pm$ 0.04 & 5.32$_{-0.08}^{+0.08}$           & -2.05 $\pm$ 0.22\tnote{c} & 8.49 $\pm$ 0.10 &  ... &  0.33 & 17.0 \\
 VCC286 & 8.26 $\pm$ 0.04 &     $<$7.93     & 5.35$_{-0.07}^{+0.07}$(${\ast}$) & -2.00 $\pm$ 0.22\tnote{c} & 8.46 $\pm$ 0.10 &  ... \si{$> 0.50$} & 32.0 \\
 VCC322 & 8.00 $\pm$ 0.04 & 8.27 $\pm$ 0.01 & 4.93$_{-0.16}^{+0.29}$           & -2.29 $\pm$ 0.25\tnote{c} & 8.58 $\pm$ 0.10 &  ... &  0.31 & 17.0 \\
 VCC324 & 8.72 $\pm$ 0.04 & 8.23 $\pm$ 0.01 & 5.50$_{-0.06}^{+0.06}$           & -0.75 $\pm$ 0.07\tnote{a} & 8.14 $\pm$ 0.10 & 8.37 &  0.40 & 17.0 \\
 VCC328 & 7.66 $\pm$ 0.04 & 7.99 $\pm$ 0.01 & 4.70$_{-0.17}^{+0.31}$           & -2.33 $\pm$ 0.25\tnote{c} & 8.46 $\pm$ 0.10 &  ... &  0.46 & 17.0 \\
 VCC334 & 8.04 $\pm$ 0.04 & 7.95 $\pm$ 0.01 & 4.94$_{-0.07}^{+0.07}$           & -1.77 $\pm$ 0.17\tnote{c} & 8.22 $\pm$ 0.10 & 7.92 &  0.17 & 17.0 \\
 VCC340 & 9.11 $\pm$ 0.04 & 8.89 $\pm$ 0.01 & 6.05$_{-0.06}^{+0.06}$           & -0.84 $\pm$ 0.07\tnote{a} & 8.26 $\pm$ 0.10 &  ... & -0.01 & 32.0 \\
 VCC367 & 8.24 $\pm$ 0.04 &     $<$ 7.99    & 5.64$_{-0.06}^{+0.07}$(${\ast}$) &       ...                 &       ...       &  ... &\si{$> 0.50$} & 32.0 \\
 VCC446 & 8.36 $\pm$ 0.04 & 7.79 $\pm$ 0.04 & 5.30$_{-0.11}^{+0.17}$           & -1.79 $\pm$ 0.17\tnote{c} & 8.25 $\pm$ 0.10 &  ... &  0.75 & 23.0 \\
 VCC562 & 7.76 $\pm$ 0.04 & 7.74 $\pm$ 0.03 & 5.00$_{-0.07}^{+0.07}$           & -1.74 $\pm$ 0.17\tnote{c} & 8.10 $\pm$ 0.10 & 8.32 &  0.44 & 17.0 \\
 VCC620 & 8.00 $\pm$ 0.04 & 8.06 $\pm$ 0.01 & 5.25$_{-0.06}^{+0.05}$(${\ast}$) & -1.97 $\pm$ 0.22\tnote{c} & 8.24 $\pm$ 0.10 &  ... &  0.52 & 17.0 \\
 VCC641 & 8.11 $\pm$ 0.04 & 7.86 $\pm$ 0.04 & 5.45$_{-0.06}^{+0.06}$(${\ast}$) & -1.83 $\pm$ 0.17\tnote{c} & 8.21 $\pm$ 0.10 &  ... &  0.59 & 23.0 \\
 VCC693 & 8.33 $\pm$ 0.04 & 8.27 $\pm$ 0.01 & 5.55$_{-0.06}^{+0.07}$           & -1.93 $\pm$ 0.22\tnote{c} & 8.43 $\pm$ 0.10 &  ... &  0.27 & 17.0 \\
 VCC699 & 9.19 $\pm$ 0.04 & 8.94 $\pm$ 0.01 & 6.26$_{-0.06}^{+0.06}$           & -0.63 $\pm$ 0.06\tnote{a} & 8.30 $\pm$ 0.10 &  ... &  0.08 & 23.0 \\
 VCC737 & 8.35 $\pm$ 0.04 & 8.66 $\pm$ 0.01 & 5.73$_{-0.07}^{+0.07}$           & -1.85 $\pm$ 0.17\tnote{c} & 8.28 $\pm$ 0.10 &  ... & -0.17 & 17.0 \\
 VCC741 & 7.82 $\pm$ 0.04 &    $<$ 8.04     & 5.17$_{-0.08}^{+0.10}$           & -2.04 $\pm$ 0.22\tnote{c} & 8.54 $\pm$ 0.10 &  ... &\si{$> 0.31$}  & 17.0 \\
 VCC802 & 7.58 $\pm$ 0.04 &    $<$ 6.70     & 4.97$_{-0.06}^{+0.06}$(${\ast}$) & -2.02 $\pm$ 0.22\tnote{c} & 8.45 $\pm$ 0.10 & 8.35 &\si{$> 1.49$} & 17.0 \\
 VCC825 & 8.30 $\pm$ 0.04 &    $<$ 7.16     & 4.86$_{-0.07}^{+0.07}$(${\ast}$) &        ...                & 8.79 $\pm$ 0.10 &  ...  &\si{$> 1.47$} & 23.0 \\
 VCC841 & 8.12 $\pm$ 0.04 & 7.68 $\pm$ 0.03 & 5.20$_{-0.07}^{+0.08}$           & -1.62 $\pm$ 0.07\tnote{a} & 8.33 $\pm$ 0.10 & 8.34 &  0.68 & 17.0 \\
 VCC848 & 8.48 $\pm$ 0.04 & 8.92 $\pm$ 0.01 & 5.30$_{-0.07}^{+0.07}$           & -1.42 $\pm$ 0.09\tnote{b} & 8.61 $\pm$ 0.10 & 8.12 & -0.20 & 23.0 \\
 VCC888 & 8.41 $\pm$ 0.04 & 8.41 $\pm$ 0.02 & 5.97$_{-0.10}^{+0.13}$           & -2.24 $\pm$ 0.25\tnote{c} &       ...       &  ... &  0.31 & 23.0 \\
 VCC985 & 8.00 $\pm$ 0.04 & 7.46 $\pm$ 0.04 & 4.93$_{-0.11}^{+0.12}$           & -2.07 $\pm$ 0.22\tnote{c} & 8.35 $\pm$ 0.10 &  ... &  0.72 & 17.0 \\
VCC1021 & 8.49 $\pm$ 0.04 & 7.77 $\pm$ 0.05 & 4.87$_{-0.07}^{+0.07}$(${\ast}$) & -2.46 $\pm$ 0.25\tnote{c} &       ...       &  ... &  0.95 & 23.0 \\
VCC1141 & 8.27 $\pm$ 0.04 & 8.12 $\pm$ 0.03 & 5.24$_{-0.08}^{+0.11}$           & -2.05 $\pm$ 0.22\tnote{c} & 8.32 $\pm$ 0.10 &  ... &  0.05 & 23.0 \\
VCC1179 & 8.34 $\pm$ 0.04 & 7.70 $\pm$ 0.05 & 5.12$_{-0.07}^{+0.08}$           & -1.66 $\pm$ 0.07\tnote{a} & 8.33 $\pm$ 0.10 &  ... &  1.02 & 23.0 \\
VCC1200 & 8.05 $\pm$ 0.04 & 8.22 $\pm$ 0.01 & 5.13$_{-0.10}^{+0.14}$           & -2.04 $\pm$ 0.22\tnote{c} & 8.57 $\pm$ 0.10 &  ... &  0.38 & 17.0 \\
VCC1273 & 8.69 $\pm$ 0.04 &     $<$ 7.16    & 5.36$_{-0.09}^{+0.10}$           & -2.67 $\pm$ 0.57\tnote{c} & 8.59 $\pm$ 0.10 &  ... &\si{$> 1.56$} & 23.0 \\
VCC1356 & 8.23 $\pm$ 0.04 & 8.38 $\pm$ 0.01 & 5.30$_{-0.08}^{+0.07}$           & -1.69 $\pm$ 0.06\tnote{a} & 8.34 $\pm$ 0.10 &  ... &  0.14 & 17.0 \\
VCC1374 & 8.46 $\pm$ 0.04 & 8.26 $\pm$ 0.01 & 5.66$_{-0.07}^{+0.07}$           & -1.40 $\pm$ 0.07\tnote{a} & 8.63 $\pm$ 0.10 & 8.26 &  0.31 & 17.0 \\
VCC1437 & 8.52 $\pm$ 0.04 & 8.03 $\pm$ 0.02 & 5.23$_{-0.06}^{+0.06}$           & -1.78 $\pm$ 0.17\tnote{c} & 8.38 $\pm$ 0.10 & 8.00 &  0.11 & 17.0 \\
VCC1455 & 7.77 $\pm$ 0.04 & 7.21 $\pm$ 0.07 & 5.02$_{-0.06}^{+0.07}$(${\ast}$) & -2.24 $\pm$ 0.22\tnote{c} & 8.40 $\pm$ 0.10 &  ... &  0.98 & 17.0 \\
VCC1554 & 9.64 $\pm$ 0.04 & 9.45 $\pm$ 0.01 & 6.81$_{-0.05}^{+0.06}$           &  0.11 $\pm$ 0.07\tnote{a} & 8.26 $\pm$ 0.10 &  ... & -0.43 & 17.0 \\
VCC1575 & 9.25 $\pm$ 0.04 & 7.97 $\pm$ 0.02 & 6.24$_{-0.06}^{+0.06}$           & -0.90 $\pm$ 0.10\tnote{a} & 8.76 $\pm$ 0.10 &  ... &  0.89 & 17.0 \\
VCC1675 & 8.60 $\pm$ 0.04 & 7.44 $\pm$ 0.03 & 5.23$_{-0.10}^{+0.11}$           & -2.22 $\pm$ 0.25\tnote{c} & 8.45 $\pm$ 0.10 &  ... &  1.15 & 17.0 \\
VCC1686 & 9.07 $\pm$ 0.04 & 8.68 $\pm$ 0.01 & 6.44$_{-0.06}^{+0.06}$           & -0.89 $\pm$ 0.07\tnote{a} &     ...         &  ... &  0.38 & 17.0 \\
VCC1699 & 8.57 $\pm$ 0.04 & 8.77 $\pm$ 0.01 & 5.46$_{-0.06}^{+0.07}$           & -1.12 $\pm$ 0.08\tnote{a} & 8.07 $\pm$ 0.12 & 7.88 & -0.06 & 17.0 \\
VCC1725 & 8.59 $\pm$ 0.04 & 8.21 $\pm$ 0.01 & 5.78$_{-0.06}^{+0.07}$           & -1.36 $\pm$ 0.07\tnote{a} & 8.25 $\pm$ 0.10 & 8.31 &  0.50 & 17.0 \\
VCC1791 & 8.52 $\pm$ 0.04 & 8.72 $\pm$ 0.01 & 5.71$_{-0.06}^{+0.07}$           & -1.08 $\pm$ 0.05\tnote{a} & 8.16 $\pm$ 0.10 &  ... & -0.11 & 17.0 \\
\hline \hline
\end{tabular}
  \begin{tablenotes}
    \item[$\dagger$] Dust masses were determined fitting a single MBB with a fixed $\beta$ = 1.5 emissivity from 100 to 350 $\mu$m. Galaxies noted with $^{(\ast)}$ correspond to MBB fits with fixed dust temperature, because of the lack of enough data points (see also Table \ref{tab:beta15}).
    \item[a] SFR calculated from Eq. \ref{eq:Wen14SFR} \citep{2014MNRAS.438...97W}
    \item[b] SFR calculated from Eq. \ref{eq:K98} \citep{2012ARA&A..50..531K}
    \item[c] SFR calculated from Eq. \ref{eq:Lee2SFR} \citep{2009ApJ...706..599L}
  \end{tablenotes}
\end{threeparttable}
\end{table*}

\subsection{Oxygen abundances}
\label{sec:ancillary_met}

The Sloan Digital Sky Survey (SDSS) provides high quality optical spectra covering the wavelength range 3800-9200~$\AA$ with a resolution of $\sim \,$3~$\AA$.
The MPA-JHU collaboration provided measurements of emission-line fluxes
and oxygen abundances for a sample of about 520000 galaxies from the SDSS\footnote{http://www.mpa-garching.mpg.de/SDSS/DR7}, that we could use to derive the metal abundances of Virgo galaxies.

Because the discrepancies between the metallicities estimated from different calibrators can be as high as 70\%  \citep{2007A&A...462..535Y,2008ApJ...681.1183K},
we decided to derive the oxygen abundances
following the method described in \citet{2013A&A...550A.115H}.
Emission-line fluxes (obtained from the MPA-JHU catalogue) were
corrected for internal and galactic extinction, H$\alpha$ and H$\beta$ lines were corrected for underlying stellar absorption, and then all line fluxes were normalised to H$\alpha$.
The method of \citet{2013A&A...550A.115H} combines the strong-line metallicity calibrations
of \citet{1991ApJ...380..140M}, \citet{1994ApJ...420...87Z}, \citet{2002ApJS..142...35K}, and two calibrations from \citet{2004MNRAS.348L..59P}: the O3N2 =
[O{\sc iii}]$\lambda$5007/[N{\sc ii}]$\lambda$6584  and the N2 = [NII]$\lambda$6584/H$\alpha$ indices.
The oxygen abundances given by the five methods are
then converted into a base metallicity -- O3N2 --
via the conversion relations in \citet{2008ApJ...681.1183K},
and the final metallicities are determined from the error-weighted average of all available estimates for each galaxy.

However, the only  applicable calibrations for our sample of dwarfs were those based on the  N2 and  O3N2 indices.
The other three methods could not be  calculated since the [OII]$\lambda$3727 line is out of the measured wavelength range of the SDSS,  and this line is required for the calibration based on the  $R_{23} =$ ([OII]$\lambda$3727 + [OIII]$\lambda \lambda$4959,5007)/H$\beta$ ratio.
The final result was then obtained from either a single oxygen abundance estimate, or the error-weighted average of two estimates.
Uncertainties in the final mean metallicities were derived using the  typical errors of the applicable calibration relations, which were  determined in \citep{2013A&A...550A.115H} from the standard deviations of the  scatter between each different calibration and the rest.

The final oxygen abundances range between 8.0 $\lesssim$ 12 + log(O/H) $\lesssim$ 8.8, and
the mean error is estimated as 0.1 dex in 12+log(O/H) units (see Table \ref{Tab:VirgoSFDs}).
The adopted solar metallicity is 12 + log(O/H)$_{\odot}$ = 8.69 \citep{2009ARA&A..47..481A}.

Although the SDSS fibers sample the inner regions of the galaxies, dwarfs
have been observed to have spatially homogeneous metallicity distribution \citep{1997ApJ...489..636K,2009ApJ...705..723C},
therefore we are confident that our estimate is representative of the global metal content of the galaxies.

Metallicity estimates can vary depending on the calibration
method used \citep{2008ApJ...681.1183K}, and if we want to compare the metal content
of different galaxy samples we need to make sure that heavy element abundances are derived with the
same method. KINGFISH and DGS metallicities are estimated following \citet[][hereafter PT05]{2005ApJ...631..231P},
based on the $R_{23}$ ratio \citep{2011PASP..123.1347K,2014A&A...563A..31R}. Therefore we also
derived PT05 oxygen abundances for 13 Virgo dwarfs 
for which [OII]$\lambda$3727 line fluxes measurements were available from the literature \citep{2003ApJS..145..225V}. We will
use these values to facilitate comparison between the different surveys (see Sect. ~\ref{sec:environment}).
The average difference between the method of \citet{2013A&A...550A.115H} and \citetalias{2005ApJ...631..231P} is 0.14 dex. The PT05 metallicities are also listed in Table \ref{Tab:VirgoSFDs}.

HeViCS BGC galaxies included in the HRS \citep{2010PASP..122..261B} have oxygen abundances calculated in \citet{2013A&A...550A.115H} and we list them in Table \ref{tab:BGC}.

\subsection{Mid- and far-infrared observations from previous surveys}
\label{sec:midir}

We also searched for mid- and far-infrared observations of Virgo SFDs in the \iras\ Faint Source Catalogue \citep[][]{1990IRASF.C......0M}
and Point Source Catalogue \citep[][]{1988iras....7.....H},
and the ISOPHOT Virgo Cluster Catalogue \citep{2002ApJS..139...37T,2002ApJ...567..221P}.
We found both 60 and 100 $\mu$m detections for a total of 14 dwarfs. {\em IRAS} and {\em ISO} flux densities can also be found in the GOLDMine database.
Therefore, we complement \hers photometry
with \iras\ data
for the following galaxies: VCC144,
VCC324, VCC340, VCC699, VCC1437, VCC1554, VCC1575.
ISOPHOT measurements
are available for VCC1, VCC10, VCC87, VCC213,
VCC1686, VCC1699, VCC1725.

\section{Spectral energy distribution fitting}
\label{sec:mbb}

Assuming that dust grains are in local thermal equilibrium,
the spectral energy distribution (SED) of galaxies in the FIR-submm regime due to dust emission is found to be well represented, in the optically thin limit,
by a modified black body (MBB):

\begin{equation}
S (\nu,\, T) \propto  \kappa_{\nu}  B (\nu,\, T)
\end{equation}

\noindent
where B ($\nu,\, T$) is the Planck function, $T$ is the dust temperature,
and $\kappa_{\nu}$ is the dust emissivity or the grain absorption cross section per unit mass,
expressed as a power-law function of frequency:
 $\kappa_{\nu} = \kappa_0 (\nu/\nu_0)^{\beta}$  \citep{1983QJRAS..24..267H}.
This simplified assumption does not take into account that a galaxy can have a range of dust temperatures,
and it cannot fully
describe the range of grain sizes of the different dust components \citep{2012MNRAS.419.1833B,2014ArXiv1409.1815B}.
Nonetheless it is able to reproduce fairly well
the observed large dust grain properties of galaxies \citep{2013A&A...552A..89B}, as long as the function is not fitted
to emission that includes stochastically-heated dust.

The emissivity index $\beta$  is a parameter that is related to the physical
properties of the dust grains, such as  the grain composition (the fraction of silicate versus graphite) and the grain structure \citep[crystalline, amorphous,][]{1995A&A...295..165M,1998A&A...332..291J}, and to the dust temperature \citep{1998ApJ...496.1058M,2007A&A...468..171M,2011A&A...535A.124C}.
Laboratory studies of the two main interstellar dust analogs have shown that: i) carbonaceous grains have spectral indices varying between 1 and 2 according to their internal structure, with well-ordered graphitic grains characterised by
$\beta \sim 2$, while lower values are found for carbonaceous grains with an amorphous structure \citep{1993A&A...279..577P,1995P&SS...43.1263C,1995A&A...295..165M,1998A&A...332..291J}; ii)
crystalline silicate grains have $\beta \sim 2$ \citep{1998ApJ...496.1058M}, and for amorphous silicates the range of variation of $\beta$ at $\lambda < 700$ $\mu$m is smaller ($1.6 \leq \beta \leq 2.2$), independently of grain temperature and composition \citep{2011A&A...535A.124C}. In a study of amorphous silicates
in the temperature range 10 $< T_d  < 300$ K at wavelengths between
0.1 $\mu$m and 2 mm, \citet{2005ApJ...633..272B} report
values of the emissivity spectral index between 1.5 and 2.5.

\citet{2011A&A...536A..19P,2014A&A...571A..11P,2014A&A...566A..55P,2014arXiv1406.5093P} examined the FIR and millimetre emission in the galactic plane, the diffuse ISM, and over the whole sky, reporting $\beta$ values in the range between 1.5 and 1.8, with a mean dust emissivity at high galactic latitudes  $\beta_{FIR} =$ 1.59 $\pm$ 0.12 at $\nu \geq 353$ GHz \citep{2014A&A...571A..11P}, and a flattening of the dust SED at lower frequencies ($\nu < 353$ GHz), with  $\beta_{FIR} - \beta_{mm} = 0.15$ \citep{2014A&A...566A..55P}.

The typical values for $\beta$ determined
in {\em global} extragalactic studies  fall within the range 1.0 - 2.5
\citep{2011A&A...532A..56G,2011A&A...536A..17P,2012A&A...540A..54B,2012ApJ...745...95D,2013A&A...557A..95R}.
Nevertheless, in global studies the indices $\beta$ inferred from
MBB fitting are luminosity-averaged {\it apparent} values, and
may not correspond to the intrinsic properties of the dust grains, but rather they can
provide a measure of the {\em apparent emissivity index} \citep{2014ApJ...789..130K,2014arXiv1406.6066G,2014arXiv1409.5916H}.
Indeed, because of the mixing of different dust temperatures along the line of sight,
the presence of a dust component colder than the peak of
the blackbody emission may produce a broader SED resulting in a fitted
emissivity index shallower  than the intrinsic $\beta$ of the dust grain population \citep{2011A&A...530A.101M,2012A&A...539A..71J}.
Fitted $\beta$ are also found to vary with the intensity of the diffuse interstellar radiation field \citep[ISRF,][]{2014arXiv1409.5916H}.
This implies that it can be difficult
to assess the intrinsic dust grain properties on the basis of a single-temperature MBB fitting procedure.

Keeping in mind these issues, we adopted two approaches for the SED fitting procedure in order to investigate the range of $\beta$ values that can better represent the FIR-submm SED
of our sample of dwarf galaxies. First, we performed a single component modified black-body (MBB) fit using fixed values of the emissivity index,
namely $\beta = $ [1.0,1.2,1.5,1.8,2.0];
second, we repeated the SED fitting testing for each galaxy different
values of $\beta$ varying within the
range [0,3], and selected the value providing the best fit with the
lowest residuals. Basically in this second approach the SED was fitted for a fixed $\beta$ and the fitting process was repeated for all the values within 0 and 3 to determine the index that minimized the reduced $\chi^2$.
The best fit to the data was obtained with the least squares fitting routines in the Interactive Data Language (IDL) MPFIT\footnote{http://purl.com/net/mpfit}
\citep{2009ASPC..411..251M}.
Our procedure is essentially a grid method for fitting temperature and normalization;
such a technique tends to reduce the well-known degeneracy between temperature and $\beta$
\citep[e.g.,][]{2009ApJ...696..676S,2009ApJ...696.2234S}. These two approaches allow us to
to test which values are needed
to better describe the FIR-submm SED of our sample of dwarfs without a priori assumptions on the dust
emissivity index value, similarly to what done in other studies of galaxies based on {\em Herschel} observations
\citep{2012A&A...540A..54B,2013A&A...557A..95R,2014A&A...561A..95T,2014MNRAS.439.2542G,2014ApJ...789..130K}.

For this analysis, we considered only a subset of the sample (30 out of 49 galaxies) detected in four {\em Herschel} bands
(100, 160, 250, and 350 $\mu$m) with a signal-to-noise ratio $S/N > 5$\footnote{Note that we also included VCC741, VCC1179, and VCC1273 despite having a lower $S/N$ detection at 100 $\mu$m.}.
We restricted the SED fitting to data-points between 100 \micron\ and 350 $\mu$m, because the
submm emission at 500 \micron\ in dwarf galaxies is usually
found to exceed that expected from the model SED \citep{2010A&A...518L..52G,2010A&A...518L..58O,2013A&A...557A..95R}.
The origin of the 500 \micron\ excess is still
not clear and we will discuss this issue in more detail in
Sect. ~\ref{sec:excess}.

\begin{figure}
   \centering
\includegraphics[bb=10 0 570 700,width=9cm]{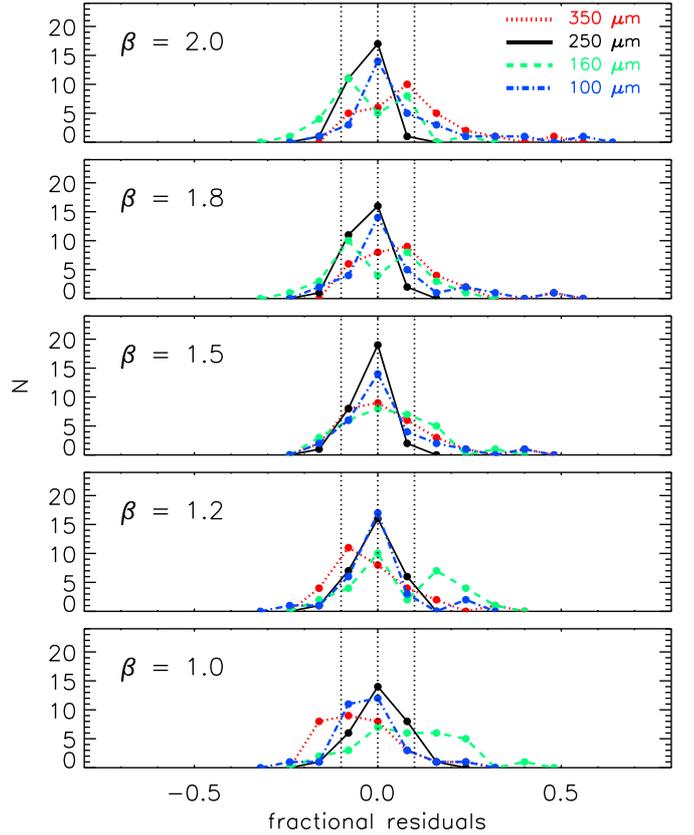} 
  \caption{Fractional residuals of the SED fitting at different wavelengths for $\beta = 1.0, 1.2, 1.5, 1.8, 2.0$. The fractional residual is calculated as the
difference at each wavelength between the measured flux density and best-fit
model divided by the best-fit model. The vertical dotted lines correspond to fractional residuals of 0 and $\pm$ 0.1.
The colours correspond to the
four wavelengths considered for the SED fitting: 350 $\mu$m (red), 250 $\mu$m (black), 160 $\mu$m (green), 100 $\mu$m (blue)}.
   \label{frac_res}%
\end{figure}

\subsection{Fixed-$\beta$ MBB fitting}
\label{sec:fixedbeta}

To establish the overall best-fit $\beta$ among the five adopted
values $\beta = $ [1.0, 1.2, 1.5, 1.8, 2.0] for the fixed-$\beta$ MBB fitting procedure,
we calculated the fractional residuals of the fits as the
difference between the measured flux density $F_{\nu}$ at 100, 160, 250, and
350 $\mu$m and the fitted function $S (\nu,\,T)$
divided by the best fit model. Then we compared the results for the
five $\beta$ values (Fig. \ref{frac_res}).
The dotted vertical lines indicates fractional residuals of 0 and $\pm$ 0.1.
The spread of the residuals for $\beta\,=\,1.5$  is smaller than
that for other emissivity indices, since most galaxies have residuals
below 0.1 in all four bands (70\%).
Moreover, unlike other $\beta$ values,
the residuals of all four bands for $\beta\,=\,1.5$, are
centred on 0.

As mentioned in the previous section, measured dust emissivity
variations among galaxies may be related to the issue of properly separating emission from warmer and
colder dust components \citep{2014ApJ...789..130K,2014ArXiv1409.1815B}, implying that a
colder diffuse dust could effectively be masked by warmer
components in single thermal component SED fits between 100 and 500 $\mu$m \citep{2012A&A...543A..74X}.
Therefore, as a further test, we repeated the fitting procedure with three data points only (160, 250 an 350 $\mu$m), using the 100 $\mu$m flux density as an upper limit, i.e. this data point was included in the SED fitting procedure only if the 160-350 $\mu$m fit resulted in an overprediction of the observed 100 $\mu$m measurement. Even in this case we obtained
that $\beta = 1.5$ provided the best output model. Both results are compared in Fig. \ref{fig:SEDfit_betafixed}, and this simple test shows that there are 7 galaxies for which performing a single-temperature MBB fit from 100 to 350 $\mu$m could hide the presence of a colder dust component blended with a warmer one \citep{2014ApJ...789..130K,2014ArXiv1409.1815B}.

Thus we will assume that for fixed $\beta$ MBB fitting,
$\beta\,=\,1.5$ is the best overall solution for the emissivity.
A modified black body with an emissivity index $\beta = 1.5$
is also found to better fit the SPIRE SED of the HRS galaxies \citep{2012A&A...540A..54B}.

\begin{table}
\center
\caption{Free-$\beta$ MBB fitting: best-fit parameters}
\label{fit_pars_bfree}
\begin{tabular}{lccr}
\hline \hline
    ID  &     $T$                &$\beta$  &  $\chi^2$    \\
        &    [K]               &         &              \\
        \hline \hline
   VCC1  &  32.7 $\pm$ 0.8  &  1.4  &   0.23  \\
  VCC10  &  18.5 $\pm$ 0.4  &  2.6  &   1.21  \\
  VCC87  &  23.3 $\pm$ 0.5  &  0.7  &   0.39  \\
 VCC135  &  23.9 $\pm$ 0.8  &  1.8  &   2.09  \\
 VCC144  &  27.6 $\pm$ 0.7  &  2.0  &   0.54  \\
 VCC172  &  17.6 $\pm$ 0.3  &  2.2  &   0.10  \\
 VCC213  &  25.6 $\pm$ 0.4  &  1.6  &   0.54  \\
 VCC223  &  22.0 $\pm$ 0.5  &  1.7  &   0.22  \\
 VCC281  &  30.8 $\pm$ 0.9  &  0.3  &   0.12  \\
 VCC324  &  35.0 $\pm$ 0.4  &  1.1  &   0.08  \\
 VCC334  &  19.9 $\pm$ 0.2  &  2.5  &   0.09  \\
 VCC340  &  34.8 $\pm$ 0.2  &  0.7  &   0.01  \\
 VCC562  &  47.4 $\pm$ 0.0  &  0.1  &   1.79  \\
 VCC693  &  19.3 $\pm$ 0.2  &  1.8  &   0.05  \\
 VCC699  &  28.4 $\pm$ 0.2  &  1.2  &   0.04  \\
 VCC737  &  26.9 $\pm$ 1.1  &  0.3  &   0.57  \\
 VCC741  &  29.2 $\pm$ 1.4  &  0.3  &   0.25  \\
 VCC841  &  23.5 $\pm$ 0.9  &  1.5  &   0.46  \\
 VCC848  &  16.8 $\pm$ 0.2  &  2.9  &   0.06  \\
VCC1179  &  21.2 $\pm$ 0.1  &  1.9  &   0.01  \\
VCC1273  &  25.5 $\pm$ 0.5  &  0.8  &   0.05  \\
VCC1356  &  29.2 $\pm$ 0.5  &  0.7  &   0.09  \\
VCC1374  &  20.2 $\pm$ 0.1  &  1.6  &   0.01  \\
VCC1437  &  21.4 $\pm$ 0.0  &  2.5  &   0.01  \\
VCC1554  &  24.9 $\pm$ 0.4  &  1.9  &   0.58  \\
VCC1575  &  20.9 $\pm$ 0.1  &  2.2  &   0.02  \\
VCC1686  &  21.4 $\pm$ 0.1  &  1.4  &   0.02  \\
VCC1699  &  32.0 $\pm$ 0.7  &  1.0  &   0.19  \\
VCC1725  &  30.7 $\pm$ 0.8  &  0.5  &   0.50  \\
VCC1791  &  16.8 $\pm$ 0.0  &  2.6  &   0.01  \\
\hline \hline
\end{tabular}
\end{table}

\subsection{Free-$\beta$ MBB fitting}
\label{sec:freebeta}

\begin{figure}
   \centering
\includegraphics[bb=70 10 550 560,width=7cm]{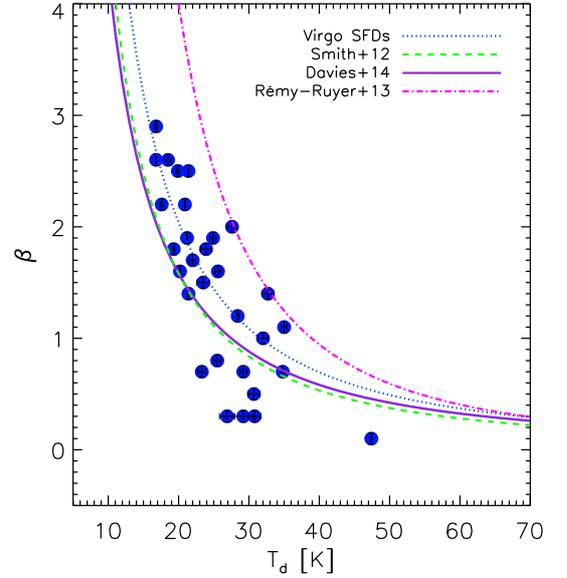}
  \caption{Emissivity index plotted against dust temperature for the Virgo SFDs (filled blue {dots}). The dotted line shows the best-fit power law to our data set. For comparison we overlay the $\beta - T_d$ relation found in Andromeda  \citep{2012ApJ...756...40S}, Virgo galaxies later than S0 \citep{2014MNRAS.438.1922D}, and DGS galaxies \citep{2013A&A...557A..95R}.}
   \label{beta_temp}%
\end{figure}

To further explore the range of possible emissivity indices
we repeated the fitting procedure for each galaxy with different values of $\beta$ within
the range 0 to 3 in steps of 0.1, selecting the index that results in the lowest $\chi^2$.
The best-fit SED models are shown in Fig. \ref{SEDfit_betafree}, and the results from the fitting  procedure
are displayed in Table \ref{fit_pars_bfree}.
Figure \ref{beta_temp} shows that the emissivity index varies substantially within the Virgo sample from $\beta = 0.1$ to 2.9. A few galaxies have a low $\beta$ value ($<$ 0.5); a flatter submm slope
may be an indicator of the presence of a submm excess \citep[see also Sect. \ref{sec:excess}]{2013A&A...557A..95R}, or of an extremely low ISRF \citep{2014arXiv1409.5916H}.
Figure \ref{beta_temp} shows the dust temperatures $T_d$
and $\beta$ indices for our sample of dwarfs (filled blue {circles}) and it indicates a clear anti-correlation between
the two parameters; the best-fit power-law which describes the relation between $\beta$ and $T_d$ is overlaid to the data\footnote{Galaxies with the lowest emissivity indices ($\beta \leq 0.3$) are not included in the fit.} (Fig. \ref{beta_temp}; blue dotted line) and it is given by

\begin{equation}
\beta = 2.04 \left( \frac{T_d}{20} \right)^{-1.55 \pm 0.06}
\end{equation}

\noindent
which is close to what was found by \citet{2012ApJ...756...40S} in the outer regions of Andromeda (red dashed line in Fig. \ref{beta_temp}),
even though the dwarfs extend to lower $\beta$ values compared to M31. A similar trend was also derived by \citet{2014MNRAS.438.1922D} combining all galaxies of the Virgo cluster later than S0 detected in the HeViCS survey (purple solid line in Fig. \ref{beta_temp}), while a steeper power-law was found by \citet{2013A&A...557A..95R} in the DGS ($\beta \propto T_d^{-2.08}$), characterised by overall higher dust temperatures compared to the Virgo SFDs ($T_d^{median} = 32$ K).
However, all these studies derived the $\beta-T_d$ relation using 100-500 $\mu$m data points in the SED fitting procedure (and even 70 $\mu$m data for some DGS galaxies), while our SED fittings were restricted to the wavelength range from 100 to 350 $\mu$m.

Although such an inverse relationship between $\beta$ and $T_d$ is found in FIR-submm studies of different environments of the Milky Way \citep{2010ApJ...713..959V},
Andromeda \citep{2012ApJ...756...40S}, and in other samples of galaxies \citep{2013A&A...557A..95R,2014MNRAS.440..942C,2014A&A...565A...4H},
 \citet{2009ApJ...696..676S} and \citet{2012ApJ...752...55K} warn against the presence of a $\beta - T_d $ correlation as a physical property
of the dust. These works suggest that there is a systematic degeneracy between $\beta$ and $T_d$ that could be due to effect of noise on the SED fitting technique,
as also shown in \citet{2014A&A...561A..95T}.
It follows that an artificial inverse $\beta - T_d$ correlation arises when a constant temperature along the line of sight is assumed to fit the properties of dust
grains which are likely to span a range of dust temperatures.

\subsection{Dust mass estimates for $\beta =$ 1.5}
\label{sec:dustmass}

Calculating dust masses of the
galaxies for the different values of the emissivity index in the case of free-$\beta$ SED fitting is not trivial.
Indeed, as recently shown by \citet{2013A&A...552A..89B}, varying $\beta$ while the value of dust opacity $\kappa_0$ is kept fixed leads to wrong dust mass estimates, because $\kappa_0$ is usually
calibrated on a dust model with a well defined $\beta$.
The correct determination of $\kappa_{\nu}$ can be assessed only if one has a consistent dust model for the corresponding value of $\beta$, or by comparing dust mass estimates obtained from SED fitting with the ones obtained
from other independent methods: e.g., using the amount of cold gas and metals, as proposed by \citet{2002MNRAS.335..753J}.

Therefore, given the difficulty of deriving the dust mass with a free emissivity index using
the scaling relation in the Milky Way for $\beta = 2$ \citep{2013A&A...552A..89B},
we decided to derive dust masses using the fixed-$\beta$ fitting result, choosing $\beta = 1.5$ as
the best compromise solution (see Sect. \ref{sec:fixedbeta}).

\begin{table}
\centering
\caption{Fixed-$\beta$ MBB fitting: dust temperatures for $\beta = 1.5$}
\label{tab:beta15}
\begin{tabular}{lclc}
\hline \hline
  ID      & $T_d$  &  ID      & $T_d$\\
          &   [K]  &          & [K] \\
\hline \hline
   VCC1  &  31.3$_{-0.9}^{+1.2}$  &     VCC737  &  18.3$_{-0.6}^{+0.6}$  \\
  VCC10  &  24.4$_{-0.4}^{+0.4}$  &     VCC741  &  19.6$_{-1.0}^{+0.9}$  \\
  VCC17  &  18.3                  &     VCC802  &  18.3                  \\
  VCC22  &  18.3                  &     VCC825  &  23.9                  \\
  VCC24  &  22.8$_{-1.0}^{+1.2}$  &     VCC841  &  23.5$_{-0.8}^{+0.8}$  \\
  VCC87  &  18.8$_{-0.4}^{+0.4}$  &     VCC848  &  23.9$_{-0.6}^{+0.8}$  \\
 VCC135  &  26.2$_{-0.3}^{+0.4}$  &     VCC888  &  16.1$_{-1.0}^{+1.1}$  \\
 VCC144  &  33.3$_{-0.8}^{+0.8}$  &     VCC985  &  20.4$_{-1.0}^{+1.2}$  \\
 VCC172  &  20.7$_{-0.6}^{+0.7}$  &    VCC1021  &  23.9                  \\
 VCC213  &  26.4$_{-0.4}^{+0.3}$  &    VCC1141  &  20.6$_{-0.9}^{+0.8}$  \\
 VCC223  &  23.3$_{-0.7}^{+0.7}$  &    VCC1179  &  23.8$_{-1.0}^{+1.0}$  \\
 VCC281  &  20.4$_{-0.7}^{+0.8}$  &    VCC1200  &  20.4$_{-1.4}^{+1.5}$  \\
 VCC286  &  21.1                  &    VCC1273  &  20.4$_{-1.0}^{+0.9}$  \\
 VCC322  &  20.5$_{-2.4}^{+2.3}$  &    VCC1356  &  22.4$_{-0.6}^{+0.8}$  \\
 VCC324  &  29.7$_{-0.5}^{+0.5}$  &    VCC1374  &  20.8$_{-0.5}^{+0.6}$  \\
 VCC328  &  22.5$_{-3.3}^{+3.9}$  &    VCC1437  &  29.0$_{-0.6}^{+0.7}$  \\
 VCC334  &  26.6$_{-0.8}^{+0.8}$  &    VCC1455  &  18.3                  \\
 VCC340  &  25.5$_{-0.4}^{+0.6}$  &    VCC1554  &  28.5$_{-0.4}^{+0.4}$  \\
 VCC367  &  18.3                  &    VCC1575  &  25.6$_{-0.3}^{+0.3}$  \\
 VCC446  &  20.1$_{-1.6}^{+1.5}$  &    VCC1675  &  20.6$_{-1.3}^{+1.5}$  \\
 VCC562  &  24.9$_{-0.9}^{+1.0}$  &    VCC1686  &  20.8$_{-0.3}^{+0.3}$  \\
 VCC620  &  18.3                  &    VCC1699  &  26.5$_{-0.7}^{+0.7}$  \\
 VCC641  &  18.3                  &    VCC1725  &  21.7$_{-0.5}^{+0.5}$  \\
 VCC693  &  20.8$_{-0.6}^{+0.5}$  &    VCC1791  &  21.9$_{-0.6}^{+0.5}$  \\
 VCC699  &  25.6$_{-0.4}^{+0.4}$  &             &                        \\
\hline \hline
\end{tabular}
\end{table}

For 10 galaxies with only two data points (at $\lambda \leq 350 \mu$m)
we performed the SED fitting with a fixed dust temperature using three values:  $T_d = 23.9$ K, the median temperature obtained
from the $\beta$\,=\,1.5 fits for the 30 galaxies with better quality photometry (Sect. ~\ref{sec:fixedbeta});
$T_d = 18.3$ K, the minimum value found in this subsample; $T_d = 21.1$ K, an intermediate value between the minimum and the median. Then we selected the temperature that provided the best fit with the lowest
 $\chi^2$.
The results for $\beta = 1.5$ are shown in Fig. \ref{fig:SEDfit_betafixed}, and the corresponding dust temperatures are displayed in Tab. \ref{tab:beta15}. The median dust temperature of the 39 galaxies for which the SED fitting could be performed leaving $T_d$ as a free parameter is $T_d^{median} =$ 22.4 K.

Dust masses for the 49 SFDs were then derived from the MBB fits according to

\begin{equation}
F_{\nu} = \frac{M_d \kappa_{0}}{D^2} \left( \frac{\nu}{\nu_0} \right)^{\beta} B (\nu,\, T)
\label{eq:dust_mass}
\end{equation}

\noindent
with $\kappa_{0}$ = 3.4 cm$^2$ g$^{-1}$ at
$\lambda$ = 250 $\mu$m, following the prescription of \citet{2013A&A...552A..89B}.
This value reproduces the average emissivity of the Milky Way dust in
the FIR-submm for $\beta$ = 1.5 \citep{2013A&A...552A..89B}.
Errors on the best-fit model parameters ($T_d$, $M_d$) were estimated via a
bootstrap technique. For each galaxy we created 300 new sets of data points randomly selected
within the error bars of the observed fluxes. Then we repeated the fitting procedure for each
new data set and determined the best fitting parameters. We calculated the 68\% confidence interval
in the parameter distributions and defined the edges of this interval as the
new upper and lower limits. The final uncertainties were given by the difference between the original best-fit solution and the upper and lower limit values from the bootstrap technique.
Dust masses of Virgo SFDs are given in Table \ref{Tab:VirgoSFDs}\footnote{For the 7 galaxies discussed in Sect. \ref{sec:fixedbeta} which might host a colder dust component blended with a warmer one, the four point fit might underestimate the dust mass by a factor of 0.1 - 0.2 dex.}.

For an average rms of 6.7 mJy/beam at 250 $\mu$m (see Sect. ~\ref{sec:stacking}) the 3$\sigma$ dust mass detection limit assuming a dust temperature $T_d^{median} = 22.4$ K and a distance of 17 Mpc is $M_d \simeq 4 \times 10^4$ M$_{\odot}$.
Regarding FIR non-detections, given the flux density derived in Sect. ~\ref{sec:stacking} ($F_{250} =$ 4.5 mJy), the average dust mass calculated with the same parameters ($\kappa_0$, $T_d^{median}$, $D = 17$ Mpc) corresponds to $M_d$ = $8.7 \times 10^3$ M$_{\odot}$. The average dust mass of the detected dwarfs is $M_d = 3 \times 10^5$ M$_{\odot}$.

To perform a homogeneous comparison of the different surveys, we
recalculated the dust masses of the DGS, KINGFISH, and BGC galaxies in the same way, i.e. we fitted a MBB with $\beta = 1.5$ to the {\em Herschel} flux densities
and we determined the uncertainties on $T_d$ and $M_d$
with the bootstrap technique as explained above.
Their values are given in the tables in Appendix \ref{app:data_tables}.
Comparison with \citet{2013A&A...557A..95R}, where DGS and KINGFISH dust masses were calculated using a free-$\beta$ emissivity, including the 500 $\mu$m data point in the SED fitting, shows that overall a fixed-$\beta$ MBB fitting provides larger dust masses. For KINGFISH the difference between ours and their estimates peaks at 0.15 dex with a dispersion of $\pm$ 0.05. Regarding the DGS, the logarithmic difference
between the two estimates is scattered between -0.2 and +1.7 dex, however for 17 out of 27 galaxies the two measurements are consistent within the uncertainties.

\section{Properties of Virgo SFDs: FIR detections versus FIR non-detections}
\label{sec:det_vs_nondet}

\begin{figure}
   \centering
\includegraphics[bb=20 20 420 460,width=8cm]{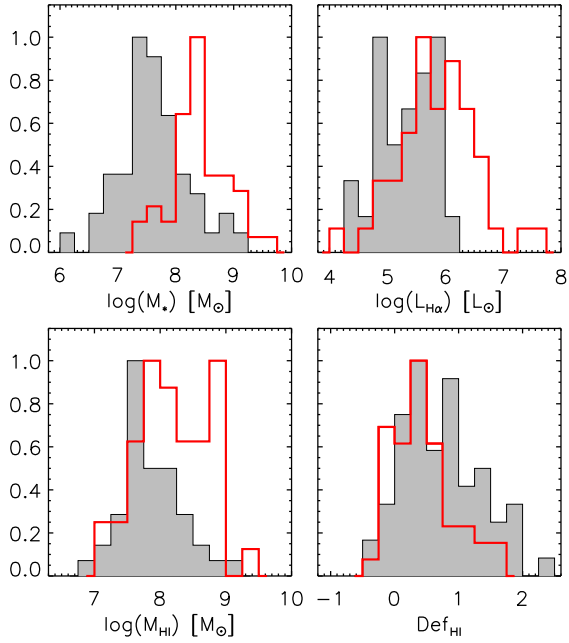} 
\caption{Stellar mass, H$\alpha$ luminosity, \hi mass, and \hi deficiency for the sample of FIR-detected (red histogram) and FIR-nondetected (filled grey histogram) Virgo dwarfs. All parameters are taken from the GOLDMine database, including the stellar masses of the {\em Herschel}-detected SFDs.}
   \label{fig:det_undet}%
    \end{figure}

Our analysis of the HeViCS data led to the selection of 49 SFDs with a FIR-submm counterpart. If we consider only dwarfs brighter than $m_B < 18$ mag, the completeness limit of the VCC catalogue, this gives a detection rate of 43\%.

The spatial distribution of Virgo SFDs can be seen in Fig. \ref{fig1}. Late-type dwarfs
are usually located at larger distances from the centre of clusters and tend to avoid the densest regions \citep{1987AJ.....94..251B}.
As expected, {\em Herschel}-detected SFDs are preferentially located in the less dense regions of the cluster.
Only five dwarfs are within 2 degrees of M87 and only two are  within 1.4 degree of M49\footnote{Low FIR detection rates in cluster A and B are also observed in the Virgo early-type dwarf population \citep{2013A&A...552A...8D}}. The other detections are distributed between the LVC,
the southern extension, the background clouds (W$^{\prime}$, W, M), and the region between cluster A and B.
The background clouds (M and W)
contain about one third of the detected SFDs,
according to the membership assignments of GOLDMine.

In this section we use global parameters of the whole sample of Virgo dwarfs
to investigate whether
FIR detections and non-detections have distinctive global properties.

Figure \ref{fig:det_undet} compares the properties of Virgo late-type dwarfs brighter than $m_B < 18$ mag, 49 with a FIR counterpart and 64 without.
Stellar masses\footnote{To facilitate the comparison, in this section we use stellar masses from GOLDMine (derived from optical photometry ($g$, $i$) as explained in Sect.  \ref{subsec:ancillary_stellmass_Virgo_SFD}) for both detected and undetected galaxies, because we did not measure {\em WISE} 3.4 $\mu$m photometry for Virgo SFDs without a FIR counterpart.}, H$\alpha$ fluxes, \hi masses, distances, and optical diameters (to derive \hi deficiencies), were taken from the GOLDMine database.  The red histograms in the figure show the {\em Herschel} detections, while the filled grey histograms correspond to the non-detections.
All histograms are normalized to their maximum values.

FIR-undetected galaxies have overall lower stellar masses, as it can be seen in the top left-hand panel of Fig. \ref{fig:det_undet}; the distribution peaks at $\log$($M_{\star}$/M$_{\odot}$) = 7.4, an order of magnitude lower compared to the detected sample.
Only 44\% of the dwarfs without a FIR counterpart have a H$\alpha$ detection,
and their H$\alpha$ luminosities do not exceed $\sim 10^6$ L$_{\odot}$.
The \hi mass distribution ranges for both samples between $10^7$ and $10^9$ M$_{\odot}$,
but FIR-emitting dwarfs have a higher fraction of \hi masses above $10^8$ M$_{\odot}$, and a higher detection rate at 21 cm (90\% against 67\%).
Finally, in the last panel we compare the \hi deficiency (including 21-cm upper limits) for both type of galaxies, showing that the sample of undetected dwarfs have a larger
fraction of objects with higher H{\sc i} deficiencies.
Most of the H{\sc i}-poor
FIR non-detections are found in cluster A and B, and in the region between these two
substructures.
Concerning the dwarf morphological types, BCDs show the highest detection rate (64\%), followed by Sm (46\%), and Im (24\%) galaxies.

The main conclusion to infer from the figure is then that our detections are "biased"
towards dwarfs with higher stellar and gas masses, less \hidef, and more star-forming.
Assuming the average dust-to-stellar mass ratio of dwarfs with a FIR counterpart ($M_d/M_{\star} \sim 10^{-3}$), galaxies
with $\log$($M_{\star}/$M$_{\odot}$) = 7.4 (the peak of the grey histogram in Fig. \ref{fig:det_undet}) would have dust masses below the 3$\sigma$ detection limit of the HeViCS survey determined in Sect.  \ref{sec:dustmass}.

There is not enough information in the SDSS spectra to derive oxygen abundances for the non-detected galaxies,
therefore we cannot assess whether dwarfs without a FIR counterpart are characterised by a lower metal content.

\begin{table}
\center
\caption{Variation of submm excess with MBB fitting procedure.}
\label{subex_tab}
\begin{tabular}{lcccc}
\hline \hline
   ID  & $\varepsilon_{500}^{\beta=2.0}$ & $\varepsilon_{500}^{\beta=1.5}$ & $\varepsilon_{500}^{\beta=1.0}$ & $\varepsilon_{500}^{\beta_{\text{free}}}$ \\
\hline \hline
  VCC10  & ...     &   ...    &  ...     &  0.36  \\
  VCC87  & 0.68    &   0.42   &  0.19    &  ...   \\
 VCC135  & 0.48    &   0.22   &  ...     &  0.37  \\
 VCC144  & 0.56    &   0.26   &  ...     &  0.56  \\
 VCC172  & 0.50    &   0.23   &  ...     &  0.62  \\
 VCC213  & 0.19    &   ...    &  ...     &  ...   \\
 VCC281  & 0.98    &   0.67   &  0.40    &  ...   \\
 VCC324  & 1.01    &   0.63   &  0.34    &  0.39  \\
 VCC340  & 0.69    &   0.39   &  0.15    &  ...   \\
 VCC562  & 2.08    &   1.52   &  1.06    &  0.45  \\
 VCC699  & 0.35    &   ...    &  ...     &  ...   \\
 VCC737  & 0.46    &   0.26   &  ...     &  ...   \\
 VCC741  & 1.41    &   1.04   &  1.07    &  0.36  \\
VCC1179  & 2.87    &   2.21   &  1.67    &  2.72  \\
VCC1356  & 1.21    &   0.83   &  0.52    &  0.37  \\
VCC1374  & 0.67    &   0.39   &  ...     &  0.44  \\
VCC1437  & 0.49    &   0.22   &  ...     &  0.82  \\
VCC1554  & 0.21    &   ...    &  ...     &  0.16  \\
VCC1686  & 0.13    &   ...    &  ...     &  ...   \\
VCC1699  & 0.43    &   ...    &  ...     &  ...   \\
VCC1725  & 0.68    &   0.38   &  0.13    &  ...   \\
VCC1791  & 0.73    &   0.47   &  0.26    &  1.10  \\
 \hline \hline
\end{tabular}
\end{table}

\section{The 500 $\mu$m excess}
\label{sec:excess}

\begin{figure}
   \centering
\includegraphics[bb=30 -10 560 710,width=8cm]{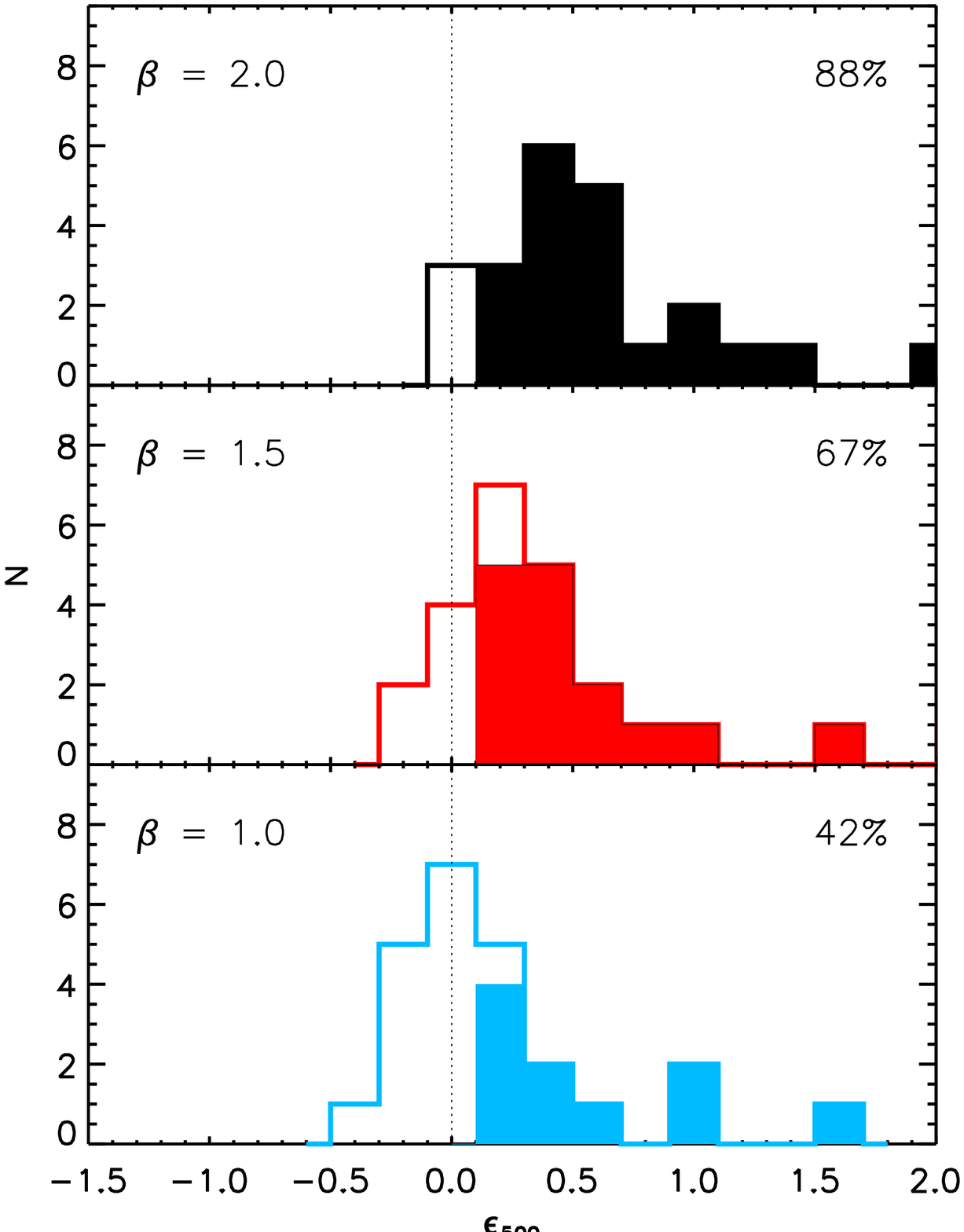} 
\includegraphics[bb=30 -15 540 360,width=8cm]{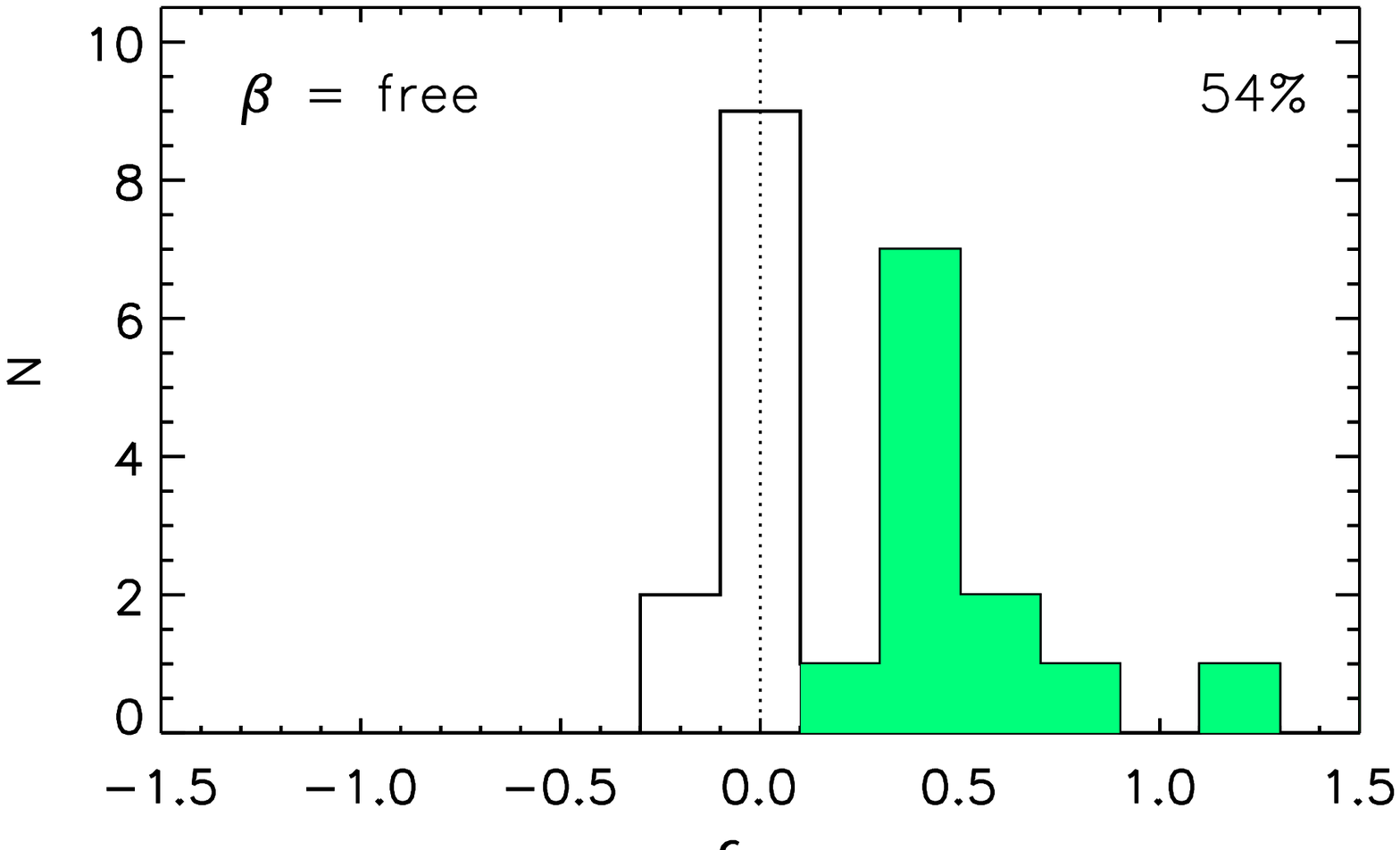} 
\caption{500 \micron\ fractional residuals for fixed- and free-$\beta$ MBB fitting. The filled histograms correspond to the
galaxies whose 500 $\mu$m excess is larger than the error on the flux density measurement. The fraction of dwarfs with a submm excess is displayed at the {top-right} corner of each panel.}
   \label{excess}%
    \end{figure}

Several works have recently found that the SEDs of late-type galaxies exhibit
emission at submm  and millimetre (mm) wavelengths in excess of what is expected when a single modified
Planck function is fitted. Such a submm excess,
is preferentially found in dwarf/irregular/Magellanic morphological types
\citep{2002A&A...382..860L,2003A&A...407..159G,2005A&A...434..867G,2009A&A...508..645G,2011A&A...532A..56G,2010A&A...523A..20B,2013A&A...557A..95R,2014A&A...565A.128C},
with only a few cases of moderately low-metallicity spiral galaxies
\citep{2004A&A...414..475D,2006ApJ...652..283B,2009ApJ...706..941Z}.

In the analysis of the Science Demonstration Phase (SDP) data set of the HeViCS survey, we found
that the 500~$\mu$m fluxes of two out of three SFDs tended to be underestimated by a single-temperature dust component fit, showing a submm excess emission \citep{2010A&A...518L..52G}.
Here we want to exploit the higher sensitivity of the completed survey, and the larger sample of detected dwarfs to derive more stringent constraints on the excess emission at 500 $\mu$m in Virgo SFDs.

We defined the 500 \micron\ excess in the same way as has been done in other studies \citep{2012ApJ...745...95D,2013A&A...557A..95R}, i.e.:

\begin{equation}
\varepsilon_{500} = \frac{F_{500} - S_{500}}{S_{500}}
\end{equation}

\noindent
where $F_{500}$ is the observed flux density and $S_{500}$ the model flux density at 500 \micron.
We determined $\varepsilon_{500}$ for both fixed- and free-$\beta$ SED fitting, including
only 500 \micron\ detections with $S/N > 5$. Thus we used 23 out of 30 galaxies with the
best FIR-submm photometry (see Sect. ~\ref{sec:mbb}).

Fig. \ref{excess} shows the variation with $\beta$ of the fractional residuals at 500 \micron.
We assume that a 500 $\mu$m excess is observed
if $F_{500} - S_{500} > \sigma_{500}$, where $\sigma_{500}$ is the error on the flux density ({filled} histograms). As expected the number of galaxies with
an excess decreases with $\beta$ (Fig. \ref{excess}). For $\beta = 1.5$, 67\% of the detections present a 500 $\mu$m excess\footnote{In the case of the three-point SED fitting procedure discussed in Sect. \ref{sec:fixedbeta} (where the 100 $\mu$m data point is used as an upper limit), the fraction of dwarfs with a submm excess decreases only marginally, with percentages of 79\%, 63\%, and 33\% for $\beta =$ 2.0, 1.5, and 1.0, respectively.}, and even when $\beta$
is allowed to vary, the fraction of galaxies with a stronger submm emission is still as significant (54\%; Fig. \ref{excess}).
As the emissivity index decreases the fitted submm spectrum flattens at long wavelengths reducing the gap between
the model SED and the observed flux density (see also Tab. \ref{subex_tab}).
Therefore the selection of lower beta values would result in an overall lower fraction of galaxies with a detected submm excess.

\begin{figure*}
   \centering
\includegraphics[bb=130 -20 730 430,width=12cm]{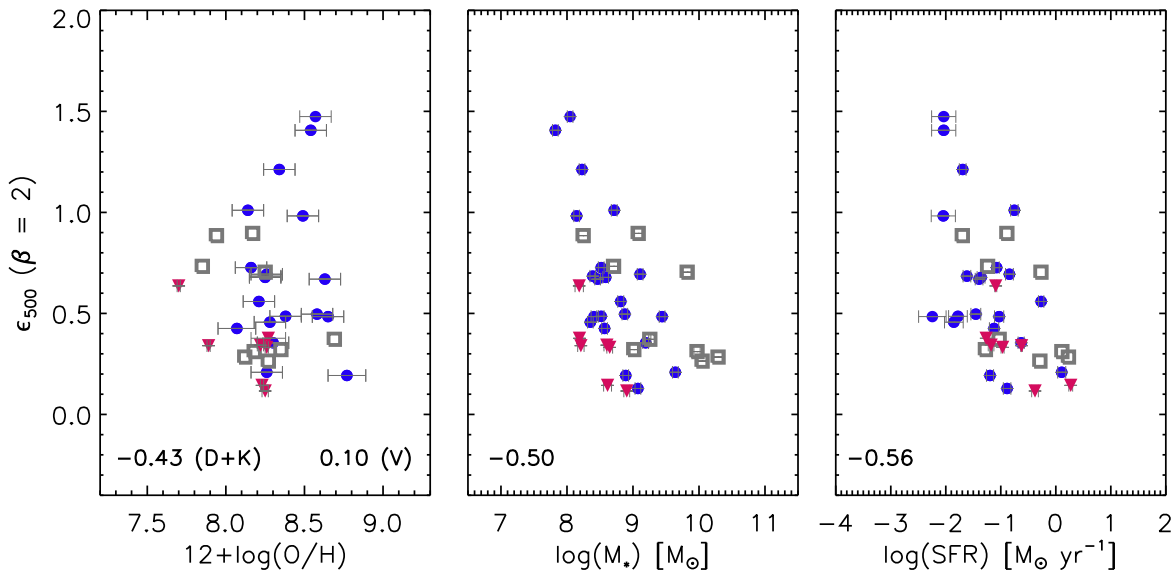}
\caption{500 \micron\ fractional residuals for $\beta = 2$ against metallicity ({\em left panel}), stellar mass ({\em central panel}), and SFR ({\em right panel}) for the three samples of dwarfs: Virgo SFDs (blue dots), DGS (purple diamonds), and KINGFISH dwarfs (grey squares).
Pearson correlation coefficients defining the degree of correlation are displayed
in each panel.
The fractional residuals of all three samples show a higher correlation with stellar mass and star formation rate.}
   \label{fig:excess_corrs}%
    \end{figure*}

If the excess emission is due to a
change in the emissivity properties of the dust, one should expect to find a correlation with metallicity or other global properties of the galaxies.
To probe whether such a link exists, we plot $\varepsilon_{500}$ for the most extreme case ($\beta = 2$) as a function of metallicity, stellar mass, and SFR (left, central, and right panel of Fig. \ref{fig:excess_corrs}, respectively).
We include also galaxies from the DGS and KINGFISH surveys showing a similar excess.
Seven objects from the DGS and 9 from KINGFISH satisfy the conditions adopted to define the presence of a submm excess in Virgo dwarfs ($(S/N)_{500} > 5$, $F_{500} - S_{500} > \sigma_{500}$). As regards the metallicity, we need to compare the Virgo dwarfs and DGS/KINGFISH (D+K) galaxies separately because of the different calibration used to derive the oxygen abundances. The left-hand panel shows that, despite the small number of objects, the excess is moderately anticorrelated with metallicity for the D+K samples but there is no correlation for Virgo SFDs. However we find a clear link between the excess with both stellar mass and star formation rate for all three surveys (Pearson coefficient correlation of -0.50 and -0.56, respectively). If stellar mass is a proxy for metallicity
\citep{2004ApJ...613..898T,2013ApJ...765..140A}, the central panel of Fig. \ref{fig:excess_corrs} suggests that
the metallicities derived from the SDSS might be poorly constrained since our estimates were based on only two calibrations (see Sect. ~\ref{sec:ancillary_met}). The last panel shows that the excess is stronger in galaxies with a lower SFR.

Previous analysis of the link between the submm excess and global galaxy properties found different results:
\citet{2006ApJ...652..283B} reported an anticorrelation with the total infrared luminosity in NGC4631, while \citet{2014MNRAS.439.2542G} did not detect a clear trend between the relative excess at 870 $\mu$m and the 24 $\mu$m surface brightness in a set of KINGFISH galaxies.

In the study of the full DGS and KINGFISH samples, \citet{2013A&A...557A..95R} found that about 45\% of the 110 galaxies have an excess emission above the SED model at 500 $\mu$m (adopting a free-$\beta$ MBB fitting),
that this feature is mainly detected in dwarfs with metal abundances Z $<$ 0.4 Z$_{\odot}$, and the most metal-poor dwarfs of the DGS sample host the strongest excesses.

Indeed, the submm/mm excess is still an open issue that challenges standard dust models in this regime.
Several hypotheses have been introduced so far to explain the peculiar dust properties at these wavelengths.
Initially, the discovery of this feature in star-forming dwarfs was interpreted as evidence for a very cold dust component
\citep{2003A&A...407..159G,2005A&A...434..867G,2009A&A...508..645G};
however, the unphysically large dust masses implied by this scenario
are difficult to reconcile with models of dust production
and with the expected dust-to-gas ratios \citep{2002A&A...382..860L,2004A&A...414..475D,2006ApJ...652..283B,2009ApJ...706..941Z,2011A&A...536A..19P}.

Alternatively, an enhanced abundance of hot, small dust grains ($T \sim$ 30 - 50 K; sizes between 1.2 and 15 nm) with a low emissivity
was suggested to explain the submm excess of the dwarf galaxy NGC1569 \citep{2002A&A...382..860L} and of NGC3310 \citep{2009ApJ...706..941Z}.
In this scenario, large grain destruction by supernovae induced shocks in the ISM would produce the enhanced abundance of small grains.

Another hypothesis suggests that the emission arises from rotating very small dust grains ($<$ 1.2 nm) with permanent electric dipole moments
located in the ionised gas ({\em spinning dust}).
Grain collisions with the ionised gas and interactions with the UV radiation field can excite rotation of dust particles \citep{1994ApJ...427..155F,1998ApJ...494L..19D}.
For example, the predicted emission spectrum of spinning dust grains was invoked to account for the mm excess in the Small Magellanic Cloud
\citep[SMC; ][]{2010A&A...523A..20B,2011A&A...536A..17P}.

A population of magnetic dust grains (magnetite, maghemite, and metallic iron) with sizes smaller then 10 nm, mixed with ``normal" dust grains has been recently
suggested as an explanation for the SMC submm excess \citep{2012ApJ...757..103D}. Low-metallicity environments such as the SMC may
provide more favorable conditions for the production and survival of iron-rich dust grains.
Other studies reported that the properties of
amorphous dust grains can depend strongly on temperature and that hotter dust grains can have low emissivity indices \citep{2007A&A...468..171M}.

An emissivity variation with wavelength parameterised by a single-temperature MBB with a broken power-law has been advocated by \citet{2014arXiv1406.6066G} as a better model to explain the submm excess in the Magellanic clouds than the introduction of an additional population of very cold dust.

Finally, submm excess emission relative to a single-temperature MBB
could not be necessarily related to peculiar dust properties, but rather it could be the
consequence of temperature mixing along the line of sight \citep{2009ApJ...696..676S}, as also discussed in Sect. \ref{sec:mbb}.
Because the measured emissivity includes both the intrinsic emissivity of the dust and the range of temperatures of the different dust components,
the temperature mixing could produce a shallower {\em apparent} $\beta$ than what one would measure in the ideal case of a single-temperature component.

If the excess is the result of different grain properties our results may support two among the scenarios discussed above.
The submm emission of small dust grains is expected to be lower than that of large dust grains, heated by both young and old stellar populations (see Sect. ~\ref{sec:dust_star_form}), thus it could be more easily detectable in galaxies with lower star formation rates and lower masses as we show in Fig. \ref{fig:excess_corrs}.
On the other hand, the excess anticorrelation with stellar mass (hence with metallicity) may favour the \citet{2012ApJ...757..103D} scenario which assumes that metal-poor ISM may host a larger fraction of iron-rich dust grains emitting at submm/mm wavelengths.

\section{Dust as a probe of galaxy evolution}
\label{sec:probe}

\subsection{Dust and star formation activity}
\label{sec:dust_star_form}

\begin{figure}
   \centering
\includegraphics[bb=100 10 520 550,width=7cm]{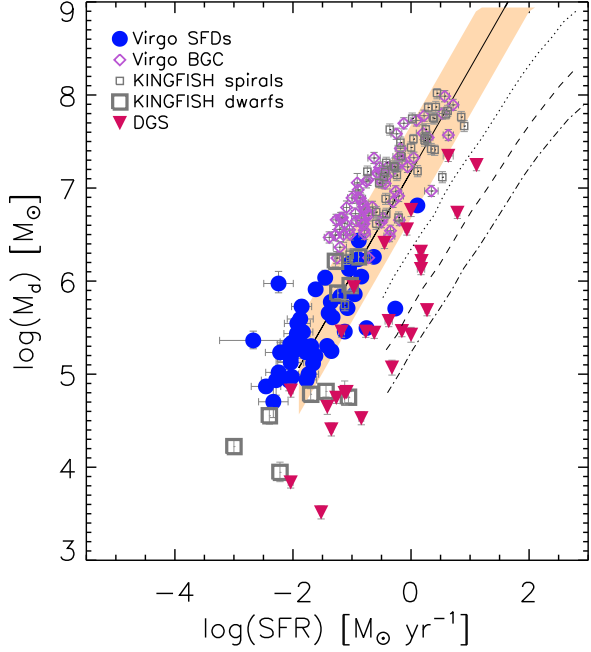}
\caption{Dust mass versus star formation rate for the Virgo SFDs (filled blue {dots}) compared to the KINGFISH
 spirals and dwarfs (small and large grey {squares}), and the DGS (red-purple {triangles}). The solid line shows the relation determined by da Cunha et al. (2010) for an \iras\ selected sample of local star-forming galaxies. The orange shaded area stands for the dispersion of the relation. Evolution of the $M_d - SFR$ relation for different amounts of dust mass destroyed by a single supernova event -- $M_{cl}$ = 100, 500, 1500 M$_{\odot}$ -- is indicated by the dotted, dashed, dot-dashed lines, respectively \citep{2014ApJ...782L..23H}.}
   \label{Mdust_sfr}%
    \end{figure}

Dust plays a fundamental role in regulating global star formation histories of galaxies and their evolution.
Here we explore the relation between dust and star formation activity, comparing the properties of the Virgo SFDs to other {\em Herschel} surveys of
dwarfs and late-type galaxies in different environments.

In Fig. \ref{Mdust_sfr}, we plot dust masses $M_{{d}}$ versus SFRs in the
Virgo dwarfs and the three comparison samples:
KINGFISH, DGS, and HeViCS BGC.
The galaxies in our sample except the DGS follow the best-fitting relation
derived from \citet{2010MNRAS.403.1894D} for an \iras-selected sample of local star-forming
galaxies ({solid line} in Fig. \ref{Mdust_sfr}).
This correlation spans four orders of magnitude in both SFR and $M_d$.
It has been shown that the slope of the $M_d - SFR$ dependence can be related to the global Schmidt-Kennicutt law exponent \citep{2014ApJ...782L..23H}.
Evolutionary models of \citet[][]{2014ApJ...782L..23H} show that starbursting galaxies are expected to be located below the relation of da Cunha et al. (2010) because of the increasing contribution of supernovae to dust destruction in such systems.
Dotted, dashed, and dot-dashed lines in Fig. \ref{Mdust_sfr} correspond to the evolution of the $M_d - SFR$ relation for different amounts of dust mass destroyed by a single supernova event : $M_{cl}$ = 100, 500, 1500 M$_{\odot}$ \citep[see for details][]{2014ApJ...782L..23H}. Although their models apply to more massive systems, this can give a hint to explain the scatter between the DGS and the other samples.

A large fraction of the radiation emitted from young stars is absorbed and re-emitted by dust.
Whereas it is commonly assumed
that warm dust is heated by young stars, the heating source of the diffuse cold dust emission in galaxies is still under debate \citep{2011AJ....142..111B,2012MNRAS.419.1833B,2012A&A...540A..54B,2013MNRAS.432.2182F,2014A&A...565A...4H}.
\hers observations of nearby spiral galaxies suggest that the cold dust is heated by evolved stars rather than star forming regions \citep{2010A&A...518L..65B,2011AJ....142..111B,2012MNRAS.419.1833B}.
On the other hand, diffuse dust might be less shielded from \hii regions in SFDs
because of their less dense ISM and low-metallicity environment, possibly making young stars
a dominant source of dust heating in these systems \citep{2010A&A...518L..55G}.

\begin{figure}
   \centering
\includegraphics[bb=30 0 710 540,width=10.4cm]{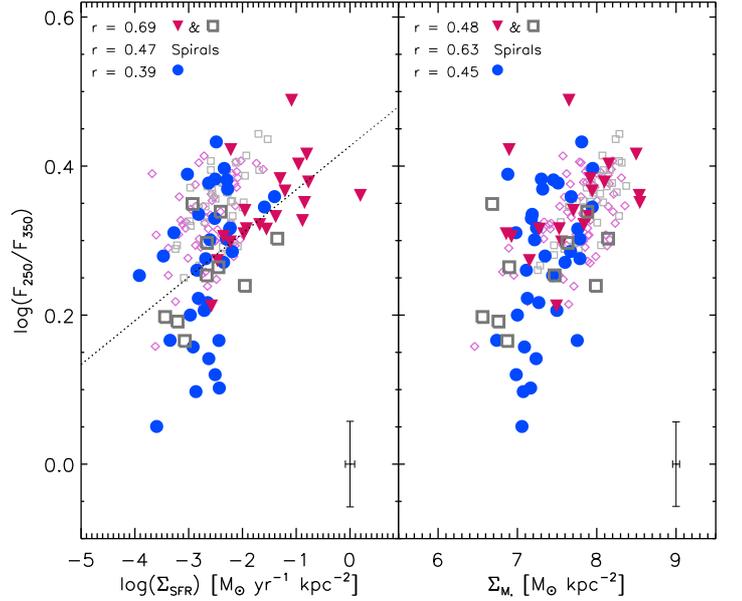}
\caption{{\em Left panel}: SPIRE colour $F_{250}/F_{350}$ against star formation rate surface density. Symbols are the same used in Fig. \ref{Mdust_sfr}.
The dotted line is a least square fit to the DGS and KINGFISH dwarf galaxies.
{\em Right panel}: SPIRE colour $F_{250}/F_{350}$ against stellar mass surface density Pearson correlation coefficients for the different samples are displayed at the top-left corner of each panel. The mean error bars are displayed at the bottom-right corner of each panel.}
\label{sfr_dust_sfe}
\end{figure}

To assess the dust heating mechanism in dwarfs
we plot in Fig. \ref{sfr_dust_sfe} the $F_{250}/F_{350}$ colour as a function of SFR and stellar mass
surface density for all samples. Spiral galaxies from the HeviCS BGC and KINGFISH samples are also included for comparison.
SPIRE colours trace the properties of the Rayleigh-Jeans tail of the cold dust emission, and it has been shown that $F_{250}/F_{350}$ can be tightly correlated to the surface brightness of both the ionising
and non-ionising interstellar radiation fields, giving hints about the dust heating mechanism \citep{2012A&A...540A..54B}.

All samples are correlated with both parameters, although to different degrees,
as shown by the Pearson correlation coefficients at the top left-hand corner of each panel.
The strongest correlation with
star formation surface density is found for the KINGFISH and DGS samples ($r = 0.69$). However the FIR colour of these dwarfs is also moderately correlated with the mass surface density ($r = 0.48$), suggesting that star formation is not the only mechanism responsible for dust heating in active SFDs. Regarding Virgo dwarfs, there is a moderate correlation between $F_{250}/F_{350}$ and the SFR ($r = 0.39$) and stellar mass ($r = 0.45$) surface densities.
Most massive galaxies are more strongly correlated with the stellar surface density ($r = 0.63$) rather than with SFR ($r = 0.47$).

This analysis suggests that,
consistent with what observed in other {\em Herschel} surveys
\citep{2012A&A...540A..54B,2014A&A...565A...4H,2014ArXiv1409.1815B}
both young and more evolved stellar populations contribute to
the heating of the cold dust component in late-type dwarf galaxies, but the contribution of ionising interstellar radiation is stronger in more active SFDs.

\subsection{Dust scaling relations: evidence for dust stripping?}

The dust-to-stellar mass ratio and the stellar mass are found to be mutually anticorrelated  \citep{2010MNRAS.403.1894D,2012A&A...540A..52C}, meaning that more massive galaxies have lower specific dust masses. This has been explained as the result of the correlation between sSFR and stellar mass:
because of the higher star formation activity a large fraction of dust is formed, exceeding the amount
of dust grains destroyed in the ISM \citep{2007ApJS..173..315S,2010MNRAS.403.1894D,2012A&A...540A..52C}. At higher stellar mass, the sSFR and gas
fraction start decreasing and dust destruction begins to dominate over dust production, affecting the total dust mass of a galaxy.
Such a trend is also confirmed by simulation of the time evolution of dust properties of late-type galaxies \citep{2013MNRAS.432.2298B}.

\begin{figure}
\centering
\includegraphics[bb=90 10 520 550,width=7cm]{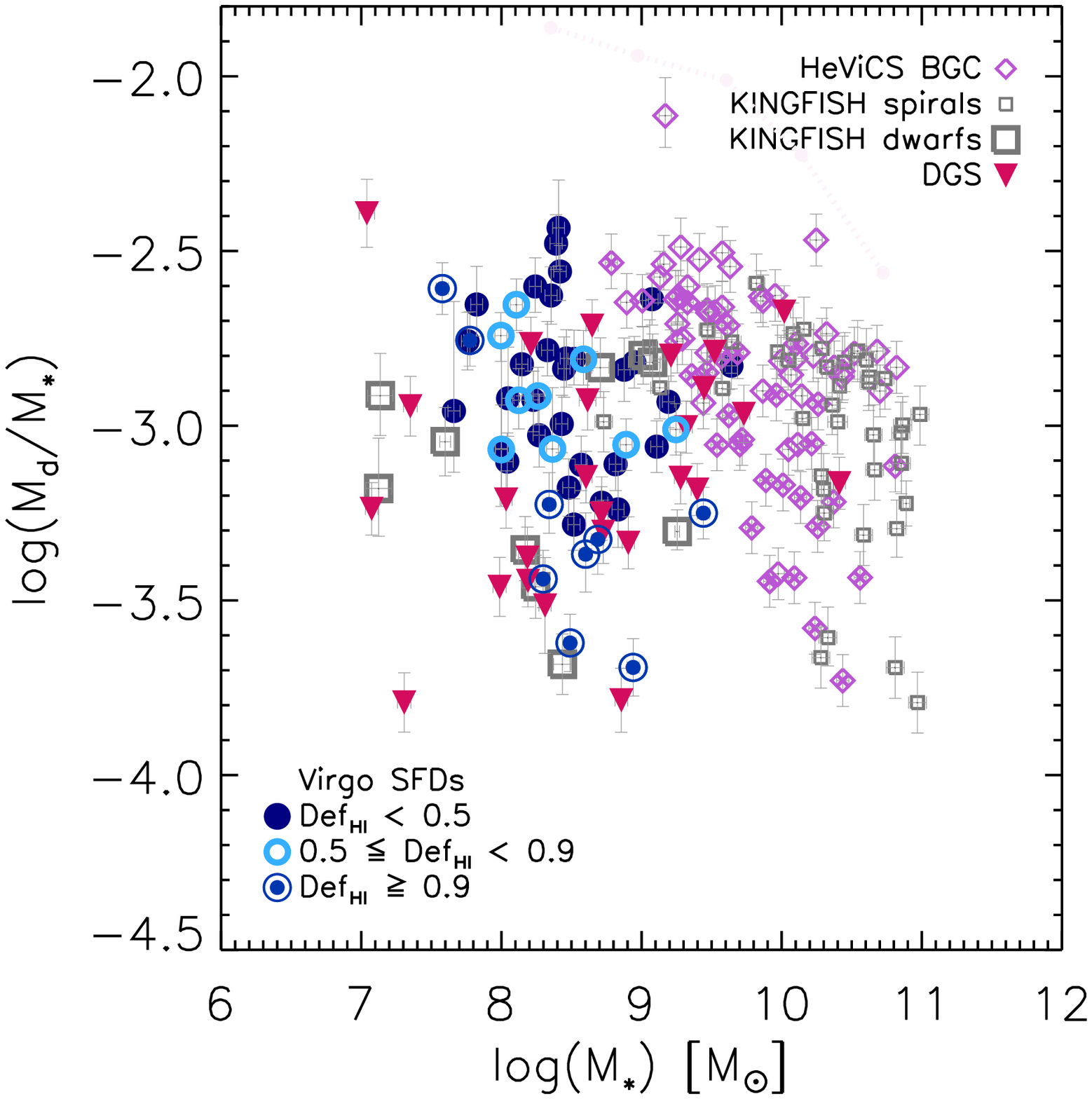}
\includegraphics[bb=40 10 625 410,width=8.2cm]{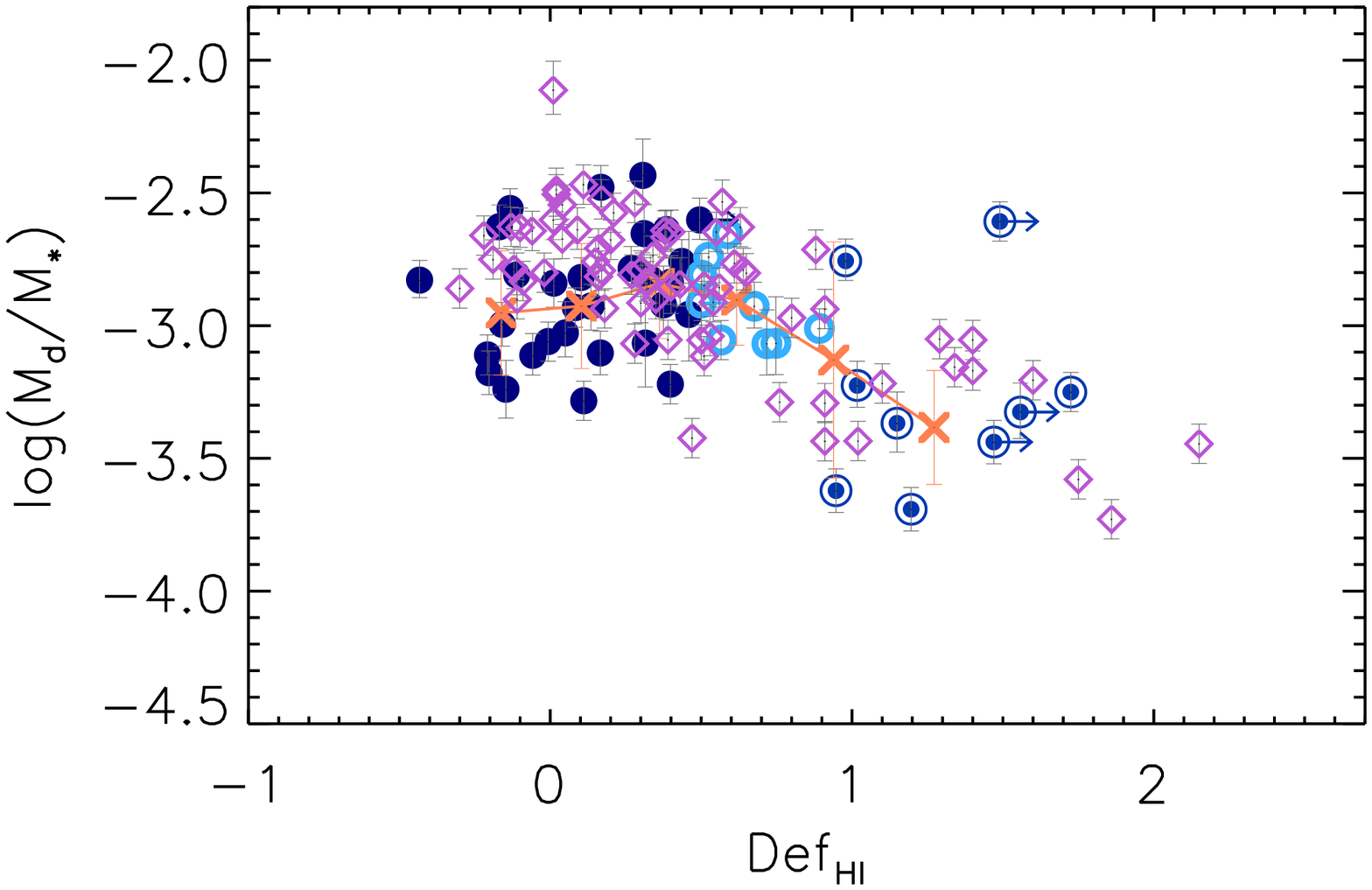}
  \caption{{\em Upper panel:} Dust-to-stellar mass ratios versus stellar masses of Virgo SFDs ({filled dots, rings, ringed dots}).
  The different shapes and gradation of blue correspond to three ranges of the H{\sc i}-deficiency parameter as defined in Fig. \ref{ssfr_mstar}. For comparison, we show
   data for HeViCS BGC (purple {diamonds}), KINGFISH dwarfs and spirals (large and small grey {squares}), and DGS (red-purple {triangles}). {H\sc i}-deficient HeViCS BGC galaxies ($Def_{HI} \geq 0.5$) are indicated by a diamond with a cross. {\em Lower panel:} Dust-to-stellar mass ratios versus \hi\ deficiency for Virgo SFDs and HeViCS BGC.}
   \label{fig:stell_dust}%
\end{figure}

We explore the relation between the dust-to-stellar mass ratio and stellar mass in Fig. \ref{fig:stell_dust} (upper panel).
The HeViCS BGC galaxies do show an anticorrelation between the two parameters.
Concerning the dwarf samples, the relationship between dust fraction and stellar mass is less clear:
Virgo SFDs form a parallel sequence to  that defined by Virgo BGC objects,
and their dust-to-stellar mass ratio is weakly anticorrelated with the stellar mass ($r = -0.34$), while for the other dwarfs
there is no correlation between these two parameters.
The DGS galaxies are mostly characterised by higher sSFRs and lower metal abundances
compared to the Virgo SFDs, implying
that they are in a different evolutionary stage.
A combination of the more intense star formation activity
\citep[responsible for a higher dust destruction rates via supernova shocks; ][]{2014ApJ...782L..23H} and outflows \citep{2001MNRAS.328..223E,2011MNRAS.417.1510D} could cause the lack of a correlation between dust fraction and stellar mass for this sample of galaxies.

\citet{2012A&A...540A..52C} found that at stellar masses larger than $10^9$~M$_{\odot}$, Virgo cluster galaxies show systematically lower values of the dust-to-star mass ratio, compared to the HRS,
suggesting that dust content has been affected by the cluster environment.
The trend between H{\sc i}-normal and H{\sc i}-deficient
HRS galaxies has been interpreted as an indication of ISM stripping \citep{2012A&A...540A..52C}.
In the lower panel of Fig. \ref{fig:stell_dust} the dust-to-stellar mass ratio is plotted against \hi\-deficiency for all Virgo galaxies.
Galaxies with the highest H{\sc i}-deficiencies (both dwarfs and spirals)
do appear to have the lowest dust fractions,
suggesting that environmental effects are affecting also the dust content. However, a larger sample of extremely H{\sc I}-deficient dwarfs would be needed to confirm that the
same processes that make dwarfs gas deficient can also
lower their dust masses.

\subsection{Environmental effects on the dust-to-gas mass ratio}
\label{sec:environment}

The dust-to-gas mass ratio \D\ (M$_{d}$/M$_{g}$) gives an indication of the enrichment of the gas
by heavy elements produced in stars (C, O, Mg, Si, Fe), the amount of metals that are locked in dust grains, and the net balance between the production and growth of dust grains
and their destruction in the ISM.

If the ratio of dust-to-metals in the ISM does not vary among galaxies,
the relation between \D\ and the oxygen abundance $O/H$ is expected to be linear \citep{2001MNRAS.328..223E,2007ApJ...663..866D}.
Several models predict the evolution of \D\ as a function of metallicity \citep{1998ApJ...501..643D,1998ApJ...496..145L,2001MNRAS.328..223E,2002A&A...388..439H}.
However, the relation between these two parameters at the very low metallicity end (12+log(O/H) $\lesssim$ 8) is still an open issue, because
metal-poor dwarf galaxies do not follow the same linear dependence of metal-rich systems
\citep{2007ApJ...663..866D,2011A&A...532A..56G,2012ApJ...752..112H,2014A&A...561A..49H,2014A&A...563A..31R}.

\begin{figure}
   \centering
\includegraphics[bb=20 -10 550 780,width=8cm]{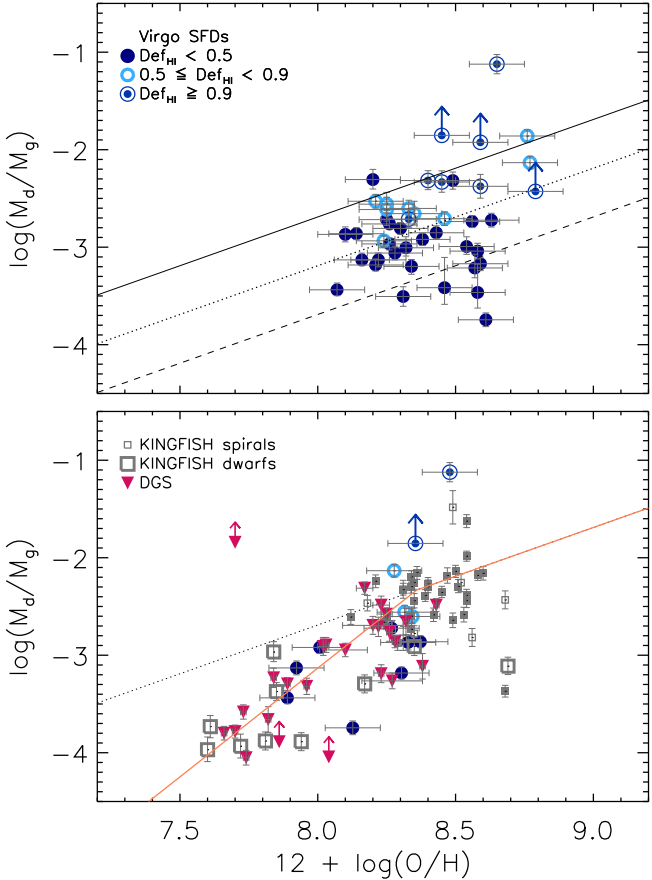}
  \caption{{\em Upper panel:} Dust-to-gas mass ratios against metallicity for Virgo SFDs. Here we plot the oxygen abundance obtained
  from N2 e O3N2 indices for the whole sample. Solid line indicates
  a linear scaling of the Milky Way \D\ and metallicity, dotted and dashed
lines correspond to a factor of 5 and 10 difference from MW, respectively.
  {\em Lower panel:} Comparison
   data for KINGFISH and DGS samples.
   Here the metallicity of Virgo SFDs was derived using the $R_{23}$ ratio only for 13 objects with [OII]$\lambda$3727 line measurements available in the literature. A broken power-law fit (orange line) with a break at [12 + log(O/H)]$_{break}$ = 8.36 and low-metallicity slope $\alpha = 2.23$, is overlaid to the data. The dotted line shows the linear scaling of the dust-to-gas ratio of the Milky Way with metallicity.}
   \label{gas_dust_met}
    \end{figure}

Figure \ref{gas_dust_met} explores the variation of  \D\ with nebular oxygen abundance for Virgo SFDs (upper panel) and comparison samples (lower panel).
We assumed that the total gas mass of the SFDs is given by the atomic component only (with a correction for neutral helium $M_g = 1.33 M_{HI}$), because of the lack of CO measurements for Virgo and KINGFISH dwarfs, and the uncertainty in assessing the amount of molecular gas in the metal-poor DGS galaxies \citep{2014A&A...563A..31R}.
Because of the different methods used to derive metal abundances (see Sect. ~\ref{sec:ancillary_met}), we need to analyse the samples of dwarfs separately.

\begin{figure}
   \centering
\includegraphics[bb=80 10 550 550,width=8cm]{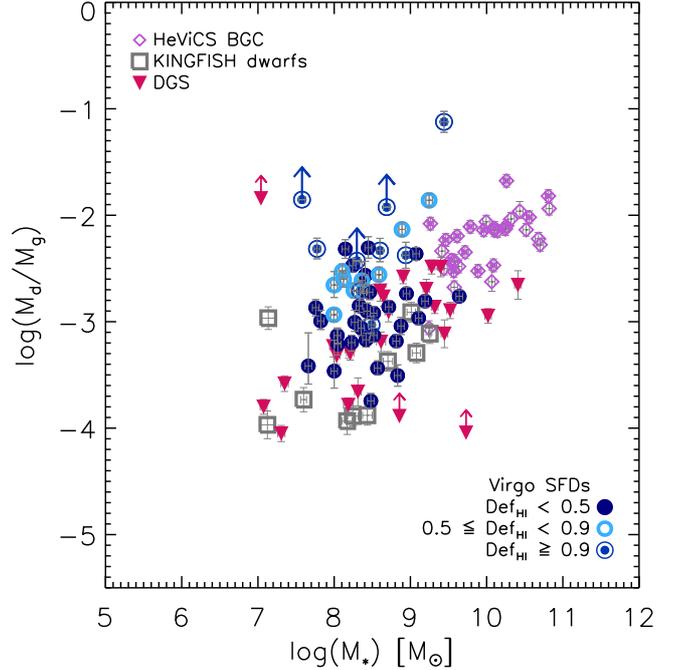} 
\caption{Dust-to-gas mass ratio versus stellar mass for Virgo SFDs, KINGFISH dwarfs, DGS, and HeViCS BGC (with available molecular gas mass estimates). Symbols are the same as Fig. \ref{fig:stell_dust}. \hidef\ HeViCS BGC galaxies are indicated by a diamond with a cross.}
   \label{dustgas_mstar}%
    \end{figure}

In the top-panel we use the oxygen abundances of the HeviCS dwarfs based on the N2 and O3N2 indices.
The Virgo SFDs are scattered along the \D -metallicity plane.
The solid black line in the figure shows the linear scaling of the Milky Way \D\ and metallicity \citep{2007ApJ...663..866D}, with dotted and dashed
lines showing a factor of 5 and 10 difference from MW, respectively.
Most of the Virgo dwarfs, especially the more gas-rich Virgo ones,
have dust-to-gas ratios lower than expected by linearly scaling the Milky Way values.
\hi envelopes in dwarf galaxies are known to be more extended than the stellar and dust components. Since all dwarfs are unresolved by the $\sim$ 3$^{\prime}$.5 Arecibo beam, the ALFALFA catalogue provides the global \hi content.
Only a few Virgo dwarfs have been mapped at 21-cm, thus there is not much information about the size of the \hi discs.
A Very Large Array (VLA) survey of Virgo BCDs \citep{2003AJ....126.2774H} which includes 5 of our dwarfs (VCC10, VCC24, VCC172, VCC340, VCC1437) found for these galaxies \hi -to-optical diameter ratios, $D_{HI}/D_{25}$, varying between
1.2 and 3.4, with a mean value around 2.
\D\ might be underestimated in some cases because of the different size of the gas compared to the apertures adopted to measure the dust content (1.4$\times R_{25}$),
and we will assess this issue in a future paper by comparing dust distribution to 21-cm maps obtained for a subset
of our galaxies (Coelho et al. 2015, in prep.).

Given the uncertainties in the correct estimate of the gas masses at this stage we cannot draw firm conclusions on the relation between \D\ and metallicity in our sample of Virgo dwarfs; nonetheless Fig. \ref{gas_dust_met} suggests that
\hidef\ dwarfs have a higher \D\ compared to those with a normal \hi content.
VCC135 for example, the highest point in the figure, has a dust-to-gas ratio which is about one order of
magnitude higher than what expected from the linear scaling of the Milky Way \D\ and metallicity.
The high \D\ is a likely consequence of gas stripping by the cluster environment in these cases.

Furthermore,
H{\sc i}-deficient dwarfs are preferentially found at higher
metallicities, and the oxygen abundance progressively increase from H{\sc i}-normal to H{\sc i}-poor systems, similarly to what \citet{2013A&A...550A.115H} found in nearby spiral galaxies. However,
analysis of a larger, statistically significant sample of H{\sc i}-deficient dwarfs is needed to confirm this trend.

In the lower panel of Fig. \ref{gas_dust_met}, we compare Virgo SFDs to
KINGFISH and DGS, for which the metallicities
have been derived with the \citetalias{2005ApJ...631..231P} method (see Sect. ~\ref{sec:ancillary_met}).
Only Virgo dwarfs with [12 + $\log$(O/H)]  estimated according to \citetalias{2005ApJ...631..231P} can be compared to the other surveys without introducing systematic offsets due to the different calibrations. Here
the increase in \D\
with the gas deficiency is still clear, despite the smaller number of galaxies shown.

Metal-poor dwarfs do not follow the linear scaling of the Milky Way metallicity and $\mathcal D$.
Models that include
the production and destruction
of dust by supernovae, removal of dust through
outflows from galaxies, and dust production in the envelopes
of stars
\citep[e.g.,][]{2002A&A...388..439H,2013MNRAS.432..637A,2014A&A...562A..76Z}
yield non-linear relations between \D\ and $O/H$, and may explain the breakdown of the trend at low metal abundances.
As an exercise we fit a broken power-law to the three samples of galaxies, similarly to \citet{2014A&A...563A..31R}. We fixed the power-law at high metallicity to 1 and found a low-metallicity slope $\alpha = 2.23 \pm 0.17$ with a break at [12 + log(O/H)]$_{break}$ = 8.36$\pm$0.06 (orange line).
For a more detailed analysis of the variation of \D\ with
metallicity in the DGS and KINGFISH galaxies we refer the reader to \citet{2014A&A...563A..31R}.

The difference in the dust-to-gas ratio within different environments is also shown in Fig. \ref{dustgas_mstar} where we plot \D\ as a function of the stellar mass for all samples.
The dwarfs with a larger gas content (DGS, KINGFISH dwarfs, and {H\sc i}-normal Virgo SFDs) show a lower \D\ for a given stellar mass,
compared to the {H\sc i}-deficient Virgo dwarfs and the HeViCS bright galaxies sample,
which can be once again interpreted as the evidence of the effects of the cluster environment
on the gas component of low-mass systems.

\section{Summary and conclusions}
\label{sec:conclusions}

We used \hers\ observations of the Virgo cluster taken as part of the {\em Herschel} Virgo Cluster Survey to investigate the FIR-submm properties of a sample of SFD galaxies and the effects of the cluster environment on the interstellar medium.
We gathered optical, mid-infrared, and centimetre ancillary data from the literature to compare the dust content to stellar and gas masses, star formation rates, and metallicity.
Among 140 late-type dwarf galaxies included in the HeViCS fields, we detected 49 objects at FIR-submm wavelengths.
If we consider only the dwarfs brighter than $m_B = 18$ mag, the completeness limit of the VCC, this gives a detection rate of 43\%.

To assess the range of apparent $\beta$ values that can better represent the shape of the FIR-submm SED of Virgo SFDs
we performed MBB fitting of a subset of 30 galaxies (i.e. with detections in at least four \hers\ bands) following two approaches. First we used a single MBB with fixed values of the emissivity index ($\beta = $ [1.0,1.2,1.5,1.8,2.0]),
secondly we repeated the SED fitting letting $\beta$ vary between 0 and 3, and selecting the value which provided the best fit.
With the first method (fixed-$\beta$), we found that the best-fit emissivity index minimising the fraction of residuals in four \hers\ bands (100 - 350 $\mu$m) is
$\beta = 1.5$. The range in dust temperature for $\beta = 1.5$ MBB fits is between 16.1 and 33.3 K, with a median of $T_d =$ 22.4 K.
In the free-$\beta$ case, the best-fit emissivities vary substantially among the sample, and we obtained values between 0.1 and 2.9.
Dust masses of the 49 {\em Herschel}-detected dwarfs were calculated with $\beta$ fixed at 1.5 following the calibration of the dust opacity of \citet{2013A&A...552A..89B}, and they range between $10^{4.7}$ and $10^{6.8}$ M$_{\odot}$.

Stacking analysis of 64 SFDs without a 250 $\mu$m counterpart resulted in a 3.5$\sigma$ detection with $<F_{250}> = $ 4.2 mJy.
Adopting $\beta = 1.5$ and $T_d = $ 22.4 K, the average dust mass of undetected dwarfs (brighter than the VCC completeness limit) corresponds to $M_d = 8.7 \times 10^3$ M$_{\odot}$ (at $d = 17$ Mpc), $\sim$30 times lower than the mean value of the detected sample. Dwarfs without a FIR counterpart have lower stellar masses, lower H$\alpha$ luminosities, and are more H{\sc i}-deficient.

Among the \hers\ detections in all five bands (23 out of 49), 67\% present an excess emission at 500 $\mu$m beyond the modified black-body model, assuming an emissivity index of $\beta = 1.5$.
The fraction of Virgo dwarfs with a 500 $\mu$m excess decreases from $\beta = 2$ (88\%) to $\beta = 1$ (42\%). Even if a $\beta$-free SED modelling is applied, this fraction is still high (54\%).
The 500 $\mu$m fractional residuals show an inverse correlation with star formation rate and stellar masses.
If the excess is due to different dust grain properties, our results may support either a scenario where the emission is produced by small dust grains \citep{2002A&A...382..860L,2006ApJ...652..283B,2009ApJ...706..941Z}, or by iron-rich dust grains which are expected to be more abundant in a metal-poor ISM \citep{2012ApJ...757..103D}.

To study the variations in the global properties of our sample due to environmental effects, we compared Virgo SFDs to other \hers\ surveys targeting dwarfs in lower density environments such as the DGS and KINGFISH. We also included spiral galaxies from the HeViCS BGC and KINGFISH to investigate variations in dust properties with the morphological type.

From the analysis of SPIRE $F_{250}/F_{350}$ colour we infer that both young stars and more evolved stellar populations
contribute to the heating of the cold dust component in Virgo SFD galaxies, and that the contribution of ionising insterstellar radiation is stronger in more active dwarfs such as those in the DGS and KINGFISH. On the other hand, old stars appear to dominate the dust heating process in the Virgo and KINGFISH spiral galaxies, consistent with previous studies.

We explored the relations between stellar mass and \hi fraction, sSFR, dust fraction, gas-to-dust ratio over a wide range of stellar masses (from $10^{7}$ to $10^{11}$ M$_{\odot}$) and morphological types.
Increasingly more massive galaxies have progressively lower \hi gas
fraction and sSFR, however Virgo galaxies are offset towards lower values of these parameters at a
a given stellar mass, compared to similar galaxies in less dense environments.

A similar scaling relation is found for the dust content of spiral galaxies, but we do not find a clear correlation
between $M_{\star}$/\mdust and \mstar\ in the dwarf samples. These two parameters are marginally correlated only in Virgo SFDs, while in more active KINGFISH and DGS SFDs any correlation is lacking. We interpreted the lack of correlation in these systems as the consequence of a higher dust destruction rate and outflows due to the more intense star formation activity of these galaxies.

The most \hidef\ dwarfs show lower sSFRs, \hi, and dust fractions providing evidence for the effects of the cluster environment on the ISM and star formation activity. However, we conclude that the amount of removed dust has to be lower compared to the stripped \hi component, to explain the large \D\ observed in
the \hidef\ systems.  This is likely due to the larger extension of the \hi discs compared to the dust distributions.
As the Virgo star-forming dwarfs are likely to be entering the cluster for the first time, longer time scales might be necessary to strip or destroy the more
centrally concentrated dust distribution and transform these dwarfs into transition-type \citep{2013MNRAS.436.1057D} or early-type dwarfs \citep{2013A&A...552A...8D}.

\section{Acknowledgments}
MG gratefully acknowledges support from the Science and
Technology Foundation (FCT, Portugal) through the research grant
PTDC/CTE-AST/111140/2009.
SB, EC, and LKH acknowledge support from PRIN-INAF 2012/2013.
SCM and ARR acknowledge support from the Agence Nationale de la Recherche (ANR) through the programme SYMPATICO (Program Blanc Projet ANR-11-BS56-0023).
TMH gratefully acknowledges the financial support from the Belgian  Science Policy Office (BELSPO) in the frame of the PRODEX project  C90370 ({\em Herschel}-PACS Guaranteed Time and Open Time Programs: Science  Exploitation).
IDL is a postdoctoral researcher
of the FWO-Vlaanderen (Belgium).

Funding for SDSS-III has been provided by the Alfred P. Sloan Foundation, the Participating Institutions, the National Science Foundation, and the U.S. Department of Energy Office of Science. The SDSS-III web site is http://www.sdss3.org/.
SDSS-III is managed by the Astrophysical Research Consortium for the Participating Institutions of the SDSS-III Collaboration including the University of Arizona, the Brazilian Participation Group, Brookhaven National Laboratory, Carnegie Mellon University, University of Florida, the French Participation Group, the German Participation Group, Harvard University, the Instituto de Astrofisica de Canarias, the Michigan State/Notre Dame/JINA Participation Group, Johns Hopkins University, Lawrence Berkeley National Laboratory, Max Planck Institute for Astrophysics, Max Planck Institute for Extraterrestrial Physics, New Mexico State University, New York University, Ohio State University, Pennsylvania State University, University of Portsmouth, Princeton University, the Spanish Participation Group, University of Tokyo, University of Utah, Vanderbilt University, University of Virginia, University of Washington, and Yale University.

This publication makes use of data products from the Wide-field Infrared Survey Explorer, which is a joint project of the University
of California, Los Angeles, and the Jet Propulsion Laboratory/California Institute of Technology, funded by the National Aeronautics
and Space Administration.

We thank the many members of the ALFALFA team
who have contributed to the acquisition and processing of
the ALFALFA dataset over the last six years.
The Arecibo Observatory is operated by SRI International under a
cooperative agreement with the National Science Foundation (AST-1100968), and in alliance with Ana
G. Mandez-Universidad Metropolitana, and the Universities
Space Research Association.

This research has made use of the NASA/IPAC Extragalactic
Database (NED) which is operated by the Jet Propulsion Laboratory,
California Institute of Technology, under contract with the National
Aeronautics and Space Administration.

\bibliographystyle{aa} 
\bibliography{SFdwarfbib} 

\clearpage

\begin{appendix}

\section{HeViCS BGC, DGS, KINGFISH stellar masses: comparison to previous estimates}
\label{app_sec:compare_mstar}

\begin{figure}[h!]
   \centering
 \includegraphics[bb=110 0 530 380,width=7.5cm]{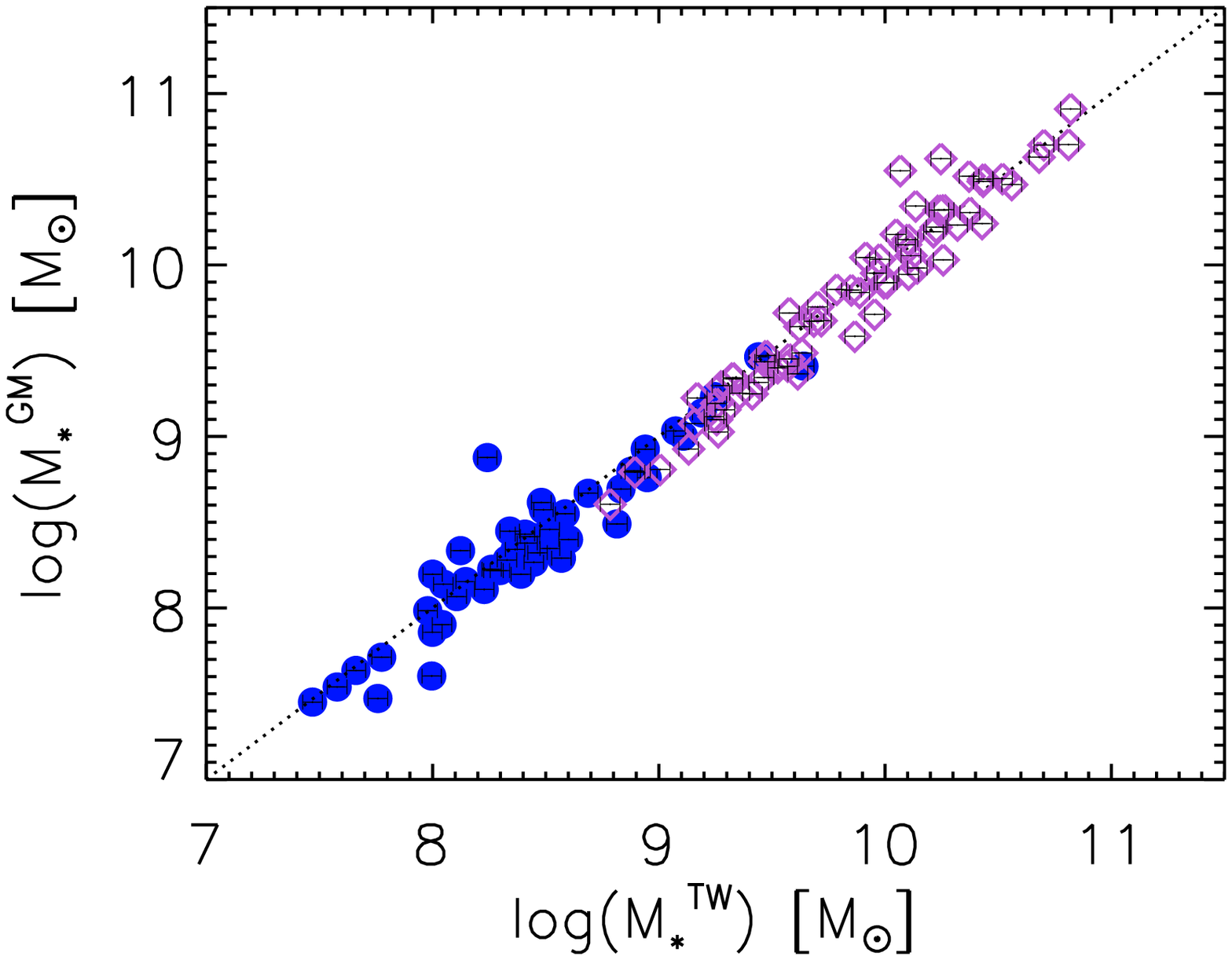}
 \includegraphics[bb=20 -40 540 280,width=8cm]{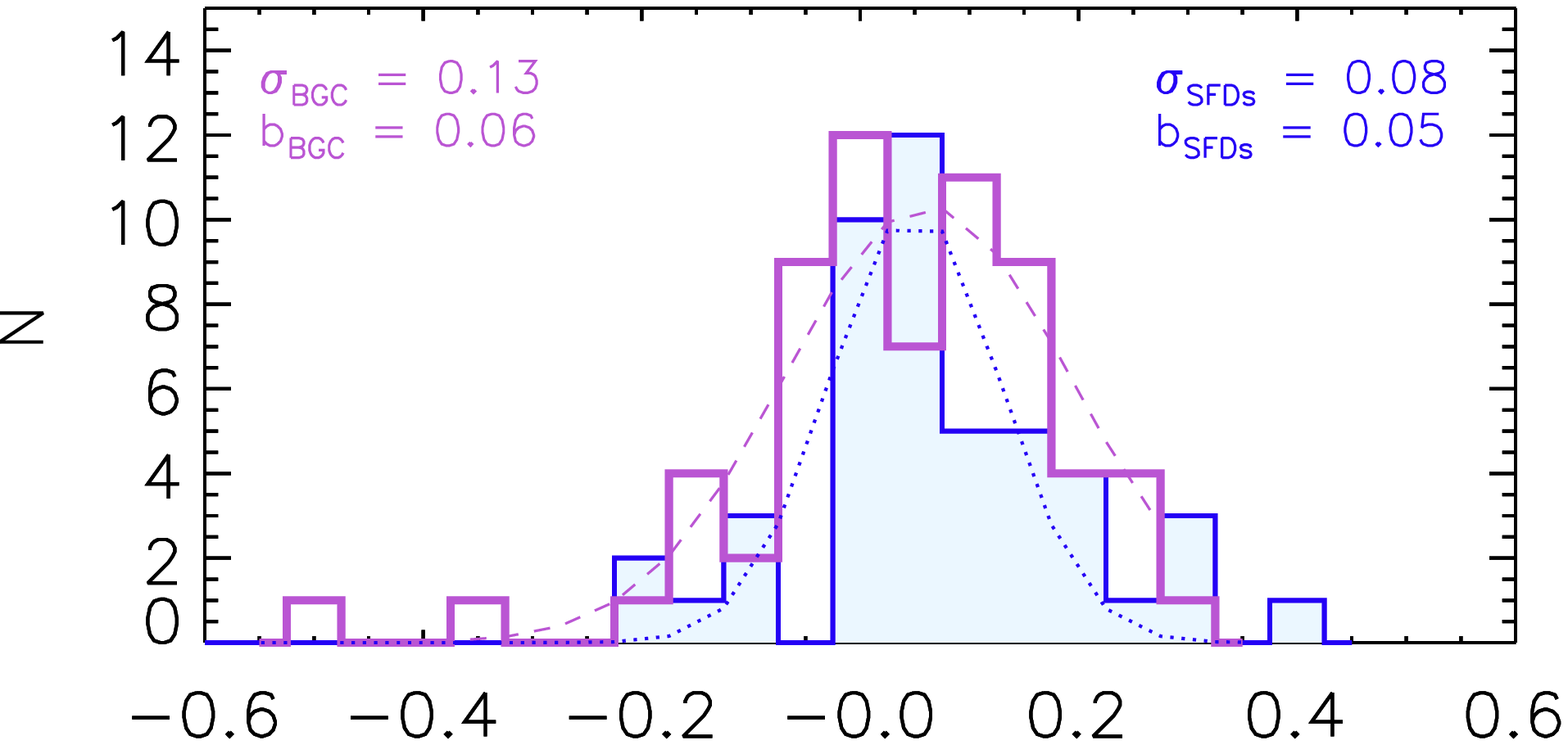}
   \caption{{\em Upper panel:} Comparison between stellar masses estimated in this work from {\em WISE} photometry, $M_{\star}^{TW}$, and those derived from the $i$ magnitude and ($g - i$)$_0$ colour following \citet{2013A&A...553A..89G}, extracted from the GOLDMine database $M_{\star}^{GM}$. Blue dots and purple diamonds correspond to the HeViCS SFDs and BGC galaxies, respectively. The dotted line shows the one-to-one relation.
{\em Lower panel:} Distribution of the residuals of the two stellar mass estimates for
 the HeViCS dwarfs (blue histogram) and BGC galaxies (purple histogram). The resulting gaussian fit is overlaid to both histograms.}
   \label{app_fig:compare_mstar}%
    \end{figure}

Different stellar-mass estimation methods can yield mass values that disagree by
factors up to $\sim$2 \citep{2007ApJ...657L...5K,2014AJ....148...77M}. To assess the reliability of our estimates based on MIR photometry, in this appendix we compare the stellar masses of the BGC, DGS, and KINGFISH samples to those derived by previous studies.
In Fig. \ref{app_fig:compare_mstar}
we show the results for the HeViCS BGC galaxies which are compared to \citet{2013A&A...553A..89G}, where stellar masses were calculated using a relation combining the $g-i$ colour and the $i$ magnitude, calibrated on the MPA-JHU sample.
We find a fair good agreement between the two estimates for this sample (purple diamonds), as we found for the HeViCS SFDs (blue dots, see Sect.  \ref{sec:ancillary_stellmass}): the residual distribution for the BGC sample (purple histogram) is slightly asymmetric, and peaks at 0.06 dex, with a dispersion of 0.13 dex.

Regarding the DGS, comparison to \citet{2013A&A...557A..95R}, where  stellar masses were derived from IRAC 3.6 and 4.5 $\mu$m photometry following the method of \citet{2012AJ....143..139E},
shows that our estimates are on average systematically smaller by a factor of $\sim$0.17 $\pm$ 0.05 dex.

Finally, comparison to the stellar masses of the KINGFISH sample calculated by \citet{2011ApJ...738...89S} based on optical colours \citep{2009MNRAS.400.1181Z} shows that our estimates are on average systematically larger by a factor of $\sim$0.5 dex. However for those galaxies with available SDSS photometry we compared our estimates with the
stellar masses determined from $i$-band luminosities $L_i$ using the $g - i$
colour-dependent stellar mass-to-light ratio relation \citep{2009MNRAS.400.1181Z}, and found an average difference of
0.11 dex with a dispersion of 0.17 dex. The discrepancy is larger when the relations using $B - V$, or $B - R$ colours
are used for those objects without SDSS photometry.

\section{Two-component modified black-body SED fitting}
\label{app:twocompfit}

\begin{figure*}
\centering
 \includegraphics[bb=0 265 520 565,width=18cm]{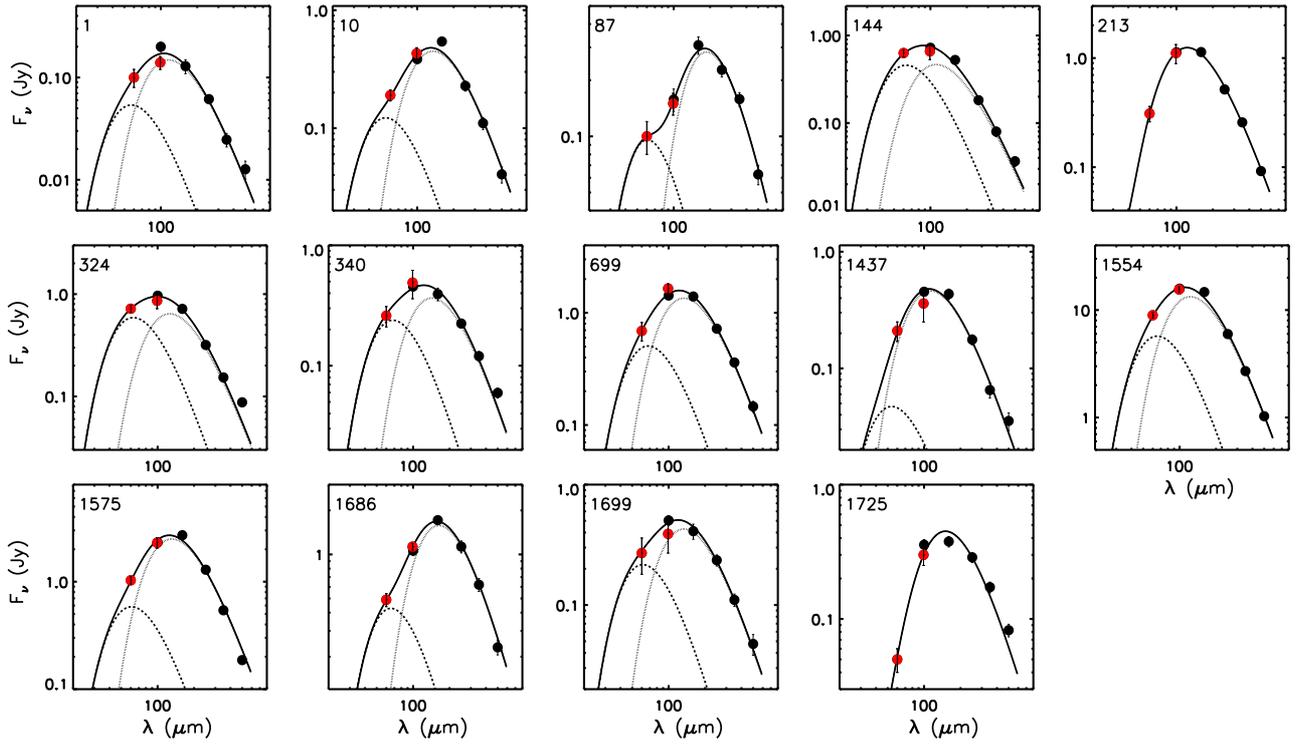}
   \caption{2-component MBB fits for 14 dwarfs with available \iras\ and {\em ISO} photometry. Filled red { dots} correspond to \iras\ or {\em ISO} data, while black {dots} show {\em Herschel} photometry. The emissivity index of the cold dust component (dotted line) is fixed at $\beta_c = 1.5$, while the warm dust component (dashed line) has $\beta_w = 2.0$.
   The VCC catalogue ID is given at the upper-left corner of each plot.
   }
   \label{fig:SED_fit_09}
    \end{figure*}

Analysing two-component MBB models is important to
 begin to assess the dust temperature mixing along the line of sight, which could in principle lead to a lower
$\beta$ value when the SED fitting takes only into account one dust component.
We combined MIR photometry from the literature with our FIR-submm observation for a subset of 14 galaxies
with available mid-infrared (MIR) observations
(see Sect. \ref{sec:midir}), and we fitted the SED
using two modified black-body models, one for the warm component and one for the cold component.
We fixed the emissivity
index of the warm component at $\beta_w = 2$, an approximation of the opacity in the
standard \citet{2001ApJ...554..778L} dust models,
and that of the cold component at $\beta_c = 1.5$.
We used the 22 $\mu$m data point in the fit as an upper limit to better constrain the
warm dust modified blackbody.
The result is shown in Fig. \ref{fig:SED_fit_09}. For two galaxies an additional dust component is not necessary to fit observations at 60 and 100 $\mu$m (VCC213, VCC1725).

The temperature of the warm component ranges between 43 and 54 K, while
the change in the cold dust temperature, compared to a single temperature MBB fit (see
Tables \ref{tab:2compMBB} and \ref{tab:beta15}), varies between -0.1 and -4.3 K.

\begin{table}[h!]
\centering
\caption{Two-component MBB SED fitting for the subset of Virgo dwarfs with $IRAS$ and $ISO$ photometry.
}
\label{tab:2compMBB}
\begin{tabular}{lcccc}
\hline \hline
  ID      & $F_{60}$  & $F_{100}$  &  $T_c$    &       $T_w$      \\
          &   [Jy]    &  [Jy]      &    [K]    &        [K]       \\
\hline \hline
   VCC1 & 0.10 $\pm$  0.02  &   0.14 $\pm$   0.02    & 27.4$\pm$3.8 &    50.9$\pm$ 9.8 \\
  VCC10 & 0.19 $\pm$  0.02  &   0.43 $\pm$   0.05    & 23.9$\pm$1.4 &    52.8$\pm$ 6.7 \\
  VCC87 & 0.10 $\pm$  0.02  &   0.15 $\pm$   0.02    & 17.3$\pm$0.5 &    47.8$\pm$ 1.6 \\
 VCC144 & 0.63 $\pm$  0.06  &   0.66 $\pm$   0.13    & 29.0$\pm$4.2 &    46.0$\pm$ 5.9 \\
 VCC213 & 0.31 $\pm$  0.05  &   1.11 $\pm$   0.22    & 26.3$\pm$0.4 &    ...           \\
 VCC324 & 0.72 $\pm$  0.06  &   0.86 $\pm$   0.14    & 25.7$\pm$1.1 &    46.0$\pm$ 3.1 \\
 VCC340 & 0.26 $\pm$  0.05  &   0.49 $\pm$   0.13    & 22.6$\pm$0.7 &    42.7$\pm$ 1.2 \\
 VCC699 & 0.69 $\pm$  0.13  &   1.65 $\pm$   0.16    & 24.2$\pm$0.7 &    42.8$\pm$ 1.9 \\
VCC1437 & 0.21 $\pm$  0.04  &   0.36 $\pm$   0.11    & 28.7$\pm$2.4 &    53.7$\pm$ 8.3 \\
VCC1554 & 8.95 $\pm$  0.54  &  15.53 $\pm$   0.97    & 26.4$\pm$1.2 &    44.4$\pm$ 1.5 \\
VCC1575 & 1.03 $\pm$  0.08  &   2.30 $\pm$   0.25    & 24.7$\pm$1.2 &    47.7$\pm$ 7.4 \\
VCC1686 & 0.49 $\pm$  0.05  &   1.13 $\pm$   0.06    & 19.8$\pm$0.6 &    44.2$\pm$ 6.5 \\
VCC1699 & 0.38 $\pm$  0.09  &   0.60 $\pm$   0.12    & 24.2$\pm$1.4 &    47.4$\pm$ 7.7 \\
VCC1725 & 0.05 $\pm$  0.01  &   0.30 $\pm$   0.05    &  21.2$\pm$1.2 &      ...  \\
\hline \hline
\end{tabular}
\end{table}

\section{Data Tables and SED fitting results}
\label{app:data_tables}

\newpage

\begin{table*}
\centering
\begin{threeparttable}
\caption{{\em Herschel} photometry of the sample of Virgo star-forming dwarf galaxies.}
\label{tab:fir_fluxes}
\begin{tabular}{lcccccrr}
\hline \hline
    ID   &  $F_{100}$    & $F_{160}$ &  $F_{250}$     &  $F_{350}$        &  $F_{500}$ & $a_{25}\,$ & $b_{25}\,$ \\
         &  [Jy]         & [Jy]      &   [Jy]         & [Jy]              &   [Jy]     & $^{\prime\prime} \:\:\:$ &  $^{\prime\prime} \:\:$  \\ 
 \hline \hline
   VCC1 &     0.200 $\pm$ 0.021          &     0.129 $\pm$ 0.020 & 0.061 $\pm$ 0.006 &     0.025 $\pm$ 0.004 & $<$ 0.009             & 24.0 &  5.4 \\
  VCC10 &     0.384 $\pm$ 0.027          &     0.542 $\pm$ 0.037 & 0.228 $\pm$ 0.022 &     0.110 $\pm$ 0.013 &     0.041 $\pm$ 0.006 & 30.9 &  6.6 \\
  VCC17 & $<$ 0.095                      & $<$ 0.070             & 0.056 $\pm$ 0.010 &     0.037 $\pm$ 0.010 &     0.023 $\pm$ 0.004 & 27.3 & 13.5 \\
  VCC22 & $<$ 0.059                      & $<$ 0.039             & 0.019 $\pm$ 0.005 &     0.015 $\pm$ 0.004 & $<$ 0.008             &  8.1 &  6.3 \\
  VCC24 &     0.073 $\pm$   0.018        &     0.117 $\pm$ 0.022 & 0.055 $\pm$ 0.011 &     0.032 $\pm$ 0.010 & $<$ 0.024             & 30.0 & 11.1 \\
  VCC87 &     0.158 $\pm$ 0.022$^{\dag}$ &     0.308 $\pm$ 0.033 & 0.227 $\pm$ 0.019 &     0.158 $\pm$ 0.013 &     0.088 $\pm$ 0.004 & 43.5 & 21.6 \\
 VCC135 &     0.686 $\pm$ 0.040$^{\dag}$ &     0.759 $\pm$ 0.049 & 0.309 $\pm$ 0.025 &     0.164 $\pm$ 0.013 &     0.074 $\pm$ 0.004 & 34.8 & 17.1 \\
 VCC144 &     0.724 $\pm$ 0.052          &     0.525 $\pm$ 0.040 & 0.182 $\pm$ 0.016 &     0.080 $\pm$ 0.010 &     0.036 $\pm$ 0.004 & 18.9 &  9.6 \\
 VCC172 &     0.104 $\pm$ 0.021$^{\dag}$ &     0.204 $\pm$ 0.025 & 0.121 $\pm$ 0.019 &     0.063 $\pm$ 0.015 &     0.034 $\pm$ 0.004 & 37.8 & 16.8 \\
 VCC213 &     1.130 $\pm$ 0.063          &     1.135 $\pm$ 0.064 & 0.516 $\pm$ 0.038 &     0.257 $\pm$ 0.020 &     0.093 $\pm$ 0.004 & 27.9 & 21.3 \\
 VCC223 &     0.090 $\pm$ 0.015$^{\dag}$ &     0.127 $\pm$ 0.022 & 0.064 $\pm$ 0.008 &     0.034 $\pm$ 0.007 &     0.019 $\pm$ 0.005 & 10.2 &  7.8 \\
 VCC281 &     0.083 $\pm$ 0.019          &     0.114 $\pm$ 0.017 & 0.077 $\pm$ 0.010 &     0.052 $\pm$ 0.007 &     0.031 $\pm$ 0.004 & 10.8 & 10.8 \\
 VCC286 & $<$ 0.043                      & $<$ 0.068             & 0.026 $\pm$ 0.006 &     0.016 $\pm$ 0.004 & $<$ 0.011             & 15.3 &  9.9 \\
 VCC322 &     0.051 $\pm$ 0.013$^{\dag}$ &     0.035 $\pm$ 0.009 & 0.040 $\pm$ 0.006 &     0.024 $\pm$ 0.008 & $<$ 0.010             & 37.8 & 15.9 \\
 VCC324 &     0.965 $\pm$ 0.061          &     0.717 $\pm$ 0.058 & 0.318 $\pm$ 0.024 &     0.153 $\pm$ 0.014 &     0.086 $\pm$ 0.004 & 40.5 & 34.5 \\
 VCC328 & $<$ 0.095                      &     0.046 $\pm$ 0.013 & 0.024 $\pm$ 0.006 &     0.016 $\pm$ 0.005 &     0.012 $\pm$ 0.003 & 30.0 & 12.9 \\
 VCC334 &     0.137 $\pm$ 0.022          &     0.163 $\pm$ 0.018 & 0.070 $\pm$ 0.009 &     0.029 $\pm$ 0.005 & $<$ 0.013             & 16.8 & 15.3 \\
 VCC340 &     0.455 $\pm$ 0.037          &     0.394 $\pm$ 0.047 & 0.224 $\pm$ 0.018 &     0.120 $\pm$ 0.011 &     0.058 $\pm$ 0.004 & 33.0 & 12.9 \\
 VCC367 & $<$ 0.065                      & $<$ 0.058             & 0.033 $\pm$ 0.007 &     0.021 $\pm$ 0.005 & $<$ 0.012             & 16.8 & 13.5 \\
 VCC446 &     0.053 $\pm$ 0.016$^{\dag}$ &     0.043 $\pm$ 0.011 & 0.042 $\pm$ 0.006 &     0.026 $\pm$ 0.006 & $<$ 0.011             & 25.5 & 12.9 \\
 VCC562 &     0.133 $\pm$ 0.021          &     0.125 $\pm$ 0.015 & 0.059 $\pm$ 0.006 &     0.045 $\pm$ 0.005 &     0.032 $\pm$ 0.004 & 18.9 & 14.7 \\
 VCC620 & $<$ 0.052                      & $<$ 0.065             & 0.049 $\pm$ 0.008 &     0.031 $\pm$ 0.006 &     0.020 $\pm$ 0.006 & 37.8 & 12.3 \\
 VCC641 & $<$ 0.114                      & $<$ 0.102             & 0.039 $\pm$ 0.005 &     0.031 $\pm$ 0.004 &     0.026 $\pm$ 0.004 & 21.9 & 12.6 \\
 VCC693 &     0.134 $\pm$ 0.025          &     0.219 $\pm$ 0.039 & 0.153 $\pm$ 0.014 &     0.075 $\pm$ 0.009 &     0.030 $\pm$ 0.004 & 34.8 & 30.0 \\
 VCC699 &     1.427 $\pm$ 0.102          &     1.398 $\pm$ 0.087 & 0.722 $\pm$ 0.054 &     0.361 $\pm$ 0.030 &     0.146 $\pm$ 0.016 & 58.5 & 41.4 \\
 VCC737 &     0.109 $\pm$ 0.019$^{\dag}$ &     0.157 $\pm$ 0.019 & 0.150 $\pm$ 0.013 &     0.093 $\pm$ 0.008 &     0.049 $\pm$ 0.004 & 32.1 & 10.5 \\
 VCC741 &     0.043 $\pm$ 0.013$^{\dag}$ &     0.069 $\pm$ 0.012 & 0.046 $\pm$ 0.006 &     0.033 $\pm$ 0.005 &     0.025 $\pm$ 0.004 & 25.2 &  7.2 \\
 VCC802 & $<$ 0.042                      & $<$ 0.036             & 0.023 $\pm$ 0.004 &     0.019 $\pm$ 0.003 & $<$ 0.011             & 19.2 &  6.3 \\
 VCC825 & $<$ 0.089                      & $<$ 0.081             & 0.023 $\pm$ 0.006 &     0.013 $\pm$ 0.003 & $<$ 0.009             & 30.0 & 30.0 \\
 VCC841 &     0.138 $\pm$ 0.025          &     0.156 $\pm$ 0.017 & 0.101 $\pm$ 0.011 &     0.042 $\pm$ 0.007 &     0.011 $\pm$ 0.003 & 25.2 &  8.7 \\
 VCC848 &     0.088 $\pm$ 0.019$^{\dag}$ &     0.142 $\pm$ 0.016 & 0.069 $\pm$ 0.007 &     0.028 $\pm$ 0.005 &     0.017 $\pm$ 0.004 & 34.8 & 29.4 \\
 VCC888 & $<$ 0.149                      &     0.105 $\pm$ 0.017 & 0.082 $\pm$ 0.009 &     0.073 $\pm$ 0.009 &     0.038 $\pm$ 0.004 & 34.8 & 16.5 \\
 VCC985 &     0.028$\pm$ 0.008$^{\ddag}$ &     0.047 $\pm$ 0.015 & 0.034 $\pm$ 0.007 & $<$ 0.018             & $<$ 0.010             & 18.9 &  8.7 \\
VCC1021 & $<$ 0.086                      & $<$ 0.062             & 0.024 $\pm$ 0.005 &     0.014 $\pm$ 0.004 & $<$ 0.009             & 34.8 & 17.1 \\
VCC1141 &     0.035$\pm$ 0.010$^{\ddag}$ &     0.058 $\pm$ 0.013 & 0.039 $\pm$ 0.006 &     0.021 $\pm$ 0.006 & $<$ 0.009             & 13.8 &  9.0 \\
VCC1179 &     0.062 $\pm$ 0.016          &     0.081 $\pm$ 0.014 & 0.044 $\pm$ 0.004 &     0.020 $\pm$ 0.004 &     0.027 $\pm$ 0.004 & 34.8 & 10.5 \\
VCC1200 &     0.064 $\pm$ 0.018$^{\dag}$ &     0.061 $\pm$ 0.015 & 0.050 $\pm$ 0.006 &     0.034 $\pm$ 0.005 &     0.025 $\pm$ 0.004 & 37.8 & 25.2 \\
VCC1273 &     0.049 $\pm$ 0.015$^{\dag}$ &     0.067 $\pm$ 0.011 & 0.050 $\pm$ 0.007 &     0.028 $\pm$ 0.005 & $<$ 0.012             & 34.8 & 12.9 \\
VCC1356 &     0.133 $\pm$ 0.020$^{\dag}$ &     0.161 $\pm$ 0.020 & 0.096 $\pm$ 0.013 &     0.058 $\pm$ 0.007 &     0.039 $\pm$ 0.004 & 33.0 & 12.9 \\
VCC1374 &     0.172 $\pm$ 0.030$^{\dag}$ &     0.283 $\pm$ 0.032 & 0.191 $\pm$ 0.022 &     0.099 $\pm$ 0.014 &     0.058 $\pm$ 0.010 & 36.0 &  8.1 \\
VCC1437 &     0.454 $\pm$ 0.037$^{\dag}$ &     0.434 $\pm$ 0.045 & 0.176 $\pm$ 0.017 &     0.065 $\pm$ 0.010 &     0.034 $\pm$ 0.004 & 17.7 & 13.5 \\
VCC1455 & $<$ 0.043                      & $<$ 0.049             & 0.028 $\pm$ 0.005 &     0.019 $\pm$ 0.005 & $<$ 0.009             & 19.2 &  7.5 \\
VCC1554 &    15.799 $\pm$ 0.852          &    14.700 $\pm$ 0.896 & 5.979 $\pm$ 0.420 &     2.701 $\pm$ 0.190 &     1.028 $\pm$ 0.073 & 78.0 & 30.0 \\
VCC1575 &     2.319 $\pm$ 0.141          &     2.706 $\pm$ 0.142 & 1.292 $\pm$ 0.094 &     0.542 $\pm$ 0.041 &     0.186 $\pm$ 0.016 & 60.0 & 42.3 \\
VCC1675 & $<$ 0.085                      &     0.103 $\pm$ 0.020 & 0.070 $\pm$ 0.009 &     0.037 $\pm$ 0.006 &     0.015 $\pm$ 0.004 & 37.8 & 22.2 \\
VCC1686 &     1.061 $\pm$0.080$^{\ddag}$ &     1.714 $\pm$ 0.105 & 1.130 $\pm$ 0.106 &     0.621 $\pm$ 0.063 &     0.232 $\pm$ 0.026 & 83.7 & 51.3 \\
VCC1699 &     0.504 $\pm$ 0.049$^{\dag}$ &     0.410 $\pm$ 0.059 & 0.236 $\pm$ 0.025 &     0.110 $\pm$ 0.013 &     0.048 $\pm$ 0.009 & 46.5 & 24.9 \\
VCC1725 &     0.357 $\pm$0.033$^{\ddag}$ &     0.377 $\pm$ 0.034 & 0.287 $\pm$ 0.025 &     0.172 $\pm$ 0.016 &     0.082 $\pm$ 0.009 & 46.5 & 29.1 \\
VCC1791 &     0.258 $\pm$ 0.050$^{\dag}$ &     0.471 $\pm$ 0.091 & 0.270 $\pm$ 0.026 &     0.115 $\pm$ 0.014 &     0.077 $\pm$ 0.011 & 38.7 & 19.2 \\
\hline \hline
\end{tabular}
  \begin{tablenotes}
    \item[a] Flux density from \citet{2014MNRAS.440..942C}
    \item[b] Time-line photometry from \citet{Pappalardo:submitted}
    \item[$\dag$] PACS 100 $\mu$m aperture size smaller by a factor $\sim$0.65 compared to other \hers bands.
    \item[$\ddag$] PACS 100 $\mu$m aperture size smaller by a factor $\sim$0.50 compared to other \hers bands.
  \end{tablenotes}
\end{threeparttable}
\end{table*}

\newpage

\begin{table*}
\centering
\begin{threeparttable}
\caption{Stellar masses, \hi masses, dust masses, star formation rates, metallicities, H{\sc i} deficiency, and distances of the 27 objects selected from the Dwarf Galaxy Survey.}
\label{tab:DGS}
\begin{tabular}{lrccrcSS}
\hline \hline
    ID   &$\log$($M_{\star}$)$\:\:\:\:$& $\log$($M_{HI}$)\tnote{a}  &  $\log$($M_d$)\tnote{b}     &  $\log$($SFR$)       &  12 + log(O/H)\tnote{a}  &   \si{$\: \: Def_{HI}$}     & \si{D\tnote{c}}           \\
         &[M$_{\odot}$]$\:\:\:\:\:\:$& [M$_{\odot}$]   & [M$_{\odot}$]    & [M$_{\odot}$ yr$^{-1}$] &          &             & \si{[Mpc]}           \\
 \hline \hline
  Haro2 &  9.40 $\pm$ 0.04 &  8.58 $\pm$ 0.07 &6.21 $\pm$ 0.06 &  0.18 $\pm$ 0.06 & 8.23 $\pm$ 0.03 &  0.13 &  21.7    \\
  Haro3 &  9.32 $\pm$ 0.04 &  9.05 $\pm$ 0.01 &6.31 $\pm$ 0.06 &  0.17 $\pm$ 0.05 & 8.28 $\pm$ 0.01 & -0.42 &  19.3    \\
 He2-10 &  9.28 $\pm$ 0.04 &  8.49 $\pm$ 0.03 &6.13 $\pm$ 0.06 &  0.17 $\pm$ 0.06 & 8.43 $\pm$ 0.01 & -0.05 &   8.7    \\
HS0052+2356 &  9.73 $\pm$ 0.04 &     $<$10.68     &6.76 $\pm$ 0.08 &...$\:\:\:\:\:\:\:\:$& 8.04 $\pm$ 0.10 &\si{\text{...}}& 191.0    \\
HS1304+3529 &  8.73 $\pm$ 0.04 &...       &5.42 $\pm$ 0.10 &...$\:\:\:\:\:\:\:\:$& 7.93 $\pm$ 0.10 &\si{\text{...}}&  78.7    \\
   IC10 &   ...$\:\:\:\:\:\:\:\:\:$ &  7.64 $\pm$ 0.02 &5.46 $\pm$ 0.14 & -0.77 $\pm$ 0.03 & 8.17 $\pm$ 0.03 &  0.06 &   0.7    \\
 IIZw40 &  8.61 $\pm$ 0.06 &  8.75 $\pm$ 0.07 &5.69 $\pm$ 0.07 &  0.27 $\pm$ 0.04 & 8.23 $\pm$ 0.01 & -0.84 &  12.1    \\
Mrk1089 & 10.02 $\pm$ 0.04 & 10.17 $\pm$ 0.03 &7.35 $\pm$ 0.07 &  0.63 $\pm$ 0.09 & 8.10 $\pm$ 0.08 & -1.29 &  56.6    \\
Mrk1450 &  7.99 $\pm$ 0.05 &  7.63 $\pm$ 0.07 &4.53 $\pm$ 0.08 & -0.84 $\pm$ 0.04 & 7.84 $\pm$ 0.01 &  0.37 &  19.8    \\
 Mrk153 &  8.86 $\pm$ 0.05 &  8.83 $\pm$ 0.00 &5.07 $\pm$ 0.10 & -0.32 $\pm$ 0.02 & 7.86 $\pm$ 0.04 & -0.23 &  40.3    \\
 Mrk209 &  7.31 $\pm$ 0.05 &  7.44 $\pm$ 0.03 &3.51 $\pm$ 0.07 & -1.52 $\pm$ 0.02 & 7.74 $\pm$ 0.01 &  0.25 &   5.8    \\
 Mrk930 &  9.52 $\pm$ 0.05 &  9.50 $\pm$ 0.05 &6.73 $\pm$ 0.07 &  0.79 $\pm$ 0.04 & 8.03 $\pm$ 0.01 & -0.65 &  77.8    \\
NGC1140 &  9.45 $\pm$ 0.04 &  9.54 $\pm$ 0.12 &6.55 $\pm$ 0.06 & -0.07 $\pm$ 0.04 & 8.38 $\pm$ 0.01 & -0.68 &  20.0    \\
NGC1569 &  8.71 $\pm$ 0.05 &  8.25 $\pm$ 0.07 &5.46 $\pm$ 0.07 & -0.16 $\pm$ 0.02 & 8.02 $\pm$ 0.02 & -0.04 &   3.1    \\
NGC1705 &  8.19 $\pm$ 0.04 &  7.88 $\pm$ 0.05 &4.75 $\pm$ 0.07 & -1.27 $\pm$ 0.02 & 8.27 $\pm$ 0.11 &  0.24 &   5.1    \\
NGC2366 &  8.19 $\pm$ 0.06 &  8.47 $\pm$ 0.03 &4.81 $\pm$ 0.07 & -1.09 $\pm$ 0.02 & 7.70 $\pm$ 0.01 &  0.24 &   3.2    \\
NGC4214 &  8.65 $\pm$ 0.04 &  8.58 $\pm$ 0.07 &5.93 $\pm$ 0.06 & -0.97 $\pm$ 0.03 & 8.26 $\pm$ 0.01 &  0.10 &   2.9    \\
NGC4449 &  9.21 $\pm$ 0.04 &  8.98 $\pm$ 0.07 &6.41 $\pm$ 0.06 & -0.45 $\pm$ 0.03 & 8.20 $\pm$ 0.11 & -0.27 &   4.2    \\
NGC4861 &  8.21 $\pm$ 0.05 &  8.61 $\pm$ 0.03 &5.44 $\pm$ 0.07 & -0.62 $\pm$ 0.01 & 7.89 $\pm$ 0.01 &  0.17 &   7.5    \\
NGC5253 &  8.91 $\pm$ 0.04 &  8.03 $\pm$ 0.02 &5.57 $\pm$ 0.06 & -0.38 $\pm$ 0.06 & 8.25 $\pm$ 0.02 &  0.52 &   4.0    \\
 NGC625 &  8.60 $\pm$ 0.04 &  8.04 $\pm$ 0.08 &5.46 $\pm$ 0.06 & -1.18 $\pm$ 0.05 & 8.22 $\pm$ 0.02 &  0.58 &   3.9    \\
NGC6822 &  8.03 $\pm$ 0.04 &  8.02 $\pm$ 0.05 &4.82 $\pm$ 0.15 & -2.04 $\pm$ 0.04 & 7.96 $\pm$ 0.01 & -0.03 &   0.5    \\
 Pox186 &  7.04 $\pm$ 0.06 &   $<$6.37        &4.65 $\pm$ 0.09 & -1.42 $\pm$ 0.02 & 7.70 $\pm$ 0.01 &\si{\text{...}}&  18.3    \\
  UM133 &  8.31 $\pm$ 0.04 &  8.33 $\pm$ 0.02 &4.80 $\pm$ 0.12 & -1.12 $\pm$ 0.06 & 7.82 $\pm$ 0.01 &  0.29 &  22.7    \\
  UM448 & 10.41 $\pm$ 0.04 &  9.78 $\pm$ 0.12 &7.25 $\pm$ 0.06 &  1.11 $\pm$ 0.01 & 8.32 $\pm$ 0.01 & -0.90 &  87.8    \\
  UM461 &  7.35 $\pm$ 0.05 &  7.86 $\pm$ 0.02 &4.41 $\pm$ 0.17 & -1.35 $\pm$ 0.03 & 7.73 $\pm$ 0.01 & -0.27 &  13.2    \\
VIIZw40 &  7.07 $\pm$ 0.05 &  7.51 $\pm$ 0.04 &3.84 $\pm$ 0.06 & -2.04 $\pm$ 0.04 & 7.66 $\pm$ 0.01 &  0.36 &   4.5    \\
\hline \hline
\end{tabular}
  \begin{tablenotes}
    \item[a] \citet{2014A&A...563A..31R}
    \item[b] Dust masses were determined fitting a single MBB with a fixed $\beta$ = 1.5 emissivity from 100 to 350 $\mu$m.
    \item[c] \citet{2013PASP..125..600M}
  \end{tablenotes}
\end{threeparttable}
\end{table*}

\begin{table*}
\centering
\begin{threeparttable}
\caption{Stellar masses, \hi masses, dust masses, star formation rates, metallicities, H{\sc i} deficiency, and distances of the KINGFISH dwarf galaxy sample.}
\label{tab:King_dwarfs}
\begin{tabular}{lrccccSc}
\hline \hline
    ID   &$\log$($M_{\star}$)$\:\:\:\:$& $\log$($M_{HI}$)\tnote{a}  &  $\log$($M_d$)\tnote{b}     &  $\log$($SFR$)\tnote{c}      &  12 + log(O/H)\tnote{c}  &   \si{$\: \: Def_{HI}$}     &  D\tnote{c}           \\
         &[M$_{\odot}$]$\:\:\:\:\:\:$& [M$_{\odot}$]   & [M$_{\odot}$]    & [M$_{\odot}$ yr$^{-1}$] &          &             & \si{[Mpc]}           \\
 \hline \hline
   HOII &  8.17 $\pm$ 0.06 &  8.62 $\pm$ 0.10 &4.81 $\pm$ 0.07 & -1.44 & 7.72 &  0.04 &   3.0    \\
 DDO053 &  7.12 $\pm$ 0.09 &  7.79 $\pm$ 0.07 &3.94 $\pm$ 0.11 & -2.22 & 7.60 & -0.01 &   3.6    \\
NGC2915 &  8.24 $\pm$ 0.07 &  8.55 $\pm$ 0.07 &4.78 $\pm$ 0.06 & -1.70 & 7.94 & -0.60 &   3.8    \\
    HoI &  7.60 $\pm$ 0.04 &  8.16 $\pm$ 0.07 &4.55 $\pm$ 0.09 & -2.40 & 7.61 &  0.18 &   3.9    \\
NGC3077 &  9.25 $\pm$ 0.02 &  8.94 $\pm$ 0.07 &5.95 $\pm$ 0.06 & -1.03 & 8.69 & -0.37 &   3.8    \\
 M81DwB &  7.14 $\pm$ 0.09 &  7.06 $\pm$ 0.07 &4.22 $\pm$ 0.08 & -3.00 & 7.84 &  0.41 &   3.6    \\
 IC2574 &  8.71 $\pm$ 0.06 &  9.12 $\pm$ 0.07 &5.87 $\pm$ 0.06 & -1.24 & 7.85 & -0.03 &   3.8    \\
NGC4236 &  9.08 $\pm$ 0.05 &  9.43 $\pm$ 0.07 &6.26 $\pm$ 0.06 & -0.89 & 8.17 &  0.06 &   4.4    \\
NGC4625 &  9.01 $\pm$ 0.06 &  9.00 $\pm$ 0.07 &6.21 $\pm$ 0.06 & -1.28 & 8.35 & -0.44 &   9.3    \\
 DDO154 &  7.19 $\pm$ 0.11 &  8.55 $\pm$ 0.07 &      ...       & -2.70 & 7.54 & -0.26 &   4.3    \\
 DDO165 &  7.87 $\pm$ 0.06 &  8.05 $\pm$ 0.06 &      ...       & -2.70 & 7.63 &  0.37 &   4.6    \\
NGC5408 &  8.44 $\pm$ 0.06 &  8.51 $\pm$ 0.07 &4.75 $\pm$ 0.06 & -1.06 & 7.81 & -0.52 &   4.8    \\
\hline \hline
\end{tabular}
  \begin{tablenotes}
    \item[a] \citet{2014A&A...563A..31R}
    \item[b] Dust masses were determined fitting a single MBB with a fixed $\beta$ = 1.5 emissivity from 100 to 350 $\mu$m.
    \item[c] \citet{2011PASP..123.1347K}
  \end{tablenotes}
  \end{threeparttable}
\end{table*}

\newpage

\begin{table*}
\centering
\begin{threeparttable}
\caption{Stellar masses, \hi masses, dust masses, star formation rates, H$_2$ masses, H{\sc i} deficiency and distances of KINGFISH spiral galaxies (from Sa to Sd).}
\label{tab:King_spirals}
\begin{tabular}{lrrcccccSS}
\hline \hline
    ID   &  $\log$($M_{\star}$) $\:\:$    & $\log$($M_{HI}$)\tnote{a}  &  $\log$($M_d$)\tnote{b}  &  $\log$($SFR$)\tnote{c}$\:\:$        &  log($M_{H_2}$)$_{MW}$\tnote{a} &  log($M_{H_2}$)$_{Z}$\tnote{a} &  12 + $\log$(O/H)\tnote{c} & \si{$\: \: Def_{HI}$}      & \si{D\tnote{c}}           \\
         &  [M$_{\odot}$]  $\:\:\:\:$ & [M$_{\odot}$]   & [M$_{\odot}$]    & [M$_{\odot}$ yr$^{-1}$]$\:\:$ &   [M$_{\odot}$]       &   [M$_{\odot}$] &   &       & [Mpc]           \\
 \hline \hline
NGC0337 &  9.97 $\pm$ 0.05 &  9.52 $\pm$ 0.05 &7.18 $\pm$ 0.06 &  0.11 & $<$  8.84 & $<$  9.86 & 8.18 & -0.37 &  19.3    \\
NGC0628 & 10.08 $\pm$ 0.06 &  9.57 $\pm$ 0.07 &7.34 $\pm$ 0.06 & -0.17 &      8.94 &      9.62 & 8.35 & -0.24 &   7.2    \\
NGC0925 &  9.82 $\pm$ 0.06 &  9.66 $\pm$ 0.07 &7.23 $\pm$ 0.06 & -0.27 &      8.79 &      9.67 & 8.25 & -0.18 &   9.1    \\
NGC1097 & 10.85 $\pm$ 0.06 &  9.88 $\pm$ 0.07 &7.83 $\pm$ 0.06 &  0.62 &      8.22 &      8.66 & 8.47 & -0.21 &  14.2    \\
NGC1291 & 10.81 $\pm$ 0.06 &  9.25 $\pm$ 0.07 &7.12 $\pm$ 0.06 & -0.46 &       ... &       ... & 8.52 &  0.26 &  10.4    \\
  IC342 & 10.41 $\pm$ 0.02 &  9.98 $\pm$ 0.07 &7.53 $\pm$ 0.06 &  0.27 &      9.20 &      9.66 & 8.49 & -0.69 &   3.3    \\
NGC1512 & 10.15 $\pm$ 0.06 &  9.87 $\pm$ 0.07 &7.17 $\pm$ 0.06 & -0.44 &       ... &       ... & 8.56 & -0.35 &  11.6    \\
NGC2146 & 10.89 $\pm$ 0.02 &  9.59 $\pm$ 0.10 &7.67 $\pm$ 0.06 &  0.90 &     10.89 &     10.91 & 8.68 & -0.07 &  17.2    \\
NGC2798 & 10.30 $\pm$ 0.08 &  9.33 $\pm$ 0.07 &7.12 $\pm$ 0.06 &  0.53 &      9.51 &     10.21 & 8.34 & -0.06 &  25.8    \\
NGC2841 & 10.85 $\pm$ 0.06 &  9.94 $\pm$ 0.07 &7.75 $\pm$ 0.06 &  0.39 &      9.47 &      9.76 & 8.54 & -0.36 &  14.1    \\
NGC2976 &  9.13 $\pm$ 0.07 &  8.10 $\pm$ 0.07 &6.24 $\pm$ 0.05 & -1.09 &      7.76 &      8.42 & 8.36 &  0.47 &   3.5    \\
NGC3049 &  9.58 $\pm$ 0.06 &  9.08 $\pm$ 0.07 &6.68 $\pm$ 0.06 & -0.21 &      8.26 &      8.57 & 8.53 & -0.09 &  19.2    \\
NGC3190 & 10.58 $\pm$ 0.06 &  8.63 $\pm$ 0.16 &7.27 $\pm$ 0.06 & -0.42 & $<$  8.59 & $<$  8.99 & 8.49 &  0.77 &  19.3    \\
NGC3184 & 10.32 $\pm$ 0.07 &  9.53 $\pm$ 0.07 &7.49 $\pm$ 0.06 & -0.18 &      9.08 &      9.44 & 8.51 & -0.11 &  11.7    \\
NGC3198 & 10.16 $\pm$ 0.07 &  9.84 $\pm$ 0.12 &7.44 $\pm$ 0.06 &  0.00 &      9.11 &      9.81 & 8.34 & -0.23 &  14.1    \\
NGC3351 & 10.28 $\pm$ 0.06 &  9.01 $\pm$ 0.07 &7.14 $\pm$ 0.06 & -0.24 &      8.68 &      8.86 & 8.60 &  0.27 &   9.3    \\
NGC3521 & 10.86 $\pm$ 0.06 &  9.94 $\pm$ 0.07 &7.87 $\pm$ 0.05 &  0.29 &      9.68 &     10.28 & 8.39 & -0.32 &  11.2    \\
NGC3621 & 10.05 $\pm$ 0.06 &  9.84 $\pm$ 0.07 &7.23 $\pm$ 0.06 & -0.29 &       ... &       ... & 8.27 & -0.47 &   6.6    \\
NGC3627 & 10.66 $\pm$ 0.06 &  8.93 $\pm$ 0.07 &7.54 $\pm$ 0.06 &  0.23 &      9.51 &     10.21 & 8.34 &  0.48 &   9.4    \\
NGC3938 & 10.45 $\pm$ 0.06 &  9.90 $\pm$ 0.07 &7.63 $\pm$ 0.06 &  0.25 &      9.64 &     10.18 & 8.42 & -0.42 &  17.9    \\
NGC4254 & 10.60 $\pm$ 0.07 &  9.58 $\pm$ 0.07 &7.79 $\pm$ 0.06 &  0.59 &      9.83 &     10.31 & 8.45 & -0.22 &  14.4    \\
NGC4321 & 10.74 $\pm$ 0.06 &  9.38 $\pm$ 0.07 &7.87 $\pm$ 0.06 &  0.42 &      9.73 &     10.11 & 8.50 &  0.16 &  14.3    \\
NGC4536 & 10.36 $\pm$ 0.06 &  9.24 $\pm$ 0.07 &7.42 $\pm$ 0.06 &  0.34 &      9.24 &     10.20 & 8.21 &  0.32 &  14.5    \\
NGC4559 &  9.64 $\pm$ 0.07 &  9.61 $\pm$ 0.07 &6.88 $\pm$ 0.06 & -0.43 &      8.28 &      9.08 & 8.29 & -0.28 &   7.0    \\
NGC4569 & 10.30 $\pm$ 0.06 &  8.19 $\pm$ 0.07 &7.05 $\pm$ 0.05 & -0.54 &      9.06 &      9.28 & 8.58 &  1.27 &   9.9    \\
NGC4579 & 10.82 $\pm$ 0.06 &  8.74 $\pm$ 0.07 &7.53 $\pm$ 0.05 &  0.04 &      9.28 &      9.58 & 8.54 &  0.74 &  16.4    \\
NGC4594 & 10.97 $\pm$ 0.06 &  8.41 $\pm$ 0.07 &7.18 $\pm$ 0.06 & -0.74 &      8.33 &      8.63 & 8.54 &  0.95 &   9.1    \\
NGC4631 & 10.29 $\pm$ 0.06 &  9.94 $\pm$ 0.07 &7.51 $\pm$ 0.06 &  0.23 &      9.04 &     10.18 & 8.12 & -0.34 &   7.6    \\
NGC4725 & 10.66 $\pm$ 0.06 &  9.56 $\pm$ 0.07 &7.63 $\pm$ 0.06 & -0.36 &      9.33 &     10.01 & 8.35 &  0.09 &  11.9    \\
NGC4736 & 10.33 $\pm$ 0.06 &  8.61 $\pm$ 0.07 &6.72 $\pm$ 0.06 & -0.42 &      8.64 &      9.40 & 8.31 &  0.51 &   4.7    \\
NGC4826 & 10.28 $\pm$ 0.06 &  8.44 $\pm$ 0.07 &6.62 $\pm$ 0.06 & -0.59 &      8.68 &      8.98 & 8.54 &  0.68 &   5.3    \\
NGC5055 & 10.62 $\pm$ 0.06 &  9.75 $\pm$ 0.07 &7.75 $\pm$ 0.06 &  0.02 &      9.44 &     10.02 & 8.40 & -0.25 &   7.9    \\
NGC5398 &  8.73 $\pm$ 0.06 &  8.39 $\pm$ 0.05 &5.74 $\pm$ 0.06 & -1.12 &       ... &       ... & 8.35 &  0.20 &   7.7    \\
NGC5457 & 10.54 $\pm$ 0.06 & 10.06 $\pm$ 0.07 &7.76 $\pm$ 0.06 &  0.37 &       ... &       ... & 8.68 & -0.18 &   6.7    \\
NGC5474 &  9.06 $\pm$ 0.05 &  8.99 $\pm$ 0.11 &6.28 $\pm$ 0.06 & -1.04 & $<$  7.77 & $<$  8.53 & 8.31 & -0.15 &   6.8    \\
NGC5713 & 10.40 $\pm$ 0.07 &  9.74 $\pm$ 0.07 &7.41 $\pm$ 0.05 &  0.40 &      9.53 &     10.43 & 8.24 & -0.54 &  21.4    \\
NGC6946 & 10.62 $\pm$ 0.06 &  9.55 $\pm$ 0.07 &7.76 $\pm$ 0.06 &  0.85 &      9.65 &     10.23 & 8.40 & -0.20 &   6.8    \\
NGC7331 & 10.99 $\pm$ 0.06 &  9.95 $\pm$ 0.07 &8.02 $\pm$ 0.05 &  0.44 &      9.83 &     10.53 & 8.34 & -0.20 &  14.5    \\
NGC7793 &  9.47 $\pm$ 0.06 &  8.94 $\pm$ 0.07 &6.74 $\pm$ 0.06 & -0.59 &       ... &       ... & 8.31 & -0.04 &   3.9    \\
\\
\hline \hline
\end{tabular}
  \begin{tablenotes}
    \item[a] \citet{2014A&A...563A..31R}
    \item[b] Dust masses were determined fitting a single MBB with a fixed $\beta$ = 1.5 emissivity from 100 to 350 $\mu$m.
    \item[c] \citet{2011PASP..123.1347K}
  \end{tablenotes}
\end{threeparttable}
\end{table*}

\newpage

\begin{table*}
\tiny
\centering
\begin{threeparttable}
\caption{Stellar masses, \hi masses, dust masses, star formation rates, H$_2$ masses, H{\sc i} deficiency, and distances of HeViCS BGC galaxies.}
\label{tab:BGC}
\begin{tabular}{lrccrcccrc}
\hline \hline
    ID   &  $\log$($M_{\star}$) $\:\:$    & $\log$($M_{HI}$)\tnote{a}  &  $\log$($M_d$)\tnote{b}     &  $\log$($SFR$)$\:\:$        &   $\log$($M_{H_2}$)$_{MW}$\tnote{c}  & $\log$($M_{H_2}$)$_{Z}$\tnote{c}  &  12 + $\log$(O/H)\tnote{d} & \si{$\: \: \: \: Def_{HI}$}     & D\tnote{a}           \\
         &  [M$_{\odot}$]  $\:\:\:\:$ & [M$_{\odot}$]   & [M$_{\odot}$]    & [M$_{\odot}$ yr$^{-1}$]$\:\:$ &   [M$_{\odot}$]       & [M$_{\odot}$]   &     &      & [Mpc]           \\
 \hline \hline
VCC47 &  9.69 $\pm$ 0.04 & 8.39 $\pm$ 0.05 &6.89$_{-0.06}^{+0.07}$ & -1.00 $\pm$ 0.08 &  ... &  ... &  ...            &  0.64 & 32.0 \\
VCC58 &  9.85 $\pm$ 0.04 & 9.48 $\pm$ 0.05 &7.22$_{-0.06}^{+0.06}$ & -0.53 $\pm$ 0.07 &  ... &  ... &  ...            & -0.10 & 32.0 \\
VCC89 & 10.37 $\pm$ 0.04 & 9.40 $\pm$ 0.05 &7.54$_{-0.06}^{+0.06}$ &  0.29 $\pm$ 0.08 &  ... &  ... & 8.70 $\pm$ 0.01 & -0.09 & 32.0 \\
VCC92 & 10.52 $\pm$ 0.04 & 9.63 $\pm$ 0.05 &7.73$_{-0.06}^{+0.06}$ &  0.09 $\pm$ 0.09 & 9.39 & 9.07 & 8.76 $\pm$ 0.10 &  0.17 & 17.0 \\
VCC97 & 10.23 $\pm$ 0.04 & 9.24 $\pm$ 0.05 &7.43$_{-0.06}^{+0.06}$ & -0.18 $\pm$ 0.09 &  ... &  ... &  ...            & -0.02 & 32.0 \\
VCC120 &  9.95 $\pm$ 0.04 & 9.71 $\pm$ 0.05 &7.33$_{-0.06}^{+0.06}$ &  0.05 $\pm$ 0.07 &  ... &  ... &  ...            & -0.13 & 32.0 \\
VCC131 &  9.13 $\pm$ 0.04 & 8.79 $\pm$ 0.05 &6.56$_{-0.06}^{+0.06}$ & -1.21 $\pm$ 0.07 &  ... &  ... & 8.65 $\pm$ 0.02 &  0.21 & 17.0 \\
VCC145 &  9.63 $\pm$ 0.04 & 9.38 $\pm$ 0.05 &7.09$_{-0.06}^{+0.06}$ & -0.73 $\pm$ 0.05 & 8.52 & 8.59 & 8.57 $\pm$ 0.03 &  0.04 & 17.0 \\
VCC157 & 10.14 $\pm$ 0.04 & 8.91 $\pm$ 0.05 &7.23$_{-0.06}^{+0.06}$ & -0.06 $\pm$ 0.09 & 9.11 & 8.99 & 8.71 $\pm$ 0.10 &  0.30 & 17.0 \\
VCC167 & 10.81 $\pm$ 0.04 & 9.25 $\pm$ 0.05 &7.70$_{-0.06}^{+0.06}$ & -0.12 $\pm$ 0.06 & 9.21 & 8.84 &  ...            &  0.51 & 17.0 \\
VCC187 &  9.41 $\pm$ 0.04 & 9.03 $\pm$ 0.05 &6.89$_{-0.06}^{+0.06}$ & -0.86 $\pm$ 0.09 & 8.06 & 8.29 & 8.42 $\pm$ 0.30 &  0.17 & 17.0 \\
VCC221 &  9.87 $\pm$ 0.04 & 8.81 $\pm$ 0.05 &6.96$_{-0.06}^{+0.06}$ & -0.26 $\pm$ 0.10 &  ... &  ... & 8.67 $\pm$ 0.07 &  0.35 & 32.0 \\
VCC226 &  9.96 $\pm$ 0.04 & 8.32 $\pm$ 0.05 &7.05$_{-0.06}^{+0.06}$ & -0.44 $\pm$ 0.09 & 9.01 & 8.98 &  ...            &  0.54 & 17.0 \\
VCC234 & 10.37 $\pm$ 0.04 & 8.45 $\pm$ 0.05 &7.15$_{-0.06}^{+0.06}$ & -0.56 $\pm$ 0.03 &  ... &  ... &  ...            &  1.10 & 32.0 \\
VCC267 &  9.28 $\pm$ 0.04 & 9.02 $\pm$ 0.05 &6.79$_{-0.07}^{+0.07}$ & -1.09 $\pm$ 0.05 &  ... &  ... &  ...            &  0.02 & 23.0 \\
VCC289 &  9.26 $\pm$ 0.04 & 9.00 $\pm$ 0.05 &6.50$_{-0.08}^{+0.08}$ & -1.14 $\pm$ 0.03 &  ... &  ... &  ...            &  0.15 & 32.0 \\
VCC307 & 10.68 $\pm$ 0.04 & 9.65 $\pm$ 0.05 &7.89$_{-0.06}^{+0.06}$ &  0.71 $\pm$ 0.09 & 10.0 & 9.73 & 8.73 $\pm$ 0.10 & -0.12 & 17.0 \\
VCC341 & 10.24 $\pm$ 0.04 & 7.63 $\pm$ 0.05 &6.66$_{-0.06}^{+0.06}$ & -1.27 $\pm$ 0.06 &  ... &  ... &  ...            &  1.75 & 23.0 \\
VCC362 & 10.14 $\pm$ 0.04 & 7.68 $\pm$ 0.05 &6.93$_{-0.06}^{+0.06}$ &...$\:\:\:\:\:\:\:\:\:$&  ... &  ... &  ...            &  1.60 & 32.0 \\
VCC382 & 10.43 $\pm$ 0.04 & 9.54 $\pm$ 0.05 &7.57$_{-0.06}^{+0.06}$ &  0.63 $\pm$ 0.09 &  ... &  ... &  ...            & -0.30 & 32.0 \\
VCC404 &  9.01 $\pm$ 0.04 & 8.39 $\pm$ 0.05 &6.36$_{-0.06}^{+0.06}$ & -1.21 $\pm$ 0.11 &  ... &  ... &  ...            &  0.38 & 17.0 \\
VCC449 &  9.47 $\pm$ 0.04 & 9.02 $\pm$ 0.05 &6.68$_{-0.06}^{+0.06}$ & -1.19 $\pm$ 0.08 &  ... &  ... &  ...            &  0.30 & 17.0 \\
VCC465 &  9.53 $\pm$ 0.04 & 9.22 $\pm$ 0.05 &6.85$_{-0.06}^{+0.06}$ & -0.31 $\pm$ 0.05 &  ... &  ... & 8.50 $\pm$ 0.10 &  0.04 & 17.0 \\
VCC483 & 10.10 $\pm$ 0.04 & 8.94 $\pm$ 0.05 &7.29$_{-0.06}^{+0.06}$ & -0.15 $\pm$ 0.07 & 9.16 & 9.07 &  ...            &  0.27 & 17.0 \\
VCC491 &  9.29 $\pm$ 0.04 & 9.04 $\pm$ 0.05 &6.54$_{-0.06}^{+0.06}$ & -0.35 $\pm$ 0.05 &  ... &  ... & 8.35 $\pm$ 0.04 & -0.19 & 17.0 \\
VCC497 & 10.32 $\pm$ 0.04 & 9.24 $\pm$ 0.05 &7.58$_{-0.06}^{+0.06}$ & -0.25 $\pm$ 0.09 & 9.29 & 9.14 &  ...            &  0.34 & 17.0 \\
VCC508 & 10.70 $\pm$ 0.04 & 9.68 $\pm$ 0.05 &7.80$_{-0.06}^{+0.06}$ &  0.53 $\pm$ 0.05 & 9.94 & 9.63 & 8.76 $\pm$ 0.10 & -0.11 & 17.0 \\
VCC524 & 10.21 $\pm$ 0.04 & 8.15 $\pm$ 0.05 &7.16$_{-0.06}^{+0.06}$ & -0.58 $\pm$ 0.10 &  ... &  ... &  ...            &  1.29 & 23.0 \\
VCC534 &  9.70 $\pm$ 0.04 & 7.64 $\pm$ 0.05 &6.65$_{-0.06}^{+0.06}$ & -0.93 $\pm$ 0.08 &  ... &  ... &  ...            &  1.40 & 23.0 \\
VCC559 &  9.89 $\pm$ 0.04 & 8.08 $\pm$ 0.05 &6.73$_{-0.06}^{+0.06}$ & -0.73 $\pm$ 0.12 & 9.10 & 9.09 &  ...            &  1.34 & 17.0 \\
VCC567 &  9.16 $\pm$ 0.04 & 8.81 $\pm$ 0.05 &6.62$_{-0.06}^{+0.07}$ & -1.12 $\pm$ 0.05 &  ... &  ... &  ...            &  0.28 & 23.0 \\
VCC570 & 10.01 $\pm$ 0.04 & 8.02 $\pm$ 0.05 &6.84$_{-0.06}^{+0.06}$ & -0.82 $\pm$ 0.09 &  ... &  ... &  ...            &  1.40 & 17.0 \\
VCC576 &  9.99 $\pm$ 0.04 & 9.01 $\pm$ 0.05 &7.18$_{-0.06}^{+0.06}$ & -0.49 $\pm$ 0.10 & 8.48 & 8.45 &  ...            &  0.16 & 23.0 \\
VCC596 & 10.82 $\pm$ 0.04 & 9.44 $\pm$ 0.05 &7.99$_{-0.06}^{+0.06}$ &  0.57 $\pm$ 0.08 & 9.91 & 9.55 & 8.75 $\pm$ 0.10 &  0.32 & 17.0 \\
VCC613 &  9.98 $\pm$ 0.04 & 8.73 $\pm$ 0.05 &6.55$_{-0.06}^{+0.06}$ & -1.04 $\pm$ 0.06 &  ... &  ... &  ...            &  0.47 & 17.0 \\
VCC630 &  9.61 $\pm$ 0.04 & 8.61 $\pm$ 0.05 &6.90$_{-0.06}^{+0.06}$ & -0.90 $\pm$ 0.10 & 8.61 & 8.72 &  ...            &  0.88 & 17.0 \\
VCC656 & 10.11 $\pm$ 0.04 & 8.78 $\pm$ 0.05 &7.06$_{-0.06}^{+0.06}$ & -0.50 $\pm$ 0.08 &  ... &  ... &  ...            &  0.39 & 23.0 \\
VCC664 &  8.78 $\pm$ 0.04 & 8.45 $\pm$ 0.05 &6.25$_{-0.06}^{+0.07}$ & -0.72 $\pm$ 0.04 &  ... &  ... & 8.32 $\pm$ 0.10 &  0.57 & 17.0 \\
VCC667 &  9.33 $\pm$ 0.04 & 8.40 $\pm$ 0.05 &6.68$_{-0.06}^{+0.06}$ & -1.16 $\pm$ 0.08 &  ... &  ... & 8.39 $\pm$ 0.01 &  0.55 & 23.0 \\
VCC692 &  9.36 $\pm$ 0.04 & 8.52 $\pm$ 0.05 &6.50$_{-0.06}^{+0.06}$ & -1.03 $\pm$ 0.06 &  ... &  ... & 8.62 $\pm$ 0.06 &  0.56 & 17.0 \\
VCC785 & 10.05 $\pm$ 0.04 & 8.84 $\pm$ 0.05 &6.98$_{-0.06}^{+0.06}$ & -0.89 $\pm$ 0.05 &  ... &  ... &  ...            &  0.28 & 17.0 \\
VCC787 &  9.26 $\pm$ 0.04 & 8.83 $\pm$ 0.05 &6.55$_{-0.06}^{+0.06}$ & -0.79 $\pm$ 0.05 &  ... &  ... & 8.58 $\pm$ 0.10 &  0.16 & 23.0 \\
VCC792 & 10.26 $\pm$ 0.04 & 8.47 $\pm$ 0.05 &7.32$_{-0.06}^{+0.06}$ & -0.63 $\pm$ 0.10 & 8.84 & 8.66 &  ...            &  0.91 & 23.0 \\
VCC827 &  9.87 $\pm$ 0.04 & 9.45 $\pm$ 0.05 &7.22$_{-0.06}^{+0.06}$ & -0.40 $\pm$ 0.09 &  ... &  ... & 8.40 $\pm$ 0.10 & -0.06 & 23.0 \\
VCC836 & 10.26 $\pm$ 0.04 & 8.65 $\pm$ 0.05 &6.97$_{-0.06}^{+0.06}$ &  0.34 $\pm$ 0.11 & 8.78 & 8.64 &  ...            &  0.76 & 17.0 \\
VCC849 &  9.47 $\pm$ 0.04 & 8.90 $\pm$ 0.05 &6.80$_{-0.06}^{+0.06}$ & -0.66 $\pm$ 0.06 &  ... &  ... & 8.43 $\pm$ 0.20 &  0.20 & 23.0 \\
VCC851 &  9.47 $\pm$ 0.04 & 8.71 $\pm$ 0.05 &6.80$_{-0.06}^{+0.06}$ & -0.80 $\pm$ 0.07 &  ... &  ... & 8.55 $\pm$ 0.01 &  0.38 & 23.0 \\
VCC873 & 10.10 $\pm$ 0.04 & 8.66 $\pm$ 0.05 &7.34$_{-0.06}^{+0.06}$ & -0.20 $\pm$ 0.11 & 9.31 & 9.23 & 8.68 $\pm$ 0.10 &  0.61 & 17.0 \\
VCC905 &  9.33 $\pm$ 0.04 & 9.23 $\pm$ 0.05 &6.73$_{-0.06}^{+0.06}$ & -0.94 $\pm$ 0.07 &  ... &  ... & 8.59 $\pm$ 0.03 &  0.01 & 23.0 \\
VCC912 &  9.62 $\pm$ 0.04 & 8.29 $\pm$ 0.05 &6.65$_{-0.06}^{+0.06}$ & -0.84 $\pm$ 0.10 &  ... &  ... & 8.68 $\pm$ 0.10 &  0.80 & 17.0 \\
VCC921 &  9.54 $\pm$ 0.04 & 8.33 $\pm$ 0.05 &6.48$_{-0.06}^{+0.06}$ & -0.37 $\pm$ 0.09 & 8.38 & 8.59 & 8.71 $\pm$ 0.20 &  0.50 & 17.0 \\
VCC938 &  9.46 $\pm$ 0.04 & 8.41 $\pm$ 0.05 &6.62$_{-0.06}^{+0.06}$ & -0.80 $\pm$ 0.07 & 8.27 & 8.44 & 8.65 $\pm$ 0.20 &  0.51 & 17.0 \\
VCC939 &  9.58 $\pm$ 0.04 & 9.34 $\pm$ 0.05 &7.07$_{-0.07}^{+0.06}$ & -0.76 $\pm$ 0.05 & 8.15 & 8.26 & 8.49 $\pm$ 0.30 &  0.02 & 23.0 \\
VCC971 &  9.24 $\pm$ 0.04 & 9.20 $\pm$ 0.05 &6.60$_{-0.06}^{+0.07}$ & -0.82 $\pm$ 0.05 & 9.04 & 9.28 & 8.30 $\pm$ 0.04 &  0.09 & 23.0 \\
VCC975 &  9.17 $\pm$ 0.04 & 9.43 $\pm$ 0.05 &7.06$_{-0.08}^{+0.10}$ & -0.91 $\pm$ 0.05 &  ... &  ... &  ...            &  0.01 & 23.0 \\
VCC979 & 10.09 $\pm$ 0.04 & 8.59 $\pm$ 0.05 &6.66$_{-0.06}^{+0.06}$ & -0.20 $\pm$ 0.09 & 8.86 & 8.79 &  ...            &  0.91 & 23.0 \\
VCC1043 & 10.56 $\pm$ 0.04 & 8.68 $\pm$ 0.05 &7.13$_{-0.06}^{+0.06}$ & -0.42 $\pm$ 0.06 & 9.01 & 8.76 &  ...            &  1.02 & 17.0 \\
VCC1118 &  9.72 $\pm$ 0.04 & 8.50 $\pm$ 0.05 &6.68$_{-0.06}^{+0.06}$ & -0.47 $\pm$ 0.08 & 8.64 & 8.68 & 8.66 $\pm$ 0.10 &  0.53 & 23.0 \\
VCC1190 & 10.44 $\pm$ 0.04 & 7.64 $\pm$ 0.05 &6.71$_{-0.06}^{+0.06}$ & -0.72 $\pm$ 0.08 &  ... &  ... &  ...            &  1.86 & 23.0 \\
VCC1193 &  8.89 $\pm$ 0.04 & 8.16 $\pm$ 0.05 &6.25$_{-0.06}^{+0.07}$ & -1.26 $\pm$ 0.05 &  ... &  ... & 8.47 $\pm$ 0.10 &  0.40 & 17.0 \\
VCC1205 &  9.44 $\pm$ 0.04 & 8.64 $\pm$ 0.05 &6.51$_{-0.06}^{+0.06}$ & -0.79 $\pm$ 0.10 & 8.35 & 8.52 & 8.52 $\pm$ 0.10 &  0.18 & 17.0 \\
VCC1330 &  9.79 $\pm$ 0.04 & 7.94 $\pm$ 0.05 &6.49$_{-0.06}^{+0.06}$ & -1.28 $\pm$ 0.06 & 8.31 & 8.33 &  ...            &  0.91 & 17.0 \\
VCC1450 &  9.26 $\pm$ 0.04 & 8.39 $\pm$ 0.05 &6.63$_{-0.06}^{+0.06}$ & -0.50 $\pm$ 0.04 & 7.89 & 8.16 & 8.60 $\pm$ 0.10 &  0.63 & 17.0 \\
VCC1508 &  9.58 $\pm$ 0.04 & 9.43 $\pm$ 0.05 &6.91$_{-0.06}^{+0.06}$ & -0.21 $\pm$ 0.08 & 8.27 & 8.37 & 8.37 $\pm$ 0.20 & -0.22 & 17.0 \\
VCC1516 &  9.57 $\pm$ 0.04 & 8.63 $\pm$ 0.05 &6.77$_{-0.06}^{+0.06}$ & -0.78 $\pm$ 0.08 & 8.90 & 9.02 & 8.51 $\pm$ 0.40 &  0.65 & 17.0 \\
VCC1552 &  9.91 $\pm$ 0.04 & 7.16 $\pm$ 0.05 &6.47$_{-0.06}^{+0.06}$ & -1.38 $\pm$ 0.09 &  ... &  ... &  ...            &  2.15 & 17.0 \\
VCC1555 & 10.25 $\pm$ 0.04 & 9.59 $\pm$ 0.05 &7.78$_{-0.06}^{+0.06}$ &  0.22 $\pm$ 0.08 & 9.55 & 9.33 & 8.77 $\pm$ 0.10 &  0.11 & 17.0 \\
VCC1673 & 10.07 $\pm$ 0.04 & 8.69 $\pm$ 0.05 &7.21$_{-0.09}^{+0.11}$ & -0.27 $\pm$ 0.10 & 9.76 & 9.67 & 8.65 $\pm$ 0.10 &  0.40 & 17.0 \\
VCC1676 & 10.43 $\pm$ 0.04 & 8.99 $\pm$ 0.05 &7.59$_{-0.07}^{+0.09}$ &  0.20 $\pm$ 0.11 & 9.43 & 9.23 & 8.77 $\pm$ 0.20 &  0.43 & 17.0 \\
\hline \hline
\end{tabular}
  \begin{tablenotes}
    \item[a] GOLDMine \citep{2003A&A...400..451G,2014arXiv1401.8123G}
    \item[b] Dust masses were determined fitting a single MBB with a fixed $\beta$ = 1.5 emissivity from 100 to 350 $\mu$m.
    \item[c] \citet{2014A&A...564A..65B}
    \item[d] \citet{2013A&A...550A.115H}
  \end{tablenotes}
\end{threeparttable}
\end{table*}
\normalsize

\begin{figure*}
   \centering
 \includegraphics[bb=95 200 445 690,width=12.5cm]{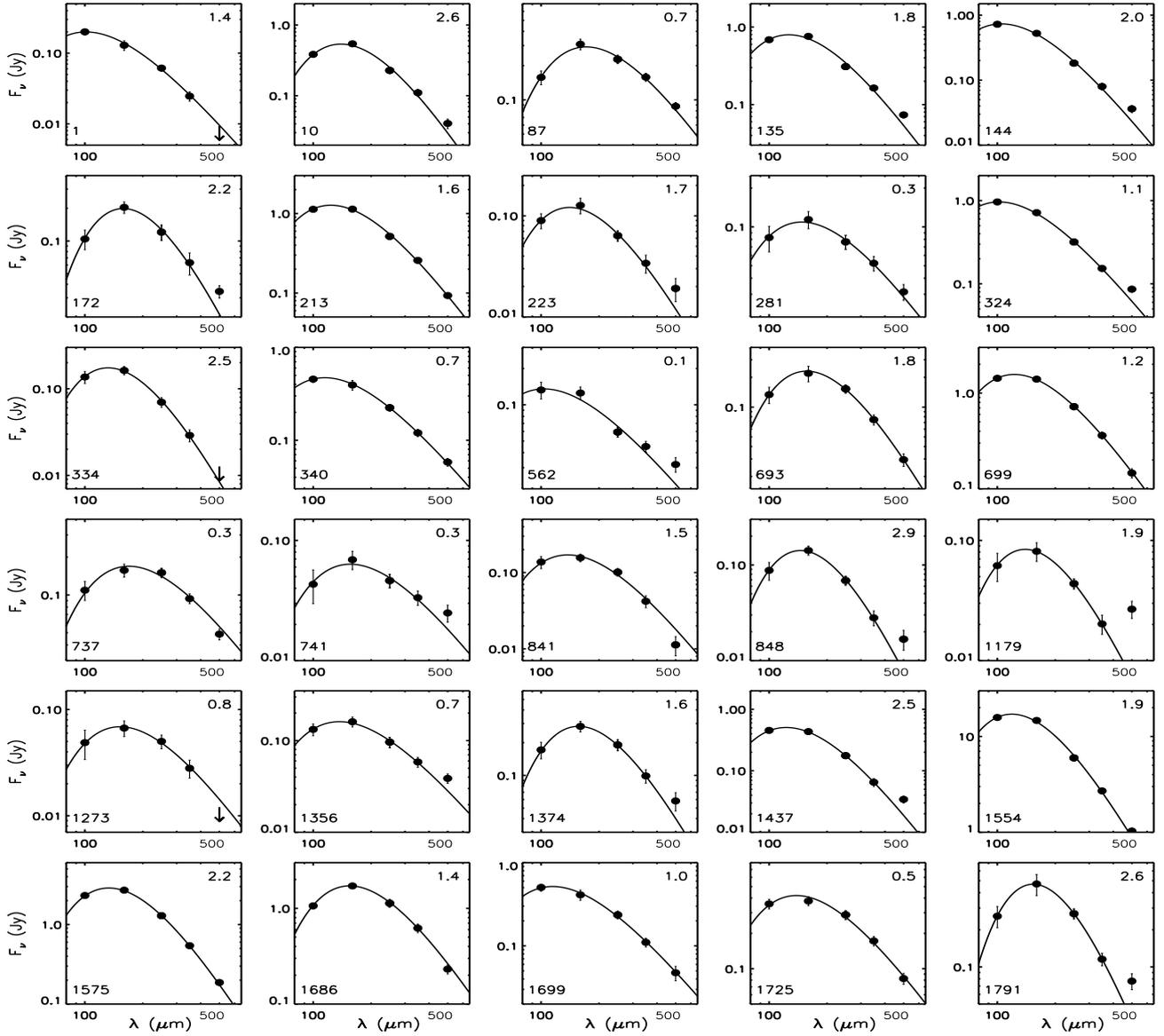}
   \caption{Free-$\beta$ MBB fitting for 30 Virgo SFDs detected in four bands
   (100, 160, 250, 350 $\mu$m).
   The best-fit emissivity index  is displayed at the upper-right corner,
   and the VCC catalogue ID is given at the lower-left corner of each plot.}
   \label{SEDfit_betafree}%
    \end{figure*}

\begin{figure*}
   \centering
 \includegraphics[bb=60 80 530 750,width=15cm]{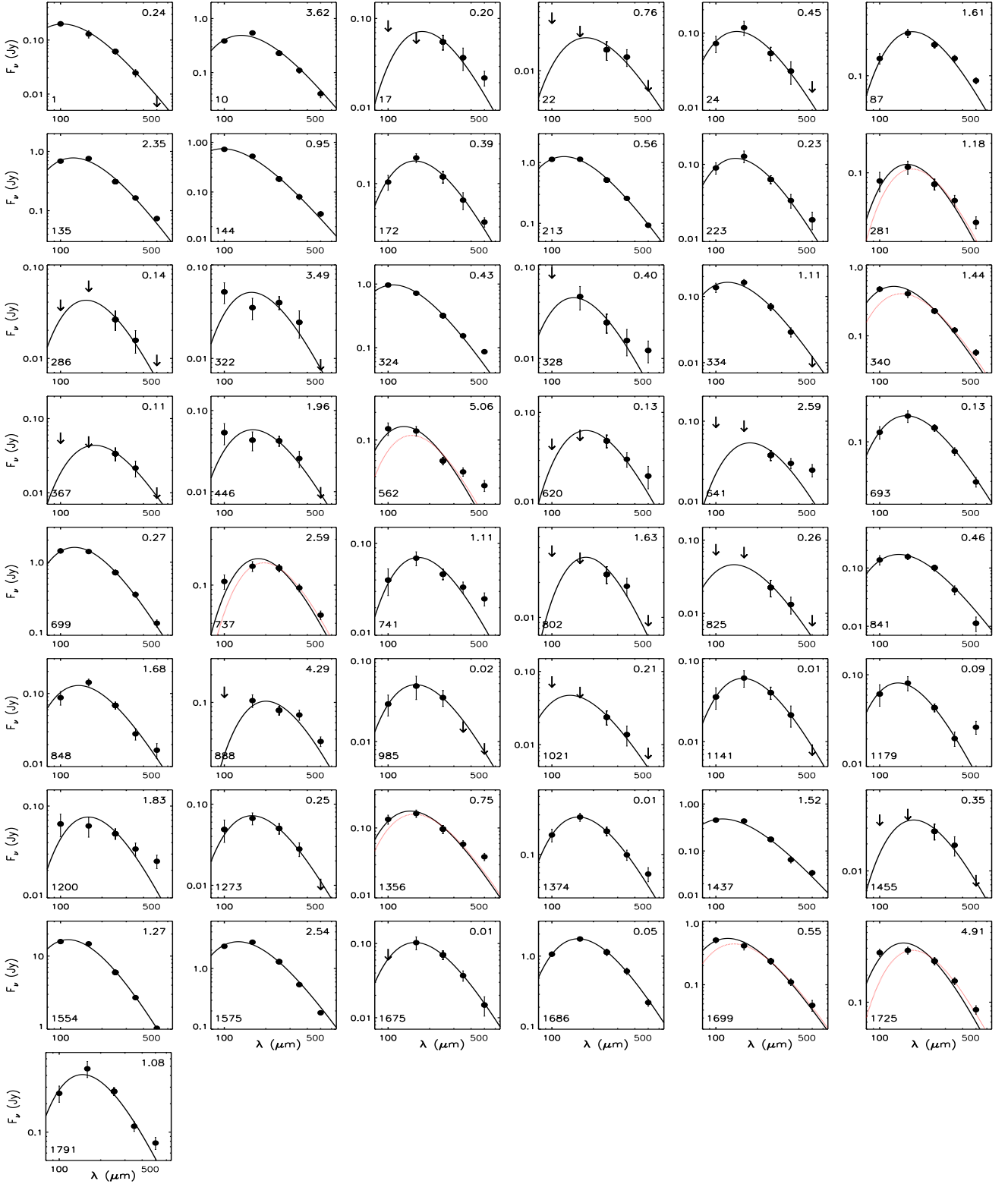} 
   \caption{Fixed-$\beta$ MBB fitting ($\beta = 1.5$) for the whole sample of Virgo dwarfs detected with {\em Herschel} (black solid line).
   The reduced $\chi^2$ value of the fitting is displayed at the upper-right corner of each plot.
   The VCC catalogue ID is given at the lower-left corner of each plot. The red dotted lines correspond to the MBB obtained by fitting only three points of the SED (160-350 $\mu$m) instead of four points (100-350 $\mu$m). The three-point fits are shown only for those galaxies where the difference between the dust masses derived with the two methods, $\Delta$$M_d = M_d^{160-350} - M_d^{100-350}$, is larger than $\sim$0.1 dex, the mean uncertainty on $M_d^{100-350}$.}
   \label{fig:SEDfit_betafixed}%
    \end{figure*}

\clearpage

\end{appendix}

\end{document}